\documentclass[twocolumn,american,floatfix, table, dvipsnames]{revtex4-1}
\usepackage[T1]{fontenc}
\usepackage[latin9]{inputenc}
\usepackage{xcolor}
\usepackage{babel}
\usepackage{amsmath}
\usepackage{amssymb}
\usepackage{graphicx}
\usepackage{esint}
\usepackage[unicode=true,
 bookmarks=false,
 breaklinks=true,pdfborder={0 0 1},backref=false,colorlinks=true]
 {hyperref}
\hypersetup{
 pdfcreator={},pdfproducer={LaTeX with hyperref},linkcolor=blue,anchorcolor=blue,citecolor=blue,filecolor=red,menucolor=red,pagecolor=red,urlcolor=blue,pdfstartview=FitV,pdfhighlight=/I,pdfpagelayout=TwoColumns,hypertexnames=true}

\makeatletter
\usepackage{lmodern}
\usepackage{bbm}
\usepackage[justification=centerlast]{caption}
\usepackage{subcaption}
\usepackage{floatrow}
\newfloatcommand{capbtabbox}{table}[][\FBwidth]
\newfloatcommand{capbtabboxL}{table}[][1.6\FBwidth]
\usepackage{color}
\usepackage[figure,table]{hypcap}

\makeatother

\begin{document}
\title{Phonon-induced rotation of the electronic nematic director in superconducting
Bi$_{2}$Se$_{3}$}
\author{Matthias Hecker }
\affiliation{School of Physics and Astronomy, University of Minnesota, Minneapolis
55455 MN, USA}
\author{Rafael M. Fernandes}
\affiliation{School of Physics and Astronomy, University of Minnesota, Minneapolis
55455 MN, USA}
\date{\today }
\begin{abstract}
The doped topological insulator $A_{x}\mathrm{Bi_{2}Se_{3}}$, with
$A=\{\mathrm{Cu},\mathrm{Sr},\mathrm{Nb}\}$, becomes a nematic superconductor
below $T_{c}\sim3-4\,\mathrm{K}$. The associated electronic nematic
director is described by an angle $\alpha$ and is experimentally
manifested in the elliptical shape of the in-plane critical magnetic
field $H_{c2}$. Because of the threefold rotational symmetry of the
lattice, $\alpha$ is expected to align with one of three high-symmetry
directions corresponding to the in-plane nearest-neighbor bonds, consistent
with a $Z_{3}$-Potts nematic transition. Here, we show that the nematic
coupling to the acoustic phonons, which makes the nematic correlation
length tend to diverge along certain directions only, can fundamentally
alter this phenomenology in trigonal lattices. Compared to hexagonal
lattices, the former possesses a sixth independent elastic constant
$c_{14}$ due to the fact that the in-plane shear strain doublet $(\epsilon_{xx}-\epsilon_{yy},-2\epsilon_{xy})$
and the out-of-plane shear strain doublet $(2\epsilon_{yz},-2\epsilon_{xz})$
transform as the same irreducible representation. We find that, when
$c_{14}$ overcomes a threshold value, which is expected to be the
case in doped $\mathrm{Bi_{2}Se_{3}}$, the nematic director $\alpha$
unlocks from the high-symmetry directions due to the competition between
the quadratic phonon-mediated interaction and the cubic nematic anharmonicity.
This implies the breaking of the residual in-plane twofold rotational
symmetry ($C_{2x}$), resulting in a triclinic phase. We discuss the
implications of these findings to the structure of nematic domains,
to the shape of the in-plane $H_{c2}$ in $A_{x}\mathrm{Bi_{2}Se_{3}}$,
and to presence of nodes inside the superconducting state.
\end{abstract}
\maketitle

\section{Introduction}

In nematic superconductors, the superconducting transition is accompanied
by the breaking of a symmetry of the crystalline lattice. As a result,
a nematic pairing state is manifested by substantial anisotropies
in thermodynamic quantities such as the upper-critical field ($H_{c2}$),
the penetration depth, and the thermal conductivity. Quite generally,
a nematic superconducting state requires a multi-component complex
order parameter $\boldsymbol{\Delta}=(\Delta_{1},\Delta_{2},\dots)^{T}$.
In one scenario, which assumes some degree of fine tuning, the components
transform as different one-dimensional irreducible representations
(IR) of the point group that order at very close transition temperatures
($T_{c}$) \citep{Chichinadze2020,Wang2021}. This would be the case,
for instance, of an $s+d_{x^{2}-y^{2}}$ state in a tetragonal lattice,
which lowers the symmetry of the system to orthorhombic \citep{Fernandes_Millis,Livanas2015,Fradkin2014}.
Another scenario, which does not require fine tuning, corresponds
to the case in which $\boldsymbol{\Delta}$ transforms as a multi-dimensional
IR \citep{Fu2014,Hecker2018}. An example is the $d_{xy}+d_{x^{2}-y^{2}}$
state in a hexagonal lattice, which breaks the sixfold rotational
symmetry of the crystal \citep{SZLin2018,Venderbos_Fernandes,Kozii2019,Scheurer2020}. 

Several materials have been found to display signatures of nematic
superconductivity, including the family of doped topological insulators
$A_{x}\mathrm{Bi_{2}Se_{3}}$, with dopants $A=\{\mathrm{Cu},\mathrm{Sr},\mathrm{Nb}\}$
\citep{Matano2015,Pan2016,Asaba2016}; few-layer transition-metal
dichalcogenide $\mathrm{Nb_{2}Se}$ \citep{Hamill2021,Cho2020}; twisted
bilayer graphene \citep{Cao2021}; and iron-pnictide superconductors
\citep{Li2017,Borisenko2020}. In this paper, we focus on the $A_{x}\mathrm{Bi_{2}Se_{3}}$
compounds, which form a trigonal lattice with point group $\mathsf{D_{3d}}$.
The fact that the superconducting state breaks the threefold rotational
symmetry ($C_{3z}$) has been well-established by measurements of
the upper critical field $H_{c2}$, the NMR Knight shift, the resistivity,
the magnetic torque, the angle-resolved specific heat, the thermal
expansion, and by scanning tunneling spectroscopy \citep{Du2016,Yonezawa2017,Pan2016,Tao2018,Kuntsevich2018,Shen2017,Willa2018,Sun2019,Kostylev2020,Nikitin2016,Smylie2017,Smylie2017a,Asaba2016,Matano2015,Cho2019,Kuntsevich2019,mi2021,Kawai2020}.
The main candidate for this pairing state is the odd-parity ``p-wave''
$E_{u}$ state, parametrized here by the two-component order parameter
$\boldsymbol{\Delta}=(\Delta_{1},\Delta_{2})^{T}$ \citep{Fu2014,Das2020,Smylie2017a,Kriener2011a,Zyuzin2017,Kuntsevich2019}.
The nematic ground state corresponds to $\boldsymbol{\Delta}=\Delta_{0}(\cos\gamma,\,\sin\gamma)^{T}$
with the directions $\gamma\in[0,\pi)$ restricted
to two sets of values, each with three possible
directions \citep{Venderbos2016,Hecker2018,How2019}.
While one set, $\gamma=\frac{\pi}{6}\{0,2,4\}$, results in a fully
gapped state, the other set, $\gamma=\frac{\pi}{6}\{1,3,5\}$, generates
point nodes. Which of the two sets is realized is still subject of
experimental studies that aim at identifying whether or not nodal
quasiparticles are present \citep{Das2020,Kriener2011a,Smylie2017,Smylie2017a}. 

Using the product decomposition $E_{u}\otimes E_{u}=A_{1g}\oplus A_{2g}\oplus E_{g}$,
one identifies two possible real-valued bilinear combinations of $\boldsymbol{\Delta}$
that transform non-trivially under the point-group $\mathsf{D_{3d}}$:
the scalar $\Phi^{A_{2g}}=\boldsymbol{\Delta}^{\dagger}\tau^{y}\boldsymbol{\Delta}$,
which breaks time-reversal symmetry and vanishes in the nematic ground
state, and the two-component order parameter:
\begin{figure*}[t]
\centering{}\ffigbox[][]{\includegraphics[scale=0.42]{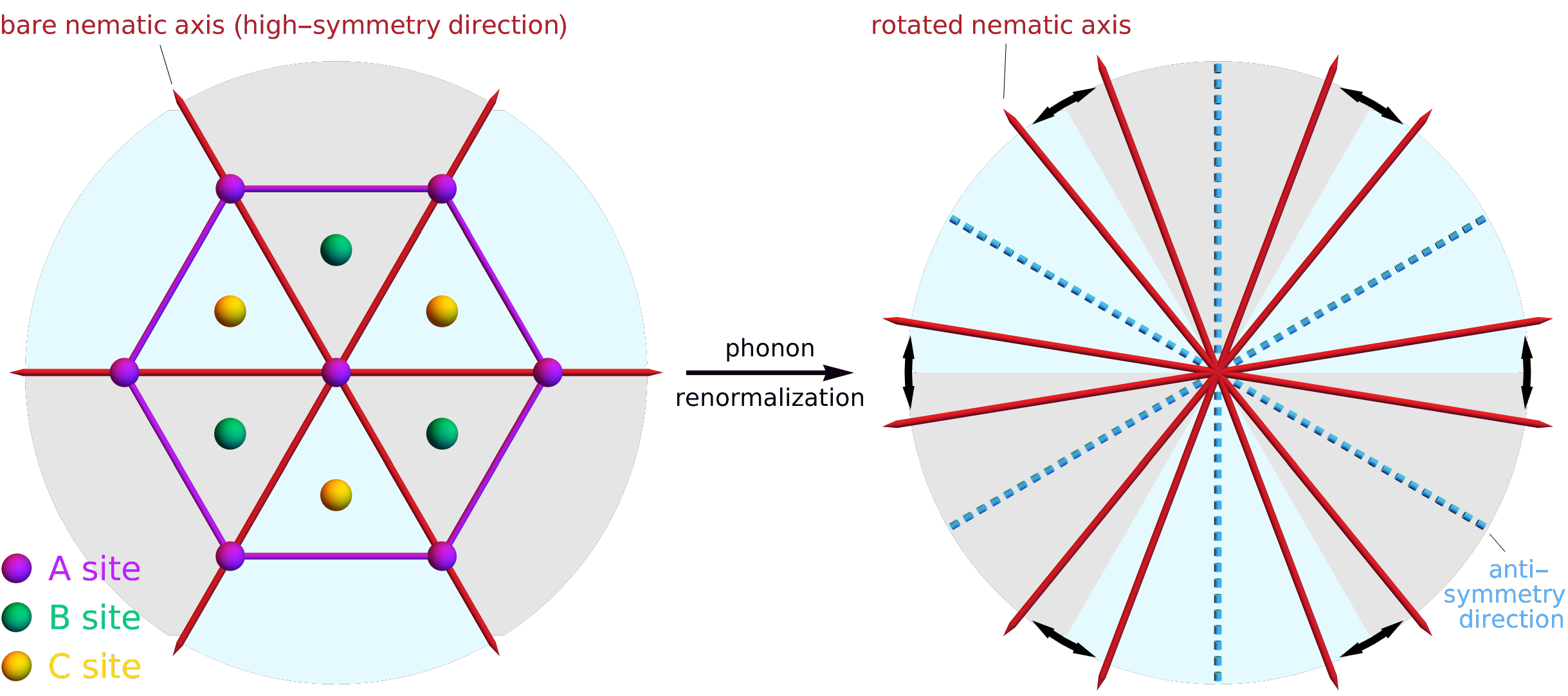}}  
{\caption{Schematic representation of the electronic nematic director in doped $\mathrm{Bi_{2}Se_{3}}$ compounds. The left panel shows the threefold degenerate nematic directors aligned with the high-symmetry directions of the crystal, superimposed with the unit cell of $\mathrm{Bi_2 Se_3}$. The latter displays a characteristic A-B-C stacking pattern for which, along a particular path along the $z$-direction, the A, B, C lattice sites are either occupied by $\mathrm{Bi}$ or $\mathrm{Se}$ atoms, see e.g. Refs.\citep{Matano2015,Hecker2020}. When the coupling to acoustic phonons overcomes a threshold value, two effects occur (right panel): (i) a rotation of the nematic director away from the high-symmetry directions, and (ii) a splitting of the original director into two, which move towards six ``anti-symmetry directions''. As a result, the number of non-identical directors doubles from three to six, and the system loses its residual in-plane twofold rotational symmetry $C_{2x}$ inside the nematic phase.} \label{fig:rot_nem_axis}}     
\end{figure*}
\begin{align}
\boldsymbol{\Phi}^{E_{g}} & =\left(\!\!\begin{array}{c}
\boldsymbol{\Delta}^{\dagger}\tau^{z}\boldsymbol{\Delta}\\
-\boldsymbol{\Delta}^{\dagger}\tau^{x}\boldsymbol{\Delta}
\end{array}\!\!\right)=\left(\!\!\begin{array}{c}
\Phi^{E_{g},1}\\
\Phi^{E_{g},2}
\end{array}\!\!\right)=|\boldsymbol{\Phi}^{E_{g}}|\left(\!\begin{array}{c}
\cos\alpha\\
\sin\alpha
\end{array}\!\right),\label{eq:C_Eg}
\end{align}
which breaks the $C_{3z}$ symmetry of the lattice and is non-zero
inside the nematic ground state. Here, the nematic
director $\alpha$ is related to $\gamma$ above via $\gamma=-\alpha/2$.
Interestingly, fluctuations can cause this bilinear to undergo its
own phase transition before the onset of superconductivity \citep{Fernandes_review},
resulting in a narrow sliver of vestigial nematicity above $T_{c}$,
as observed experimentally in $A_{x}\mathrm{Bi_{2}Se_{3}}$ \citep{Hecker2018,Cho2019,Sun2019}. 

The bilinear $\boldsymbol{\Phi}^{E_{g}}$ is thus identified as a
nematic order parameter, whose ``orientation''\textemdash encoded
in the nematic director $\alpha\in[0,2\pi)$\textemdash is directly
manifested in properties such as the anisotropy of the in-plane $H_{c2}$
or the direction of elongation (or contraction) of the crystallographic
unit cell inside the monoclinic phase. Importantly, the symmetries
of the lattice render $\boldsymbol{\Phi}^{E_{g}}$ a $Z_{3}$ (i.e.
$3$-state) Potts variable \citep{Hecker2018,Xu2020,Fernandes2020,Jin2021}.
Consequently, in three dimensions, it is expected to undergo a first-order
transition into a threefold degenerate ground state where the director
$\alpha=\alpha_{s}$ is aligned with one of the high-symmetry directions
of the lattice, $\alpha_{s}\in\{0,2,4\}\frac{\pi}{3}$ or $\alpha_{s}\in\{1,3,5\}\frac{\pi}{3}$,
as illustrated in Fig. \ref{fig:rot_nem_axis} (left panel). However,
experiments have observed an apparent discrepancy between $\alpha$
and $\alpha_{s}$ \citep{Du2016,Kuntsevich2018,Tao2018}.

In this paper, we revisit the issue of the orientation of the electronic
nematic director in trigonal lattices by considering the coupling
to the elastic degrees of freedom. It is well known that when an order
parameter couples bilinearly to strain, as it is always the case for
nematic order, the low-energy elastic fluctuations (i.e. the acoustic
phonons) mediate long-range order-parameter interactions \citep{Cowley1976,Folk1976,Chou1996}.
This results in the emergence of non-analytical terms in the susceptibility,
implying that the order parameter fluctuations are only soft along
certain momentum-space directions \citep{Xu2009,Karahasanovic2016,Paul2017,Carvalho2019,Fernandes2020}.
In the context of electronic nematic phases, this important effect
has been studied in the cases of tetragonal and hexagonal lattices,
where it was shown to promote mean-field behavior at finite temperatures
and to suppress non-Fermi liquid behavior near the putative zero-temperature
transition. We find that the case of trigonal lattices is qualitatively
different, as the nemato-elastic coupling can unlock the nematic director
from the high-symmetry directions, resulting in $\alpha\neq\alpha_{s}$
as illustrated in Fig. \ref{fig:rot_nem_axis} (right panel). More
specifically, the three possible nematic directors split into six,
each associated with four momentum-space directions where the nematic
fluctuations are the largest. 

Formally, this result is a consequence of the competition between
a phonon-mediated non-analytic quadratic term in the nematic free
energy, which prefers to align $\alpha$ with the ``anti-symmetry
directions'' $\alpha_{as}\in\{1,3,5,7,9,11\}\frac{\pi}{6}$ (i.e.
the directions farthest away from the high-symmetry directions), and
the intrinsic nematic anharmonic cubic term, which favors $\alpha$
parallel to $\alpha_{s}$. Crucially, the former appears in trigonal
lattices, but is absent in hexagonal lattices, although both types
of lattices have $C_{3z}$ symmetry. This is because only in trigonal
lattices the in-plane shear-strain doublet $\boldsymbol{\epsilon}^{E_{g},1}=(\epsilon_{11}-\epsilon_{22},-2\epsilon_{12})^{T}$
and the out-of-plane shear-strain doublet $\boldsymbol{\epsilon}^{E_{g},2}=(2\epsilon_{23},-2\epsilon_{31})^{T}$
belong to the same IR of the point group, as manifested by the existence
of an additional elastic constant $c_{14}$. Here, $\epsilon_{ij}$
denotes the strain tensor and the subscripts $\left(1,\,2,\,3\right)$
correspond to $\left(x,\,y,\,z\right)$. We find that when $c_{14}$
(or the nemato-elastic coupling) overcomes a threshold value, the
unlocking of the nematic director from the high-symmetry directions
occurs. This unlocking, which we expect to happen in $A_{x}\mathrm{Bi_{2}Se_{3}}$
compounds, results in the breaking of a residual in-plane twofold
rotational symmetry of the lattice ($C_{2x}$) in the nematic phase,
which can be experimentally detected in the shape of the in-plane
$H_{c2}$ curve or in the emergence of a triclinic phase. Furthermore,
the loss of the $C_{2x}$ symmetry lifts the possible point nodes
that are otherwise allowed to exist inside the superconducting phase,
such that the pairing state becomes fully gapped \citep{Fu2014}.

This paper is organized as follows. In Sec. \ref{sec:acoustic-Phonon-renormalization-}
we formally derive the phonon-renormalized nematic action. In Sec.
\ref{sec:Mean-field-analysis}, we minimize the effective action first
numerically and then analytically in three limits: (i) $c_{14}=0$,
(ii) $c_{14}=|c_{14}|_{\mathrm{max}}$ and (iii) an expansion for
small $c_{14}$. In Sec. \ref{sec:Implications-for-doped} we discuss
possible experimental implications that an unlocked director $\alpha\neq\alpha_{s}$
has on nematic superconductors. Sec. \ref{sec:Concluding-remarks}
contains our concluding remarks. In Appendix \ref{sec:Strain-doublet-degeneracy},
we show that the aforementioned strain doublet degeneracy only occurs
in trigonal point groups. Appendices \ref{sec:Dynamical-matrix},
\ref{sec:Symm_based_Deg} and \ref{sec:Supplemental-to-analytical}
contain mathematical details of calculations presented in section
\ref{sec:Mean-field-analysis}. In Appendix \ref{sec:UpperCritField},
we outline the derivation of the expression for the in-plane upper
critical field $H_{c2}$. In Appendix \ref{sec:-model-Hamiltonian}
we present the model Hamiltonian used to determine the superconducting
gap structure.

\section{acoustic-phonon renormalization of the nematic director\label{sec:acoustic-Phonon-renormalization-}}

We employ a phenomenological field-theoretical approach to derive
the effective nematic action renormalized by acoustic phonons. The
derivation follows the same approach as in Refs. \citep{Paul2017,Chou1996,Karahasanovic2016,Carvalho2019,Fernandes2020},
the main difference being the trigonal symmetry of the underlying
lattice. Due to its phenomenological nature, our analysis holds regardless
of the microscopic origin of the nematic order parameter. We emphasize
that, in the particular case of doped $\mathrm{Bi_{2}Se_{3}}$, the
nematic order parameter $\boldsymbol{\Phi}^{E_{g}}$ is related to
the underlying superconducting order parameter $\boldsymbol{\Delta}$
via Eq. (\ref{eq:C_Eg}). In the vicinity of the nematic phase transition,
the behavior of the order parameter $\boldsymbol{\Phi}^{E_{g}}$,
parametrized in terms of an amplitude and an angle in Eq. (\ref{eq:C_Eg}),
is captured by the action \citep{Hecker2018}
\begin{align}
\mathcal{S}_{\mathrm{nem}}\! & =\!\int_{x}\!\!\left\{ \frac{r}{2}|\boldsymbol{\Phi}^{E_{g}}|^{2}\!+\!g|\boldsymbol{\Phi}^{E_{g}}|^{3}\cos(3\alpha)\!+\!u|\boldsymbol{\Phi}^{E_{g}}|^{4}\right\} \!\!,\label{eq:S_nem}
\end{align}
where $x=(\boldsymbol{r},\tau)$ comprises space and imaginary time
and $\int_{x}=\int_{0}^{1/T}d\tau\int d\boldsymbol{r}$, with $T$
denoting the temperature and $k_{B}=1$. The quadratic coefficient
$r=a_{\mathsf{c}}(T-T_{\mathrm{nem}}^{0})$ with
$a_{\mathsf{c}}>0$ determines the distance from the nematic reference
temperature $T_{\mathrm{nem}}^{0}$. The quartic coefficient $u>0$
guarantees the stability of the functional, while the sign of the
cubic parameter $g$ determines which set of threefold degenerate
ground states is favored\textemdash either $\alpha_{s}\in\{0,2,4\}\frac{\pi}{3}$
or $\alpha_{s}\in\{1,3,5\}\frac{\pi}{3}$ for $g$ negative or positive,
respectively. We denote these nematic director angles by $\alpha_{s}$,
which correspond to the high-symmetry directions of the lattice, see
Fig. \ref{fig:rot_nem_axis}. The form of the action (\ref{eq:S_nem})
is equivalent to the $Z_{3}$-Potts model, which in three dimensions
undergoes a mean-field first-order transition into a threefold degenerate
ground state \citep{Wu1982}. 

To incorporate the effect of the acoustic phonons, we include the
coupling between the nematic order parameter and the elastic degrees
of freedom:

\begin{align}
\mathcal{S} & =\mathcal{S}_{\mathrm{nem}}+\mathcal{S}_{\mathrm{el}}+\mathcal{S}_{\mathrm{el-nem}}.\label{eq:S_0}
\end{align}
Here, the elastic action is given via 
\begin{align}
\mathcal{S}_{\mathrm{el}} & =\frac{1}{2}\int_{x}\left(\left(\partial_{\tau}\boldsymbol{u}\right)^{2}+\boldsymbol{\epsilon}^{T}\mathcal{C}\boldsymbol{\epsilon}\right),\label{eq:S_el}
\end{align}
with the lattice displacement field $\boldsymbol{u}$, and the strain
tensor elements $\epsilon_{ij}=\frac{1}{2}(\partial_{i}u_{j}+\partial_{j}u_{i})$
where $i,j=\{1,2,3\}$. The directions $i=\{1,2,3\}$ correspond to
the $\{x,y,z\}$-directions, respectively. We employ the Voigt notation
$\boldsymbol{\epsilon}=\left(\epsilon_{11},\epsilon_{22},\epsilon_{33},2\epsilon_{23},2\epsilon_{31},2\epsilon_{12}\right)^{T}$
with the elastic stiffness tensor \begin{align} 
\mathcal{C} & =\left(\begin{smallmatrix} c_{11} & c_{12} & c_{13} & c_{14} & 0 & 0\\ 
c_{12} & c_{11} & c_{13} & -c_{14} & 0 & 0\\ c_{13} & c_{13} & c_{33} & 0 & 0 & 0\\ 
c_{14} & -c_{14} & 0 & c_{44} & 0 & 0\\ 0 & 0 & 0 & 0 & c_{44} & c_{14}\\ 
0 & 0 & 0 & 0 & c_{14} & \frac{1}{2}(c_{11}-c_{12}) 
\end{smallmatrix}\right),\label{eq:stiffness_tensor} 
\end{align}containing six independent components in the $\mathsf{D_{3d}}$ point
group. Note the existence of an additional elastic constant $c_{14}$,
when compared to a standard hexagonal point group. The values that
we use in this work\textemdash unless stated otherwise\textemdash are
those reported in Ref. \citep{Gao2016} for $\mathrm{Bi_{2}}\mathrm{Se}_{3}$
through first principle calculations. At ambient pressure, they are
$c_{11}=103.2\,\mathrm{GPa}$, $c_{12}=27.9\,\mathrm{GPa}$, $c_{33}=78.9\,\mathrm{GPa}$,
$c_{44}=37.7\,\mathrm{GPa}$, $c_{13}=35.4\,\mathrm{GPa}$, and $c_{14}=-26.5\,\mathrm{GPa}$.
In the $\mathsf{D_{3d}}$ point group, the strain components can be
combined into IRs as:
\begin{align}
\epsilon^{A_{1g},1} & =\epsilon_{11}+\epsilon_{22}, & \epsilon^{A_{1g},2} & =\epsilon_{33},\label{eq:A1g_strain}\\
\boldsymbol{\epsilon}^{E_{g},1} & =\left(\begin{array}{c}
\epsilon_{11}-\epsilon_{22}\\
-2\epsilon_{12}
\end{array}\right), & \boldsymbol{\epsilon}^{E_{g},2} & =\left(\begin{array}{c}
2\epsilon_{23}\\
-2\epsilon_{31}
\end{array}\right).\label{eq:strain_doublets}
\end{align}
For later convenience, we also rewrite the elastic action (\ref{eq:S_el})
with respect to the basis $\boldsymbol{\epsilon}^{D_{3d}}=(\epsilon^{A_{1g},1},\epsilon^{A_{1g},2},(\boldsymbol{\epsilon}^{E_{g},1})^{T},(\boldsymbol{\epsilon}^{E_{g},2})^{T})$,
for which the stiffness tensor becomes \begin{align} 
\mathcal{C}^{\mathsf{D_{3d}}} & =\left(\begin{smallmatrix} c_{A1} & \,c_{A3} & \boldsymbol{0}^{T} & \boldsymbol{0}^{T}\\ 
c_{A3} & \,c_{A2} & \boldsymbol{0}^{T} & \boldsymbol{0}^{T}\\ 
\boldsymbol{0} & \boldsymbol{0} & c_{E1}\,\mathbbm{1}_{\! 2} & \;c_{E3}\,\mathbbm{1}_{\! 2}\\ 
\boldsymbol{0} & \boldsymbol{0} & c_{E3}\,\mathbbm{1}_{\! 2} & \;c_{E2}\,\mathbbm{1}_{\! 2} 
\end{smallmatrix}\right),\label{eq:C_D3d} 
\end{align}with $\boldsymbol{0}=(0,0)^{T}$ and $\mathbbm{1}_{\! 2}$, the $2\times2$
identity matrix. The relationship to the original constants is
\begin{align}
c_{A1} & =\frac{1}{2}(c_{11}+c_{12}), & c_{A2} & =c_{33}, & c_{A3} & =c_{13},\nonumber \\
c_{E1} & =\frac{1}{2}(c_{11}-c_{12}), & c_{E2} & =c_{44}, & c_{E3} & =c_{14}.\label{eq:cE3_defs}
\end{align}
The stability of the elastic action (\ref{eq:S_el}) requires the
conditions (see also Ref. \citep{Cowley1976})
\begin{align}
\mathsf{d}_{A}\equiv c_{A1}c_{A2}-c_{A3}^{2} & >0, & \mathsf{d}_{E}\equiv c_{E1}c_{E2}-c_{E3}^{2} & >0,\label{eq:stab_cond1}
\end{align}
i.e. for $\mathsf{d}_{A}=0$ or $\mathsf{d}_{E}=0$ the system reaches
a structural phase transition in the respective symmetry channel.
Additionally, it holds that $c_{A1},c_{A2},c_{E1},c_{E2}>0$. Since the $E_{g}$-strain components (\ref{eq:strain_doublets})
and the nematic order parameter (\ref{eq:C_Eg}) transform according
to the same irreducible representation $E_{g}$, a linear coupling
term is allowed:
\begin{align}
\mathcal{S}_{\mathrm{el-nem}} & =\int_{x}\left\{ \boldsymbol{\Phi}^{E_{g}}\cdot\left(\kappa_{1}\boldsymbol{\epsilon}^{E_{g},1}+\kappa_{2}\boldsymbol{\epsilon}^{E_{g},2}\right)\right\} .\label{eq:S_el_nem}
\end{align}
The nemato-elastic coupling coefficients are denoted by $\kappa_{1}$
and $\kappa_{2}$. This linear coupling is the origin for the monoclinic
crystal distortion inside the nematic phase \citep{Cho2019}. As mentioned
above, the fact that the two in-plane and out-of-plane shear strain
doublets in Eq. (\ref{eq:strain_doublets}) transform as the same
IR plays a crucial role in the unlocking of the nematic director from
the high-symmetry directions. This is a defining property of trigonal
point groups, which is absent in hexagonal point groups, as explained
in detail in Appendix \ref{sec:Strain-doublet-degeneracy}. For our
purposes, this property leads to two important consequences: a finite
elastic constant $c_{E3}$ (recall that $c_{E3}=c_{14}$) and the
presence of two nemato-elastic coupling constants, $\kappa_{1}$ and
$\kappa_{2}$, in Eq. (\ref{eq:S_el_nem}). This is to be contrasted
with the case of the $\mathsf{D_{6}}$ point group analyzed in Ref.
\citep{Fernandes2020}, where only one coupling constant is allowed. 

Having set up all the action terms, the next step is to integrate
out the fluctuating acoustic phonon modes (\ref{eq:S_el}). Then,
the partition function $Z$ becomes
\begin{align}
Z & =\int D\boldsymbol{\Phi}^{E_{g}}\int D\boldsymbol{u}\,e^{-\mathcal{S}[\boldsymbol{\Phi}^{E_{g}},\boldsymbol{u}]}\label{eq:partition_1}\\
 & =\int D\boldsymbol{\Phi}^{E_{g}}\,e^{-\mathcal{S}_{\mathrm{eff}}[\boldsymbol{\Phi}^{E_{g}}]}.\label{eq:partition2}
\end{align}
Our goal is then to determine the ground state of the effective nematic
action $\mathcal{S}_{\mathrm{eff}}[\boldsymbol{\Phi}^{E_{g}}]$. To
integrate out the elastic degrees of freedom, the elastic action (\ref{eq:S_el})
is first transformed into Fourier space, using $\boldsymbol{u}(x)=\sum_{q}e^{\mathsf{i}qx}\boldsymbol{u}_{q}$
with the notation $q=(\boldsymbol{q},\omega_{n})$ comprising the
momentum $\boldsymbol{q}$ and the bosonic Matsubara frequency $\omega_{n}=2n\pi T$.
The scalar product reads $qx=\boldsymbol{q}\cdot\boldsymbol{r}-\omega_{n}\tau$.
The elastic action then becomes 
\begin{align}
\mathcal{S}_{\mathrm{el}} & =\frac{V}{2T}\sum_{q}\boldsymbol{u}_{-q}^{T}\left[\omega_{n}^{2}\mathbbm{1}+D(\boldsymbol{q})\right]\boldsymbol{u}_{q},\label{eq:S_el_2}
\end{align}
where the dynamic matrix $D_{ij}(\boldsymbol{q})=\sum_{i^{\prime},j^{\prime}}\mathcal{C}_{ii^{\prime}jj^{\prime}}q_{i^{\prime}}q_{j^{\prime}}$
has been introduced. The matrix elements $D_{ij}(\boldsymbol{q})$
are given explicitly in Appendix \ref{sec:Dynamical-matrix}. It is
convenient to diagonalize the dynamic matrix before proceeding. Thus,
we introduce the orthogonal matrix $U_{\hat{\boldsymbol{q}}}=(\hat{\boldsymbol{e}}_{\hat{\boldsymbol{q}}}^{(1)},\hat{\boldsymbol{e}}_{\hat{\boldsymbol{q}}}^{(2)},\hat{\boldsymbol{e}}_{\hat{\boldsymbol{q}}}^{(3)})$
containing the eigenvectors $\hat{\boldsymbol{e}}_{\hat{\boldsymbol{q}}}^{(j)}$,
with $\hat{\boldsymbol{e}}_{-\hat{\boldsymbol{q}}}^{(j)}=-\hat{\boldsymbol{e}}_{\hat{\boldsymbol{q}}}^{(j)}$,
which correspond to the phonon polarization vectors. Given the definition
of the dynamic matrix, it is clear that the eigenvectors depend only
on the momentum directions $\hat{\boldsymbol{q}}=\boldsymbol{q}/|\boldsymbol{q}|$.
The resulting diagonalized dynamic matrix reads
\begin{align}
D^{\prime}(\boldsymbol{q}) & =U_{\hat{\boldsymbol{q}}}^{-1}D(\boldsymbol{q})U_{\hat{\boldsymbol{q}}}=\text{diag}(\omega_{1,\boldsymbol{q}}^{2},\omega_{2,\boldsymbol{q}}^{2},\omega_{3,\boldsymbol{q}}^{2}),\label{eq:D_prime}
\end{align}
with the three eigenvalues $\omega_{j,\boldsymbol{q}}^{2}$, corresponding
to the squared acoustic phonon frequencies. They can be rewritten
as $\omega_{j,\boldsymbol{q}}^{2}=|\boldsymbol{q}|^{2}\omega_{j,\hat{\boldsymbol{q}}}^{2}$,
with $\omega_{j,-\boldsymbol{q}}^{2}=\omega_{j,\boldsymbol{q}}^{2}$.
Finally, the elastic contribution becomes 
\begin{align}
\mathcal{S}_{\mathrm{el}} & =\frac{V}{2T}\sum_{q}\tilde{\boldsymbol{u}}_{-q}^{T}\left[\omega_{n}^{2}\mathbbm{1}+D^{\prime}(\boldsymbol{q})\right]\tilde{\boldsymbol{u}}_{q},\label{eq:S_el_3}
\end{align}
with $\boldsymbol{u}_{\boldsymbol{q}}=\mathsf{i}U_{\hat{\boldsymbol{q}}}\tilde{\boldsymbol{u}}_{\boldsymbol{q}}=\mathsf{i}\sum_{j}\hat{\boldsymbol{e}}_{\hat{\boldsymbol{q}}}^{(j)}\tilde{u}_{j,\boldsymbol{q}}$.
The imaginary $\mathsf{i}$ ensures that the new displacement field
$\tilde{\boldsymbol{u}}_{\boldsymbol{q}}^{*}=\tilde{\boldsymbol{u}}_{-\boldsymbol{q}}$
is real. Transforming the elasto-nematic coupling term (\ref{eq:S_el_nem})
into the same basis leads to the expression 
\begin{align}
\mathcal{S}_{\mathrm{el-nem}} & =\mathsf{i}\frac{V}{T}\sum_{q}\left\{ \Phi_{-q}^{E_{g},1}\boldsymbol{a}_{\boldsymbol{q}}^{(1)}+\Phi_{-q}^{E_{g},2}\boldsymbol{a}_{\boldsymbol{q}}^{(2)}\right\} \cdot\boldsymbol{u}_{\boldsymbol{q}},\nonumber \\
 & =-\frac{V}{T}\sum_{q,j}\Big\{\sum_{l=1,2}\Phi_{-q}^{E_{g},l}\boldsymbol{a}_{\boldsymbol{q}}^{(l)}\cdot\hat{\boldsymbol{e}}_{\hat{\boldsymbol{q}}}^{(j)}\Big\}\tilde{u}_{j,\boldsymbol{q}},\label{eq:S_el_nem_2}
\end{align}
with system volume $V$ and form factors defined as
\begin{align}
\boldsymbol{a}_{\boldsymbol{q}}^{(1)} & =\!\!\left(\!\!\!\begin{array}{c}
\kappa_{1}q_{x}\\
-\kappa_{1}q_{y}+\kappa_{2}q_{z}\\
\kappa_{2}q_{y}
\end{array}\!\!\!\right)\!, & \boldsymbol{a}_{\boldsymbol{q}}^{(2)} & =\!\!\left(\!\!\!\begin{array}{c}
-\kappa_{1}q_{y}-\kappa_{2}q_{z}\\
-\kappa_{1}q_{x}\\
-\kappa_{2}q_{x}
\end{array}\!\!\!\right)\!,\label{eq:a_s}
\end{align}
which satisfy $\boldsymbol{a}_{-\boldsymbol{q}}^{(l)}=-\boldsymbol{a}_{\boldsymbol{q}}^{(l)}$.
In the next step, the lattice displacement fields are integrated out
according to 
\begin{align*}
\int D\tilde{\boldsymbol{u}}\,e^{-\frac{1}{2}\sum_{q}\tilde{\boldsymbol{u}}_{q}^{\dagger}A_{q}\tilde{\boldsymbol{u}}_{q}+\sum_{q}\tilde{\boldsymbol{u}}_{q}^{\dagger}\boldsymbol{J}_{q}} & \sim e^{\frac{1}{2}\sum_{q}\boldsymbol{J}_{q}^{\dagger}A_{q}^{-1}\boldsymbol{J}_{q}},
\end{align*}
with $A_{q}=\frac{V}{T}[\omega_{n}^{2}\mathbbm{1}+D^{\prime}(\boldsymbol{q})]$
and $J_{j,q}=\frac{V}{T}\!\sum_{l}\Phi_{q}^{E_{g},l}\boldsymbol{a}_{\boldsymbol{q}}^{(l)}\cdot\hat{\boldsymbol{e}}_{\hat{\boldsymbol{q}}}^{(j)}$,
where $J_{j,-q}=J_{j,q}^{*}$. The integration leads to the effective
action
\begin{align}
\mathcal{S}_{\mathrm{eff}} & =\mathcal{S}_{\mathrm{nem}}+\mathcal{S}^{\prime},\label{eq:S_prime}
\end{align}
with the phonon-induced contribution 
\begin{align}
\mathcal{S}^{\prime} & =-\frac{1}{2}\frac{V}{T}\sum_{q}\sum_{l,l^{\prime}=1,2}\Phi_{-q}^{E_{g},l}\,\Pi_{q}^{l,l^{\prime}}\,\Phi_{q}^{E_{g},l^{\prime}},\label{eq:S_Pi}
\end{align}
and the polarization function: 
\begin{align}
\Pi_{q}^{l,l^{\prime}} & =\sum_{j}\frac{\big(\boldsymbol{a}_{\boldsymbol{q}}^{(l)}\cdot\hat{\boldsymbol{e}}_{\hat{\boldsymbol{q}}}^{(j)}\big)\big(\boldsymbol{a}_{\boldsymbol{q}}^{(l^{\prime})}\cdot\hat{\boldsymbol{e}}_{\hat{\boldsymbol{q}}}^{(j)}\big)}{\omega_{n}^{2}+\omega_{j,\boldsymbol{q}}^{2}}.\label{eq:func_Pi}
\end{align}

In agreement with previous works \citep{Karahasanovic2016,Paul2017,Labat2017,Carvalho2019,Fernandes2020},
the incorporation of the acoustic phonons leads to a renormalization
of the nematic susceptibility, which in our case becomes non-diagonal
in the $E_{g}$ subspace of the nematic order parameter:
\begin{align}
\chi_{\mathrm{nem}}^{l,l^{\prime}}(q) & =(r-\Pi_{q})_{l,l^{\prime}}^{-1}.\label{eq:nem_susy}
\end{align}
The polarization function becomes non\textendash analytic in the static
limit $\omega_{n}=0$:
\begin{align}
\Pi_{(\boldsymbol{q},0)}^{l,l^{\prime}} & =\sum_{j}\frac{\big(\boldsymbol{a}_{\hat{\boldsymbol{q}}}^{(l)}\cdot\hat{\boldsymbol{e}}_{\hat{\boldsymbol{q}}}^{(j)}\big)\big(\boldsymbol{a}_{\hat{\boldsymbol{q}}}^{(l^{\prime})}\cdot\hat{\boldsymbol{e}}_{\hat{\boldsymbol{q}}}^{(j)}\big)}{\omega_{j,\hat{\boldsymbol{q}}}^{2}},\label{eq:func_Pi-1}
\end{align}
where we defined the quantities $\omega_{j,\hat{\boldsymbol{q}}}\equiv\omega_{j,\boldsymbol{q}}/|\boldsymbol{q}|$
(which correspond to the sound velocities) and $\boldsymbol{a}_{\hat{\boldsymbol{q}}}^{(l)}\equiv\boldsymbol{a}_{\boldsymbol{q}}^{(l)}/|\boldsymbol{q}|$
that depend only on the direction $\hat{\boldsymbol{q}}$. As a consequence,
the nematic susceptibility (\ref{eq:nem_susy}) tends to diverge only
along particular momentum directions $\hat{\boldsymbol{q}}$ as the
system approaches the phase transition. As we will show later, in
our problem, the nematic order parameter actually undergoes a first-order
transition, such that the susceptibility gets enhanced along these
directions but it does not diverge. The impact of such momentum-space
restriction on the nematic phase has been previously investigated
in Refs. \citep{Paul2017,Fernandes2020} for the cases of tetragonal
and hexagonal lattices. In those cases, this effect did not alter
the allowed angles of the nematic director. As we will show here,
the situation is qualitatively different in the case of a trigonal
lattice.

The determination of the phase transition requires a free energy minimization.
Before doing so, we rewrite the action contribution (\ref{eq:S_Pi})
in a symmetry-guided way. It is convenient to define the components
of the polarization function
\begin{align}
\Pi_{q}^{A_{1g}} & =(\Pi_{q}^{1,1}+\Pi_{q}^{2,2})/2,\label{eq:P1_A1g}\\
\boldsymbol{\Pi}_{q}^{E_{g}} & =\left(\!\!\begin{array}{c}
\Pi_{q}^{E_{g},1}\\
\Pi_{q}^{E_{g},2}
\end{array}\!\!\right)=\frac{1}{2}\!\left(\!\!\begin{array}{c}
\Pi_{q}^{1,1}-\Pi_{q}^{2,2}\\
-\Pi_{q}^{1,2}-\Pi_{q}^{2,1}
\end{array}\!\!\right)\!.\label{eq:Pi_Eg}
\end{align}
Then, the action (\ref{eq:S_Pi}) can be rewritten conveniently as
\begin{align}
\mathcal{S}^{\prime} & =-\frac{V}{2T}\sum_{q}\!\boldsymbol{\Phi}_{-q}^{E_{g}}\big\{\tau^{A_{1g}}\Pi_{q}^{A_{1g}}+\boldsymbol{\tau}^{E_{g}}\cdot\boldsymbol{\Pi}_{q}^{E_{g}}\big\}\boldsymbol{\Phi}_{q}^{E_{g}},\label{eq:S_Pi_2}
\end{align}
in terms of the Pauli matrices $\tau^{A_{1g}}=\tau^{0}$ and $\boldsymbol{\tau}^{E_{g}}=(\tau^{z},-\tau^{x})$.
The representation (\ref{eq:S_Pi_2}) demonstrates that for a two-component
nematic order parameter, the mass renormalization does not only occur
in the trivial $A_{1g}$, but also in the $E_{g}$ channel. More importantly,
the $E_{g}$-channel contribution is sensitive on the nematic director
angle $\alpha$. 
\begin{figure*}[t]
\centering{}\ffigbox[\FBwidth][]
{\includegraphics[scale=0.50]{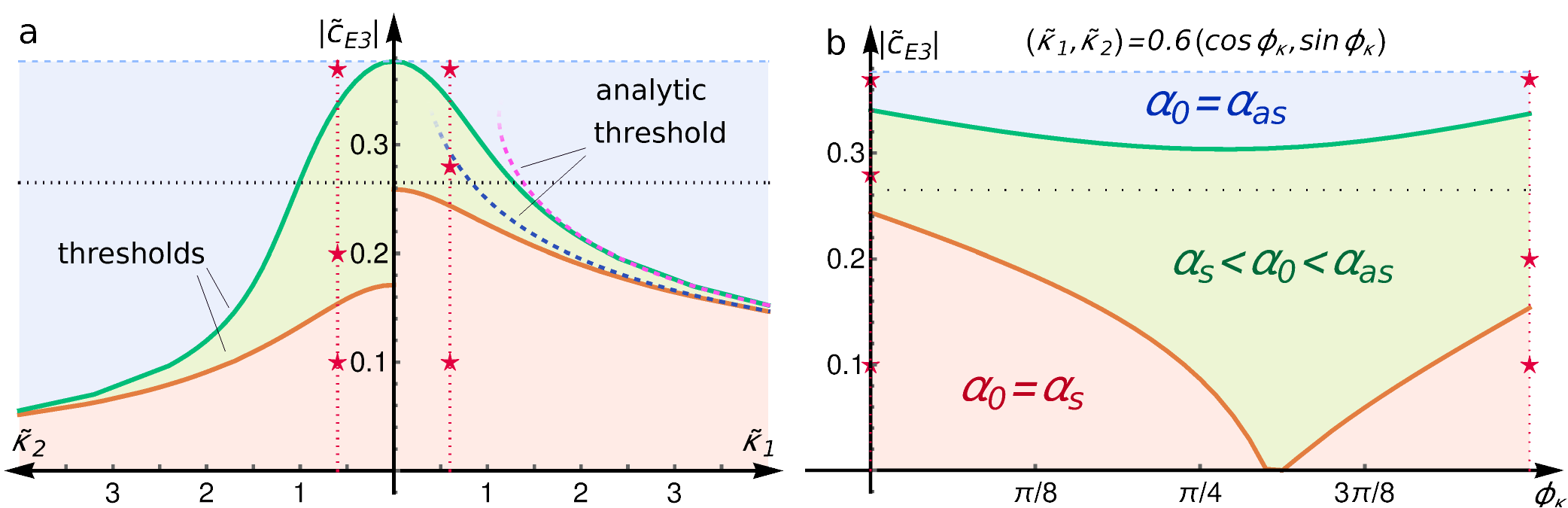}}
{\caption{   Phase diagram for the nematic director angle $\alpha_{0}$ with respect
to the parameters $\left|\tilde{c}_{E3}\right|\propto\left|c_{14}\right|$
and $(\tilde{\kappa}_{1},\tilde{\kappa}_{2})=\tilde{\kappa}_{0}(\cos\phi_{\kappa},\sin\phi_{\kappa})$.
In panel (a), the absolute value $\tilde{\kappa}_{0}$ changes for
fixed $\phi_{k}=\pi/2$ (left horizontal axis) and $\phi_{k}=0$ (right
horizontal axis). In panel (b), the absolute value is fixed at $\tilde{\kappa}_{0}=0.6$
and the relative angle $\phi_{\kappa}\in[0,\pi/2]$ is varied. There
are three distinct regimes: (i) For small values of $|c_{E3}|$, and
when $\kappa_{1}$ (or $\kappa_{2}$) is much larger than $\kappa_{2}$
(or $\kappa_{1}$), the nematic director is locked at the high-symmetry
directions $\alpha_{s}$ (red region). (ii) For large values of $|c_{E3}|$,
regardless of $\tilde{\kappa}_{i}$, the nematic director aligns with
the ``anti-symmetry directions'' $\alpha_{as}$ (blue region). (iii)
For intermediate values of $|c_{E3}|$, or when $\kappa_{1}$ and
$\kappa_{2}$ are comparable, the nematic director evolves smoothly
between $\alpha_{s}<\alpha_{0}<\alpha_{as}$ (green region). The full
dependence of $\alpha_{0}$ and $\hat{\boldsymbol{q}}_{0}$ on $\left|\tilde{c}_{E3}\right|$,
for a given maximum of $R(\hat{\boldsymbol{q}}_{0},\alpha_{0})$,
is shown in Fig. \ref{fig:max_evol} for the $(\tilde{\kappa}_{1},\tilde{\kappa}_{2})$
values corresponding to the two red dashed lines. For the six parameter
values corresponding to the red stars, the function $R(\hat{\boldsymbol{q}}_{0},\alpha_{0}(\hat{\boldsymbol{q}}_{0}))$
is plotted in Fig. \ref{fig:sol_kap1}. The horizontal black dotted
line denotes the expected value of $\tilde{c}_{E3}=-0.265$ for doped
$\mathrm{Bi_{2}Se_{3}}$ \citep{Gao2016}. The topmost light-blue
horizontal dashed line denotes the limit of structural stability as
given by the condition $|c_{E3}|_{\mathrm{max}}=\sqrt{c_{E1}c_{E2}}$,
see Eq. (\ref{eq:stab_cond1}). The analytical thresholds stem from
the calculations presented in Section \ref{subsec:expansion}.}\label{fig:thresholds}}
\end{figure*}

\section{Mean-field analysis of the effective nematic action \label{sec:Mean-field-analysis}}

We now analyze the full effective action (\ref{eq:S_prime}) that
includes both the pure nematic action $\mathcal{S}_{\mathrm{nem}}$,
Eq. (\ref{eq:S_nem}), and the phonon-induced contribution $\mathcal{S}^{\prime}$,
Eq. (\ref{eq:S_Pi_2}). Because the upper critical dimension of the
three-state Potts model is below $3$, see Ref. \citep{Wu1982}, our
model is expected to be well-described by mean-field theory in three
dimensions. The mean-field nematic order parameter is given by $\boldsymbol{\Phi}_{q}^{E_{g}}=\delta_{\omega_{n},0}\delta_{q,0}\,|\boldsymbol{\Phi}_{0}^{E_{g}}|(\cos\alpha_{0},\sin\alpha_{0})^{T}$
with homogeneous field values $|\boldsymbol{\Phi}_{0}^{E_{g}}|$ and
$\alpha_{0}$. The effective mean-field action $\mathcal{S}_{\mathrm{eff}}=\frac{V}{T}\int_{\hat{\boldsymbol{q}}}\mathcal{S}_{\mathrm{eff},\hat{\boldsymbol{q}}}$ becomes 
\begin{align}
\mathcal{S}_{\mathrm{eff},\hat{\boldsymbol{q}}} & =\frac{r\!-\!M(\hat{\boldsymbol{q}},\alpha_{0})}{2}|\boldsymbol{\Phi}_{0}^{E_{g}}|^{2}+\!g|\boldsymbol{\Phi}_{0}^{E_{g}}|^{3}\!\cos(3\alpha_{0})\!+\!u|\boldsymbol{\Phi}_{0}^{E_{g}}|^{4},\nonumber \\
 & =\frac{1}{2}|\boldsymbol{\Phi}_{0}^{E_{g}}|^{2}\left[r-\tilde{R}(\hat{\boldsymbol{q}},\alpha_{0},|\boldsymbol{\Phi}_{0}^{E_{g}}|)\right],\label{eq:S_prime_2}
\end{align}
where we introduced the momentum-dependent nematic mass function 
\begin{align}
M(\hat{\boldsymbol{q}},\alpha_{0}) & =\Pi_{(\boldsymbol{q},0)}^{A_{1g}}+\boldsymbol{\Pi}_{(\boldsymbol{q},0)}^{E_{g}}\cdot\left(\!\!\!\begin{array}{c}
\cos(2\alpha_{0})\\
-\sin(2\alpha_{0})
\end{array}\!\!\!\right)\!,\label{eq:M_func}
\end{align}
and the auxiliary function 
\begin{align}
\tilde{R}(\hat{\boldsymbol{q}},\alpha_{0},|\boldsymbol{\Phi}_{0}^{E_{g}}|) & =M(\hat{\boldsymbol{q}},\alpha_{0})+\frac{g^{2}\cos^{2}(3\alpha_{0})}{2u}\nonumber \\
 & \quad-\!2u\Big[|\boldsymbol{\Phi}_{0}^{E_{g}}|+\frac{g\cos(3\alpha_{0})}{2u}\Big]^{2}.\label{eq:R_func2}
\end{align}
We highlight the key role played by the cubic nematic term in Eq.
(\ref{eq:S_prime_2}). In a harmonic approximation, where this term
is absent, and in the special case where $c_{14}=0$, the nematic
director angle $\alpha_{0}$ can assume any value and all in-plane
directions in momentum space are equivalent. This is consistent with
the fact that the pure transverse acoustic phonon dispersion is in-plane
isotropic in this case \citep{kimura2021}. However, the cubic term
is relevant in the renormalization-group sense, and lowers the symmetry
of $\boldsymbol{\Phi}^{E_{g}}$ from SO(2) to $Z_{3}$-Potts \citep{Lou2007}.
Moreover, in three dimensions, it induces a first-order transition,
in which case the cubic term is not necessarily subleading compared
to the quadratic term. It is the competition between these two terms
that restricts both the nematic director and the soft momentum-space
directions. In a phonon description, this cubic term is equivalent
to an anharmonic phonon term, which causes the phonon properties to
no longer be isotropic in the plane (see, for instance, Ref. \citep{Paulatto2013}).

The nematic phase transition occurs when $\mathcal{S}_{\mathrm{eff},\hat{\boldsymbol{q}}}=0$,
which due to the cubic term happens when $|\boldsymbol{\Phi}_{0}^{E_{g}}|$
jumps to a non-zero value. Thus, the first-order transition temperature
can be identified from the maximum of $\tilde{R}(\hat{\boldsymbol{q}},\alpha_{0},|\boldsymbol{\Phi}_{0}^{E_{g}}|)$.
Maximizing Eq. (\ref{eq:R_func2}) leads to the non-zero nematic value
at the first-order transition
\begin{align}
|\boldsymbol{\Phi}_{0}^{E_{g}}| & =\frac{|g|}{2u}|\cos(3\alpha_{0})|,\label{eq:AbsC_val}
\end{align}
 and to the condition
\begin{align}
\text{sign}\left(g\cos(3\alpha_{0})\right) & <0.\label{eq:sign_cond}
\end{align}
Note that the case of a pure nematic order parameter, for which $\alpha_{0}=\{0,2,4\}\frac{\pi}{3}$
for $g<0$ and $\alpha_{0}=\{1,3,5\}\frac{\pi}{3}$ for $g>0$ satisfies
this condition. Hence, the last line in (\ref{eq:R_func2}) vanishes
at the maximum, and the auxiliary function that remains to be maximized
becomes:
\begin{align}
R(\hat{\boldsymbol{q}},\alpha_{0}) & =M(\hat{\boldsymbol{q}},\alpha_{0})+\frac{g^{2}\cos^{2}(3\alpha_{0})}{2u}.\label{eq:real_R}
\end{align}
Importantly, the maximization is with respect to the three variables
$\{\hat{\boldsymbol{q}},\alpha_{0}\}$, corresponding to the two independent
directions in momentum space and to the nematic director angle $\alpha_{0}$.
Hereafter, we denote the momentum direction along which (\ref{eq:real_R})
is maximized by $\hat{\boldsymbol{q}}_{0}=(\cos\varphi_{0}\sin\theta_{0},\sin\varphi_{0}\sin\theta_{0},\cos\theta_{0})^{T}$.
The nematic transition temperature is given by $T_{\mathrm{nem}}=T_{\mathrm{nem}}^{0}+r^{\mathrm{nem}}/a_{\mathsf{c}}$
with:
\begin{align}
r^{\mathrm{nem}} & =\max_{\hat{\boldsymbol{q}},\alpha_{0}}\left[R(\hat{\boldsymbol{q}},\alpha_{0})\right].\label{eq:rcnem_max}
\end{align}

As demonstrated in Appendix \ref{sec:Symm_based_Deg}, the maxima
of $R(\hat{\boldsymbol{q}},\alpha_{0})$ occur in multiples of $12$.
Indeed, if $\{\hat{\boldsymbol{q}}_{0},\alpha_{0}\}$ is a maximum
of $R$, symmetry enforces the following relationships: 
\begin{align}
R(-\hat{\boldsymbol{q}}_{0},\alpha_{0}) & =R(\hat{\boldsymbol{q}}_{0},\alpha_{0}),\label{eq:trafo_inversion}\\
R\big[\mathcal{R}_{v}^{\pm}(C_{3z})\hat{\boldsymbol{q}}_{0},\alpha_{0}\mp\frac{2\pi}{3}\big] & =R(\hat{\boldsymbol{q}}_{0},\alpha_{0}),\label{eq:trafo_threefold_rot}\\
R\big[\mathcal{R}_{v}(IC_{2n_{s}})\hat{\boldsymbol{q}}_{0},\alpha_{s}-\delta\big] & =R(\hat{\boldsymbol{q}}_{0},\alpha_{s}+\delta),\label{eq:trafo_reflection}
\end{align}
with the definitions of the symmetry elements and transformation matrices
provided in the appendix. Importantly, the relationship (\ref{eq:trafo_reflection})
implies that a finite deviation $\delta\neq0$ away from a high-symmetry
direction $\alpha_{s}$ necessarily induces two maxima $\alpha_{0}=\alpha_{s}\pm\delta$,
i.e. the nematic director splits into two, doubling the number of
non-identical directors from $3$ to $6$, see Fig.~\ref{fig:rot_nem_axis}.
As we show in the following sections, each direction $\alpha_{0}$
is associated with $4$ soft momentum-space directions $\hat{\boldsymbol{q}}_{0}$.
This implies that the function $R$ has either $12$ or $24$ degenerate
maxima depending on whether $\alpha_{0}=\alpha_{s}$ or $\alpha_{0}\neq\alpha_{s}$,
respectively. 

\begin{figure*}[t]
\centering{}\ffigbox[\FBwidth][]
{\includegraphics[scale=0.50]{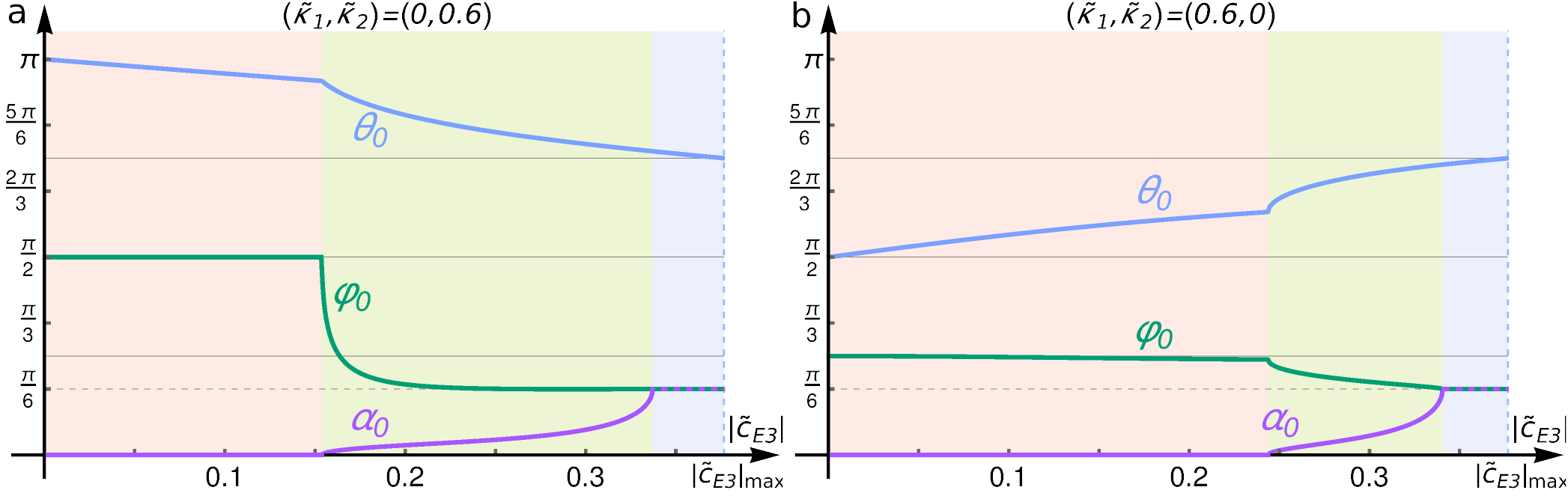}}
{\caption{   The soft momentum-space direction $\hat{\boldsymbol{q}}_{0}$ (parametrized
by the polar angle $\theta_{0}$ and the azimuthal angle $\varphi_{0}$)
and the nematic director angle $\alpha_{0}$ associated with a particular
maximum of $R(\hat{\boldsymbol{q}}_{0},\alpha_{0})$ are plotted as
a function of $\left|\tilde{c}_{E3}\right|$ for the two indicated
$(\tilde{\kappa}_{1},\tilde{\kappa}_{2})$ values, which correspond
to the red dashed lines in Fig. \ref{fig:thresholds}.} \label{fig:max_evol}}
\end{figure*}

\subsection{Numerical results\label{subsec:Numerical-results}}

We proceed with a numerical investigation of the maxima of $R(\hat{\boldsymbol{q}},\alpha_{0})$.
We consider three independent ``tuning'' parameters: the two effective
nemato-elastic coupling constants $\tilde{\kappa}_{i}=\kappa_{i}\sqrt{u}/g$,
with $i=\{1,2\}$, and the dimensionless $\tilde{c}_{E3}=c_{E3}/c_{0}$,
with a reference elastic constant $c_{0}=100\,\mathrm{GPa}$. The
other elastic constants are set to the values of $\mathrm{Bi_{2}}\mathrm{Se}_{3}$.
In Fig. \ref{fig:thresholds}, we present the ``phase diagram''
for the nematic director in this parameter space. Parametrizing the
two coupling constants as $(\tilde{\kappa}_{1},\tilde{\kappa}_{2})=\tilde{\kappa}_{0}(\cos\phi_{\kappa},\sin\phi_{\kappa})$,
panel (a) shows the phase diagram when $\phi_{k}$ is fixed ($\phi_{k}=\pi/2$
on the left side and $\phi_{k}=0$ on the right side) whereas panel
(b) presents the phase diagram for fixed $\tilde{\kappa}_{0}$. We
identify three distinct phases: (i) the nematic director aligns with
the high-symmetry directions, $\alpha_{0}=\alpha_{s}$ (red region),
where $\alpha_{s}\in\{0,2,4\}\frac{\pi}{3}$ or $\alpha_{s}\in\{1,3,5\}\frac{\pi}{3}$;
(ii) the nematic director evolves smoothly between the high-symmetry
and the ``anti-symmetry'' directions (green region); (iii) the nematic
director aligns with one of the ``anti-symmetry'' directions, $\alpha_{0}=\alpha_{as}$
(blue region), where $\alpha_{as}\in\{1,3,5,7,9,11\}\frac{\pi}{6}$.
We conclude that for the nematic director to unlock from the high-symmetry
directions, it requires a threshold value for $c_{E3}$ or the simultaneous
presence of both $\tilde{\kappa}_{1}$ and $\tilde{\kappa}_{2}$.
The horizontal black dotted line in both panels of Fig. \ref{fig:thresholds}
marks the value of $c_{E3}$ expected for $\mathrm{Bi_{2}Se_{3}}$.
Therefore, regardless of the values of the coupling constants, the
nematic director in doped $\mathrm{Bi_{2}Se_{3}}$ is expected to
be unlocked from the high-symmetry directions. The light-blue dashed
horizontal line (the top line) denotes the limit of structural stability,
defined by $|c_{E3}|_{\mathrm{max}}=\sqrt{c_{E1}c_{E2}}$ {[}or $\mathsf{d}_{E}=0$,
see Eq. (\ref{eq:stab_cond1}){]}. For this value of $c_{E3}$, the
system would undergo a structural transition on its own, even without
the coupling to nematic degrees of freedom. Upon approaching this
boundary, the system tends to align the nematic director with the
``anti-symmetry'' directions. 

\begin{figure*}[t] 
\centering{}\ffigbox[\FBwidth][]{   
\includegraphics[scale=0.28]{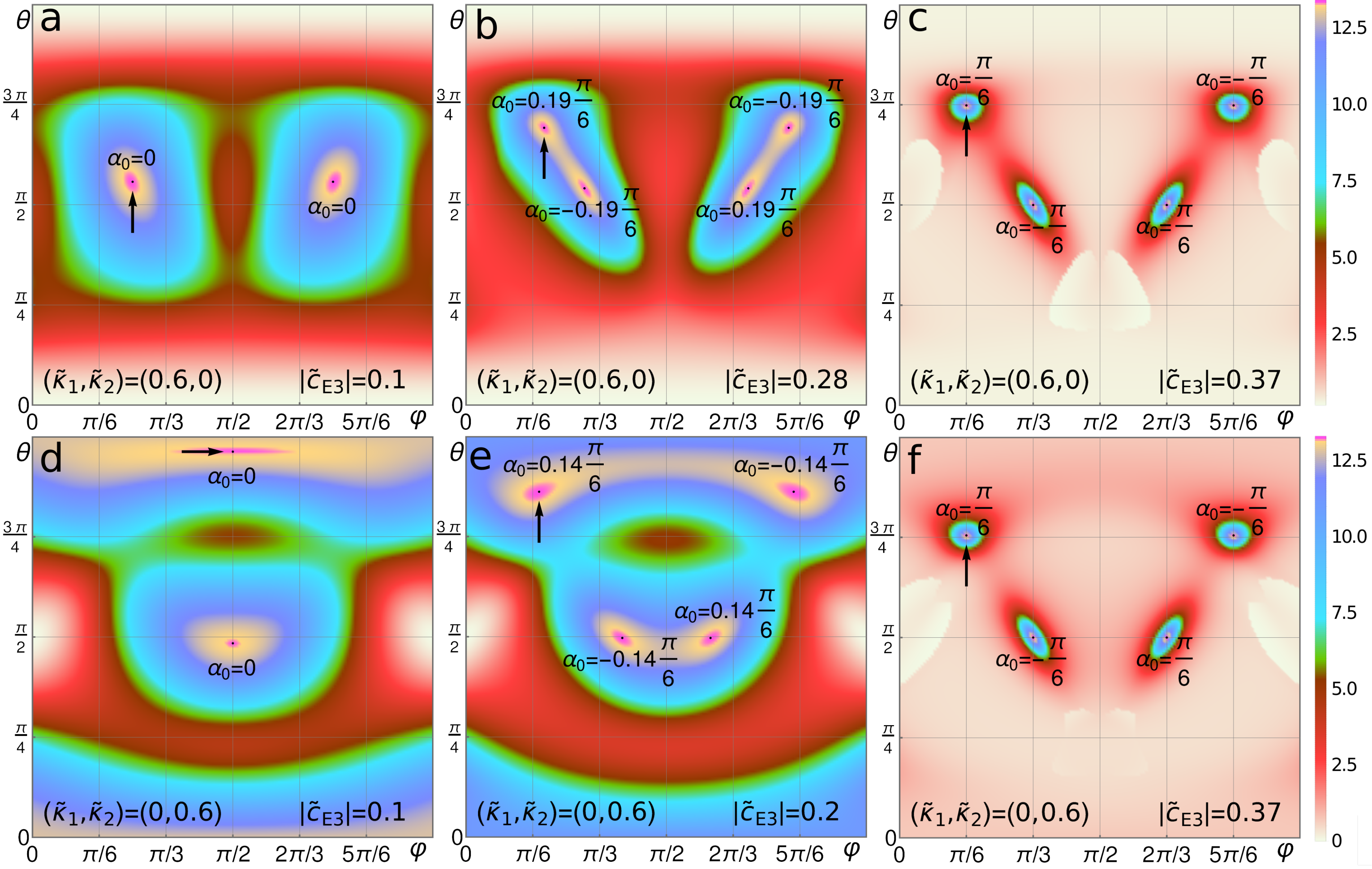}}
{\caption{ The function $R(\hat{\boldsymbol{q}},\alpha_{0}(\hat{\boldsymbol{q}}))$
is plotted as a function of the momentum-space polar angle $\theta$
and azimuthal angle $\varphi$. Because of the relationship in Eq.
(\ref{eq:trafo_inversion}), we consider only the ranges $\varphi\in[0,\pi)$
and $\theta\in[0,\pi]$. The extremal director angle value $\alpha_{0}(\hat{\boldsymbol{q}})\in[-\frac{\pi}{3},\frac{\pi}{3}]$
is shown at every maxima; note that the other nematic director angles
outside of this interval can be obtained from Eq. (\ref{eq:trafo_threefold_rot}).
The value of $\left|\tilde{c}_{E3}\right|$ increases upon moving
from the left to the right panels, encompassing the three regions
of the phase diagram of Fig. \ref{fig:thresholds}(a), as indicated
by the red stars. The top (bottom) panels correspond to a non-zero
$\kappa_{1}$ ($\kappa_{2}$). The black arrows indicate the maxima
previously presented in Fig. \ref{fig:max_evol}.}\label{fig:sol_kap1}} \end{figure*}

The two vertical red dotted lines in Fig. \ref{fig:thresholds} mark
the $(\tilde{\kappa}_{1},\tilde{\kappa}_{2})$-values for which the
complete $\alpha_{0}$ and $\hat{\boldsymbol{q}}_{0}=(\cos\varphi_{0}\sin\theta_{0},\sin\varphi_{0}\sin\theta_{0},\cos\theta_{0})^{T}$
evolutions as a function of $c_{E3}$ are shown in Fig. \ref{fig:max_evol}.
Additionally, for the six $c_{E3}$ values indicated by the red stars,
we present in Fig. \ref{fig:sol_kap1} the $R(\hat{\boldsymbol{q}},\alpha_{0}(\hat{\boldsymbol{q}}))$
dependence on $\varphi$ and $\theta$. In all panels, there are clear
maxima at well-defined $\left(\varphi_{0},\theta_{0}\right)$ points;
the corresponding value for the nematic director angle $\alpha_{0}(\hat{\boldsymbol{q}}_{0})$
at these maxima is indicated in the figure. For clarity, we only show
the nematic director $\alpha_{0}(\hat{\boldsymbol{q}})$ that falls
within the interval $\alpha_{0}\in[-\frac{\pi}{3},\frac{\pi}{3}]$.
The other symmetry-equivalent nematic directors can be obtained in
a straightforward way from Eqs. (\ref{eq:trafo_inversion})-(\ref{eq:trafo_reflection}).
Panels (a) and (d) of Fig. \ref{fig:sol_kap1} show the case in which
the nematic director is locked at the high-symmetry directions $\alpha_{0}=\alpha_{s}$
(red region in the phase diagram of Fig. \ref{fig:thresholds}(a)).
Each nematic director is associated with four distinct ``soft''
momentum directions $\left(\varphi_{0},\theta_{0}\right)$, two of
which are in the interval $\varphi\in[0,\pi)$ (shown in the figure)
and two of which are in the interval $\varphi\in[\pi,2\pi)$ (not
shown in the figure). Panels (b) and (e) show the behavior of $R(\hat{\boldsymbol{q}},\alpha_{0}(\hat{\boldsymbol{q}}))$
in the green region of the phase diagram of Fig. \ref{fig:thresholds}(a),
for which $\alpha_{s}<\alpha_{0}<\alpha_{as}$. We note that the number
of maxima of $R(\hat{\boldsymbol{q}},\alpha_{0}(\hat{\boldsymbol{q}}))$
doubles as soon as the nematic director unlocks from $\alpha_{s}$.
Even in this case, it still holds that every nematic director angle
is associated with four soft directions $\hat{\boldsymbol{q}}_{0}$
in momentum space. In panels (c) and (f), we show how the function
$R(\hat{\boldsymbol{q}},\alpha_{0}(\hat{\boldsymbol{q}}))$ looks
like in the blue region of the phase diagram in Fig. \ref{fig:thresholds}(a),
corresponding to $\alpha_{0}=\alpha_{as}$. In this case, the soft
directions in momentum-space approach the value dictated by the structural
instability, see section \ref{subsec:Limit_struc_instab}. Note that
the light-shaded regions in figures \ref{fig:sol_kap1}(c) and (f)
far away from the maxima are an artifact of restricting $\alpha_{0}\in[-\frac{\pi}{3},\frac{\pi}{3}]$. 

\subsection{Analytical approach}

To gain further insight into the numerical results, we perform analytical
approximations to study the maxima of the function $R(\hat{\boldsymbol{q}},\alpha_{0})$
defined in Eq. (\ref{eq:real_R}). We start by rewriting the momentum-dependent
mass function (\ref{eq:M_func}) as 
\begin{align}
M(\hat{\boldsymbol{q}},\alpha_{0}) & =\sum_{j=1}^{3}\frac{\sin^{2}\vartheta_{\hat{\boldsymbol{q}}}^{(j)}}{\tilde{\omega}_{j,\hat{\boldsymbol{q}}}^{2}}\Bigg\{\!\kappa_{1}\cos\!\big(\alpha_{0}\!+\!\varphi\!+\!\phi_{\hat{\boldsymbol{q}}}^{(j)}\big)\nonumber \\
 & \!\!\!\!\!\!\!\!\!\!\!\!\!\!\!\!\!\!-\kappa_{2}\!\left[\cot\!\vartheta_{\hat{\boldsymbol{q}}}^{(j)}\sin\!\big(\alpha_{0}\!-\!\varphi\big)\!+\!\cot\!\theta\sin\!\big(\alpha_{0}\!-\!\phi_{\hat{\boldsymbol{q}}}^{(j)}\big)\!\right]\!\!\!\Bigg\}^{\!\!2}\!,\label{eq:M_func2}
\end{align}
with $\tilde{\omega}_{j,\hat{\boldsymbol{q}}}\equiv\omega_{j,\hat{\boldsymbol{q}}}/\sin\theta$
and the eigenvector and momentum parametrizations
\begin{align}
\hat{\boldsymbol{e}}_{\boldsymbol{q}}^{(j)} & =\left(\cos\phi_{\hat{\boldsymbol{q}}}^{(j)}\sin\vartheta_{\hat{\boldsymbol{q}}}^{(j)},\sin\phi_{\hat{\boldsymbol{q}}}^{(j)}\sin\vartheta_{\hat{\boldsymbol{q}}}^{(j)},\cos\vartheta_{\hat{\boldsymbol{q}}}^{(j)}\right)^{T}\!\!,\label{eq:eigenvector_main}\\
\hat{\boldsymbol{q}} & =\left(\cos\varphi\sin\theta,\sin\varphi\sin\theta,\cos\theta\right)^{T}.\label{eq:momentum_rep}
\end{align}
While the analytical expressions for $\omega_{j,\hat{\boldsymbol{q}}}$
and $\hat{\boldsymbol{e}}_{\boldsymbol{q}}^{(j)}$ in terms of the
elastic constants are given in Appendix \ref{sec:Dynamical-matrix},
we note that the eigenvalues depend on the momentum direction ($\varphi,\theta$)
only through three distinct combinations that involve the elastic
constant $c_{E3}=c_{14}$:
\begin{align*}
\tilde{\omega}_{j,\hat{\boldsymbol{q}}} & =f\left(\cot^{2}\theta,c_{E3}\cot\theta\sin(3\varphi),c_{E3}^{2}\cos(6\varphi)\right).
\end{align*}
Therefore, in what follows, we consider three different asymptotic
limits of $c_{E3}$ that allow us to find simplified analytical expressions
for the eigenvalues and eigenvectors of the dynamic matrix and, consequently,
for the functions $M(\hat{\boldsymbol{q}},\alpha_{0})$ and $R(\hat{\boldsymbol{q}},\alpha_{0})$.
Before delving into these calculations, it is instructive
to consider the two different types of strain fluctuations that contribute
to the effective mass (\ref{eq:M_func2}).

\subsubsection{Contributions from static and dynamic fluctuations\label{subsec:Static-limit}}

The elastic fluctuations $\epsilon_{ij}\left(\mathbf{q}\right)$
present in the system can be thought of as arising from two distinct
contributions (see also Ref. \citep{Paul2017}): one corresponding
to uniform and static strain fluctuations, $\epsilon_{ij}\left(\mathbf{q}=0\right)$,
and the other corresponding to dynamic fluctuations, $\epsilon_{ij}\left(\mathbf{q}\neq0\right)$.
The first one gives simply a trivial shift in the mass term, which
is the same for all momentum-space directions. The second one is responsible
for generating the non-trivial directional dependence of the mass
term. Upon performing the partition function integration in Eq. (\ref{eq:partition_1})
in terms of the lattice displacement fields $D\boldsymbol{u}$, both
contributions are accounted for. To see this, and to disentangle these
two contributions, it is convenient to consider the hypothetical limit
in which only uniform and static strain fluctuations are allowed,
which corresponds to computing the partition function integration
only over the homogeneous strain field $D\boldsymbol{\epsilon}^{D_{3d}}$.
Then, the integration over the static limit of the action (\ref{eq:S_el})
with the nemato-elastic coupling (\ref{eq:S_el_nem}) leads to the
following uniform renormalization of the nematic action:
\begin{align}
\mathcal{S}_{\mathrm{stat}} & =-\frac{1}{2}\frac{V}{T}M_{\mathrm{stat}}|\boldsymbol{\Phi}_{0}^{E_{g}}|^{2},\label{eq:S_stat}
\end{align}
with the static mass
\begin{align}
M_{\mathrm{stat}} & =\frac{c_{E2}\kappa_{1}^{2}+c_{E1}\kappa_{2}^{2}-2c_{E3}\kappa_{1}\kappa_{2}}{\mathsf{d}_{E}}.\label{eq:M_stat}
\end{align}

\begin{figure}[t] 
\raggedright{}
\ffigbox[][]
{\includegraphics[scale=0.35]{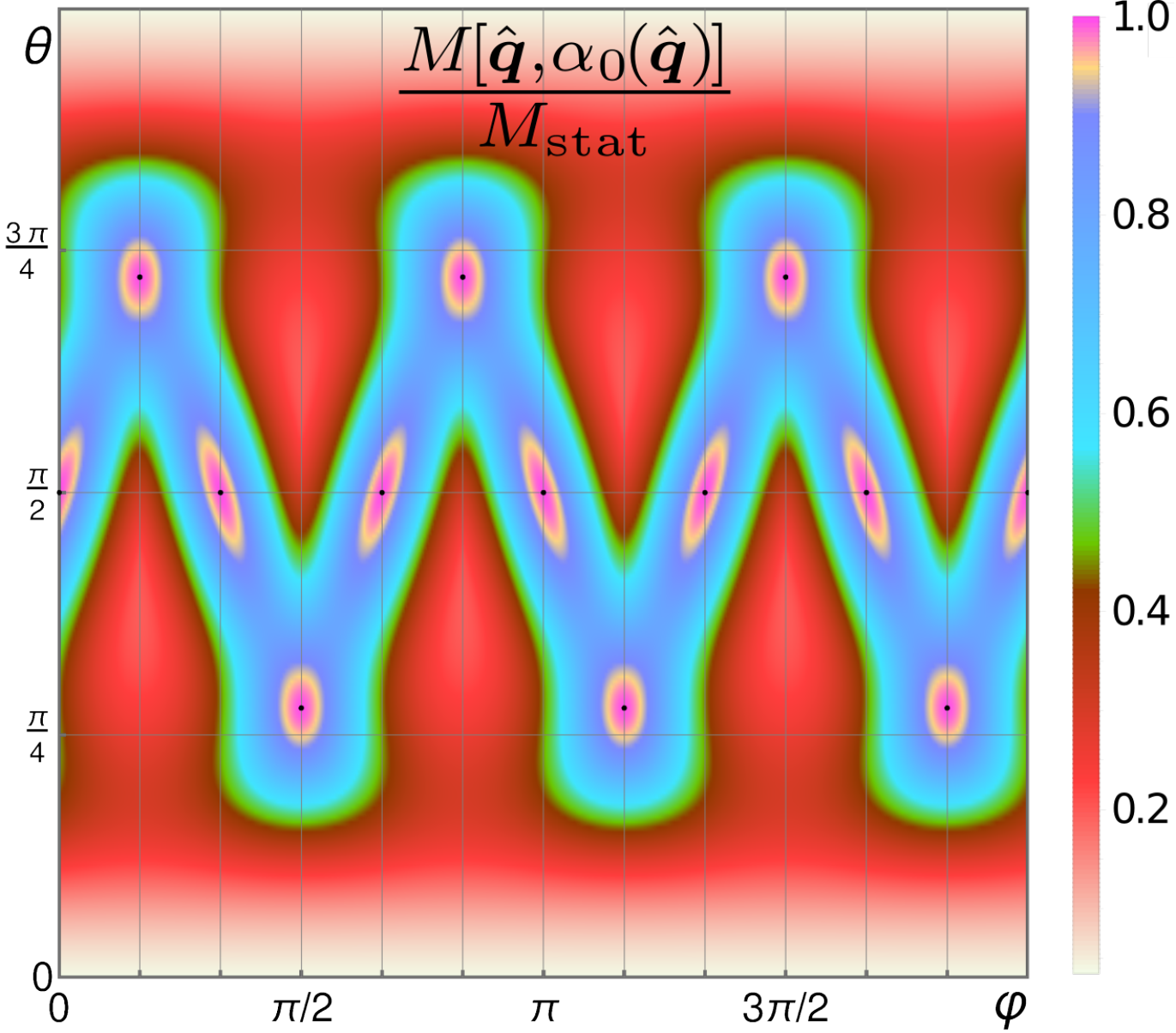}}   
{\caption{The renormalized mass (\ref{eq:M_func2}) as a function of the momentum directions $\hat{\boldsymbol{q}}=\hat{\boldsymbol{q}}(\varphi,\theta)$ with the respective maximized director angle $\alpha_0(\hat{\boldsymbol{q}})$ inserted according to Eq.~(\ref{eq:tan_alpha}). The renormalized mass attains its maximum value $M[\hat{\boldsymbol{q}},\alpha_0(\hat{\boldsymbol{q}})]=M_\mathrm{stat}$ for the twelve directions $\hat{\boldsymbol{q}}_{1,2}$, see Eqs. (\ref{eq:q1})-(\ref{eq:q2}), indicated by the black dots. The displayed features do not depend on the chosen parameters; for this plot, we set $(\tilde{\kappa}_1,\tilde{\kappa}_2)=(2.7,0.9)$ and used the elastic constant values for $\mathrm{Bi_2Se_3}$. The static mass $M_\mathrm{stat}$ is defined in Eq. (\ref{eq:M_stat}).} 
\label{fig:MoverMstat1}}      
\end{figure}

As expected, the mass renormalization is larger
the closer the system is to a pure structural transition, corresponding
to $\mathsf{d}_{E}\rightarrow0$. The key point is that the static
mass (\ref{eq:M_stat}) corresponds to the maximum value that the
renormalized mass $M(\hat{\boldsymbol{q}},\alpha_{0})$ can possibly
attain \textendash{} which only happens for specific momentum directions
$\hat{\boldsymbol{q}}_{0}$ and nematic director angles $\alpha_{0}$.
In more mathematical terms, the quadratic part of the effective action
(\ref{eq:S_prime_2}) can be rewritten as 
\begin{align}
\mathcal{S}_{\mathrm{eff},\hat{\boldsymbol{q}}}^{(2)} & =\frac{1}{2}\big[r\!-M_{\mathrm{stat}}+\!\delta M(\hat{\boldsymbol{q}},\alpha_{0})\big]|\boldsymbol{\Phi}_{0}^{E_{g}}|^{2},\label{eq:Seff_2}
\end{align}
where $\delta M(\hat{\boldsymbol{q}},\alpha_{0})\equiv M_{\mathrm{stat}}-M(\hat{\boldsymbol{q}},\alpha_{0})\geq0$
denotes the energy cost associated with the angle arrangements, and
$\delta M(\hat{\boldsymbol{q}},\alpha_{0})=0$ can only be attained
for specific directions. Since a rigorous analytic deduction of these
directions is not feasible (except for the special case of $c_{E3}=0$
that we study below), we present the numerically evaluated ratio $M(\hat{\boldsymbol{q}},\alpha_{0})/M_{\mathrm{stat}}$
in momentum space in Fig. \ref{fig:MoverMstat1} where we have inserted
the maximum angle $\alpha_{0}$ following from Eq. (\ref{eq:M_func}):
\begin{align}
\tan(2\alpha_{0}) & =-\Pi_{(\boldsymbol{q},0)}^{E_{g},2}/\Pi_{(\boldsymbol{q},0)}^{E_{g},1}.\label{eq:tan_alpha}
\end{align}
As shown in Fig. \ref{fig:MoverMstat1}, there are twelve distinct
momentum directions for which $M(\hat{\boldsymbol{q}},\alpha_{0})$
acquires its maximum value, which is equal to $M_{\mathrm{stat}}$.
We verified that the qualitative features of $M(\hat{\boldsymbol{q}},\alpha_{0})$
do not depend on the choices of $\left|\tilde{c}_{E3}\right|$, $\kappa_{1}$,
and $\kappa_{2}$. Having identified all these maxima to be located
at integer multiples of $\pi/6$ with respect to the azimuthal angle
$\varphi$, we can analytically determine the twelve directions. We
denote the six in-plane directions as $\hat{\boldsymbol{q}}_{1}$
and the six out-of-plane directions as $\hat{\boldsymbol{q}}_{2}$
which are defined through
\begin{align}
\hat{\boldsymbol{q}}_{1} & : & \varphi_{1} & =\!\frac{\pi}{6}n_{1}, & \cot\theta_{1} & =0,\label{eq:q1}\\
\hat{\boldsymbol{q}}_{2} & : & \varphi_{2} & =\!\frac{\pi}{6}n_{2}, & \cot\theta_{2} & =(\text{-}1)^{\frac{n_{2}+\!1}{2}}\frac{c_{E1}\kappa_{2}\!-\!c_{E3}\kappa_{1}}{c_{E2}\kappa_{1}\!-\!c_{E3}\kappa_{2}},\label{eq:q2}
\end{align}
with $n_{1}\in\{0,2,4,6,8,10\}$ and $n_{2}\in\{1,3,5,7,9,11\}$.
In the following, we demonstrate that along these momentum-directions\textemdash and
for the appropriately chosen director angle\textemdash the renormalized
mass indeed attains its maximum value $M_{\mathrm{stat}}$.

For the in-plane directions $\hat{\boldsymbol{q}}_{1}$,
the eigenvalues $\tilde{\omega}_{j,\hat{\boldsymbol{q}}}\equiv\omega_{j,\hat{\boldsymbol{q}}}/\sin\theta$
simplify to
\begin{align}
\tilde{\omega}_{1,\boldsymbol{q}_{1}}^{2} & =c_{A1}\!+\!c_{E1},\label{eq:omega1_q1}\\
\tilde{\omega}_{2,\boldsymbol{q}_{1}}^{2} & =\frac{c_{E1}\!+\!c_{E2}}{2}\left[1\!-\!\sqrt{1\!-\!4\mathsf{d}_{E}/(c_{E1}\!+\!c_{E2})^{2}}\,\right],\label{eq:omega2_q1}\\
\tilde{\omega}_{3,\boldsymbol{q}_{1}}^{2} & =\frac{c_{E1}\!+\!c_{E2}}{2}\left[1\!+\!\sqrt{1\!-\!4\mathsf{d}_{E}/(c_{E1}\!+\!c_{E2})^{2}}\,\right],\label{eq:omega3_q1}
\end{align}
whereas the eigenvectors (\ref{eq:eigenvector_main}) are parametrized
by:
\begin{align}
\phi_{\hat{\boldsymbol{q}}_{1}}^{(1)}\! & =\varphi_{1}, & \sin\vartheta_{\hat{\boldsymbol{q}}_{1}}^{(1)}\! & =1, & \cos\vartheta_{\hat{\boldsymbol{q}}_{1}}^{(1)}\! & =0,\nonumber \\
\phi_{\hat{\boldsymbol{q}}_{1}}^{(2)}\! & =\pi\!+\!\frac{\pi}{2}\!\frac{c_{E3}}{|c_{E3}|}\!-\!2\varphi_{1}, & \sin\vartheta_{\hat{\boldsymbol{q}}_{1}}^{(2)}\! & =\gamma^{-}, & \cos\vartheta_{\hat{\boldsymbol{q}}_{1}}^{(2)}\! & =\gamma^{+},\nonumber \\
\phi_{\hat{\boldsymbol{q}}_{1}}^{(3)}\! & =\pi\!-\!\frac{\pi}{2}\!\frac{c_{E3}}{|c_{E3}|}\!-\!2\varphi_{1}, & \sin\vartheta_{\hat{\boldsymbol{q}}_{1}}^{(3)}\! & =\gamma^{+}, & \cos\vartheta_{\hat{\boldsymbol{q}}_{1}}^{(3)}\! & =\gamma^{-}.\label{eq:eigs_q1}
\end{align}
In these expressions, we defined:
\begin{align*}
\gamma^{\pm} & =\frac{1}{\sqrt{2}}\Big(1\pm\frac{c_{E1}\!-\!c_{E2}}{\sqrt{\big(c_{E1}\!+\!c_{E2}\big)^{2}\!-\!4\mathsf{d}_{E}}}\Big)^{\frac{1}{2}}.
\end{align*}
The insertion into the renormalized mass (\ref{eq:M_func2}) leads
to
\begin{align}
M(\hat{\boldsymbol{q}}_{1},\alpha_{0})= & M_{1}\cos^{2}\left(\alpha_{0}\!+\!\frac{\pi}{3}n_{1}\right)+M_{\mathrm{stat}}\sin^{2}\left(\alpha_{0}\!+\!\frac{\pi}{3}n_{1}\right),\label{eq:M_q1}
\end{align}
with $M_{1}=\kappa_{1}^{2}/(c_{A1}\!+\!c_{E1})<M_{\mathrm{stat}}$,
see App. \ref{sec:Supplemental-to-analytical}. 

Before we further analyze Eq. (\ref{eq:M_q1}),
we derive a similar expression for the out-of-plane directions $\hat{\boldsymbol{q}}_{2}$.
To do this, we introduce the direction $\hat{\boldsymbol{q}}_{\bar{2}}=\hat{\boldsymbol{q}}_{\bar{2}}[\varphi_{2},\theta]$
which shares the same azimuthal angle with $\hat{\boldsymbol{q}}_{2}$
but keeps $\theta$ arbitrary, such that $\hat{\boldsymbol{q}}_{2}=\hat{\boldsymbol{q}}_{\bar{2}}[\varphi_{2},\theta_{2}]$.
For the directions $\hat{\boldsymbol{q}}_{\bar{2}}$ the eigensystem
can be derived as 
\begin{align}
\tilde{\omega}_{1,\hat{\boldsymbol{q}}_{\bar{2}}}^{2} & =c_{E2}\!+\!c_{A2}\cot^{2}\!\theta+\frac{1}{2}\lambda_{1,\theta}+\!\frac{1}{2}\lambda_{0,\theta},\label{eq:omega1_q2}\\
\tilde{\omega}_{2,\hat{\boldsymbol{q}}_{\bar{2}}}^{2} & =c_{E1}+2c_{E3}(\text{-}1)^{\frac{n_{2}+1}{2}}\cot\theta+c_{E2}\cot^{2}\theta\label{eq:omega2_q2}\\
\tilde{\omega}_{3,\hat{\boldsymbol{q}}_{\bar{2}}}^{2} & =c_{E2}\!+\!c_{A2}\cot^{2}\!\theta+\frac{1}{2}\lambda_{1,\theta}-\!\frac{1}{2}\lambda_{0,\theta}\!,\label{eq:omega3_q2}
\end{align}
with
\begin{align}
\phi_{\hat{\boldsymbol{q}}_{\bar{2}}}^{(1)}\! & =\!\varphi_{2}\!+\!\frac{\pi}{2}\!-\!\frac{\pi}{2}\!\frac{\lambda_{2,\theta}}{|\lambda_{2,\theta}|}, & \sin\vartheta_{\hat{\boldsymbol{q}}_{\bar{2}}}^{(1)}\! & =\!\beta_{\theta}^{+}, & \cos\vartheta_{\hat{\boldsymbol{q}}_{\bar{2}}}^{(1)}\! & =\!\beta_{\theta}^{-},\nonumber \\
\phi_{\hat{\boldsymbol{q}}_{\bar{2}}}^{(2)}\! & =\!\varphi_{2}-\frac{\pi}{2}, & \sin\vartheta_{\hat{\boldsymbol{q}}_{\bar{2}}}^{(2)}\! & =\!1, & \cos\vartheta_{\hat{\boldsymbol{q}}_{\bar{2}}}^{(2)}\! & =\!0,\nonumber \\
\phi_{\hat{\boldsymbol{q}}_{\bar{2}}}^{(3)}\! & =\!\varphi_{2}\!+\!\frac{3\pi}{2}\!-\!\frac{\pi}{2}\!\frac{\lambda_{2,\theta}}{|\lambda_{2,\theta}|}, & \sin\vartheta_{\hat{\boldsymbol{q}}_{\bar{2}}}^{(3)}\! & =\!\beta_{\theta}^{-}, & \cos\vartheta_{\hat{\boldsymbol{q}}_{\bar{2}}}^{(3)}\! & =\!\beta_{\theta}^{+},\label{eq:eigs_q2}
\end{align}
and the following auxiliary functions:
\begin{align}
\lambda_{0,\theta} & =\sqrt{\lambda_{1,\theta}^{2}+\lambda_{2,\theta}^{2}},\label{eq:lambda0}\\
\lambda_{1,\theta} & =c_{A1}\!+\!c_{E1}\!-\!c_{E2}\!-\!2c_{E3}(\text{-}1)^{\frac{n_{2}+1}{2}}\cot\!\theta\!\nonumber \\
 & \quad+\!(c_{E2}\!-\!c_{A2})\cot^{2}\!\theta,\label{eq:lambda1}\\
\lambda_{2,\theta} & =2(c_{A3}+c_{E2})\cot\theta-2c_{E3}(\text{-}1)^{\frac{n_{2}+1}{2}},\label{eq:lambda2}\\
\beta_{\theta}^{\pm} & =\frac{1}{\sqrt{2}}\sqrt{1\pm\frac{\lambda_{1,\theta}}{\lambda_{0,\theta}}}\,.\label{eq:betapm}
\end{align}
Again, we insert these expressions into the renormalized mass (\ref{eq:M_func2})
to find \begin{figure}[t] 
\raggedright{}
\ffigbox[][]
{\includegraphics[scale=0.45]{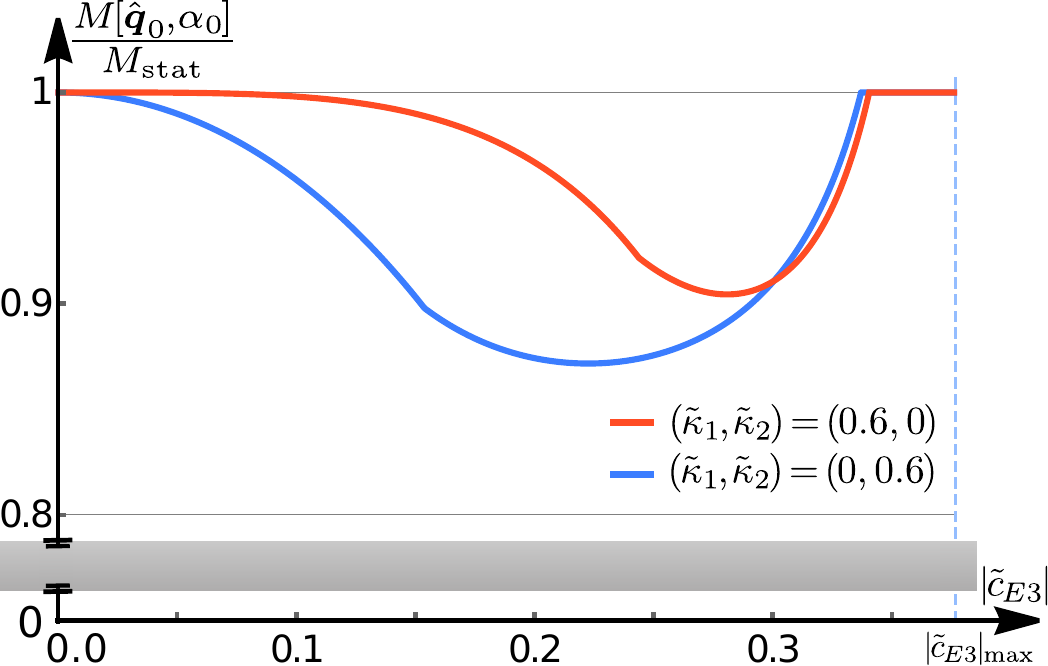}}   
{\caption{The ratio of the renormalized mass (\ref{eq:M_func2}) and the static mass (\ref{eq:M_stat}) as a function of $c_{E3}$ for the two maxima of the function $R(\hat{\boldsymbol{q}},\alpha_{0})$ presented in Fig. \ref{fig:max_evol}. The renormalized mass is maximum, i.e. $M[\hat{\boldsymbol{q}}_0,\alpha_0]=M_\mathrm{stat}$, for $c_{E3}=0$, and above the upper threshold value beyond which the acoustic phonon contribution is dominant over the anharmonic contribution.} 
\label{fig:MoverMstat2}}      
\end{figure}
\begin{align}
M(\hat{\boldsymbol{q}}_{\bar{2}},\alpha_{0})= & M_{\theta}^{c}\cos^{2}\left(\alpha_{0}\!+\!\frac{\pi}{3}n_{1}\right)+M_{\theta}^{s}\sin^{2}\left(\alpha_{0}\!+\!\frac{\pi}{3}n_{1}\right),\label{eq:M_q2}
\end{align}
with $M_{\theta}^{s}$, and $M_{\theta}^{c}$ defined in Appendix
\ref{sec:Supplemental-to-analytical}. The maximum amplitude of Eq.
(\ref{eq:M_q2}) is $M_{\mathrm{stat}}$ and it
is reached at the polar angle $\theta_{2}$ {[}Eq. (\ref{eq:q2}){]}
for which it holds $M_{\theta_{2}}^{s}=M_{\mathrm{stat}}$.

Having derived the renormalized mass expressions
for the twelve momentum-directions, $M(\hat{\boldsymbol{q}}_{1},\alpha_{0})$
in Eq. (\ref{eq:M_q1}) and $M(\hat{\boldsymbol{q}}_{2},\alpha_{0})=M(\hat{\boldsymbol{q}}_{\bar{2}}[\varphi_{2},\theta_{2}],\alpha_{0})$
in Eq. (\ref{eq:M_q2}), it is straight-forward to identify the corresponding
nematic director angles $\alpha_{0}$ for which $M(\hat{\boldsymbol{q}}_{1,2},\alpha_{0})=M_{\mathrm{stat}}$.
In both cases, the condition becomes

\begin{align}
\alpha_{0}\!+\!\frac{\pi}{3}n_{1,2} & =\frac{\pi}{2}\{\pm1,\pm3,\dots\}.\label{eq:alpha_condition}
\end{align}
Equation (\ref{eq:alpha_condition}) is only satisfied for director
angles that align with the ``anti-symmetry'' directions $\alpha_{0}=\alpha_{as}\in\frac{\pi}{6}\{1,3,5,7,9,11\}$.
Each of these nematic directors $\alpha_{0}=\alpha_{as}$ entails
four soft directions in momentum space, two within the $\hat{\boldsymbol{q}}_{1}$
manifold and two within the $\hat{\boldsymbol{q}}_{2}$ manifold.
For example, for $\alpha_{0}=\frac{\pi}{6}$, the four momentum directions
that maximize the renormalized mass are parametrized by $n_{1}=\{4,10\}$
and $n_{2}=\{1,7\}$. 

This analysis provides an interesting insight about
the two contributions to the mass term $M(\hat{\boldsymbol{q}},\alpha_{0})$.
While the uniform and static strain fluctuations enhance the tendency
towards nematic order for all momentum directions ($M_{\mathrm{stat}}$),
the dynamic fluctuations penalizes those directions that do not conform
to the constraints imposed by the anisotropy of the phonon dispersions
\textendash{} i.e. those for which $\delta M(\hat{\boldsymbol{q}},\alpha_{0})>0$.
As a result, only certain momentum directions become soft at the transition.

It is important to note that the total function
$R(\hat{\boldsymbol{q}},\alpha_{0})$ in Eq. (\ref{eq:real_R}), which
needs to be maximized to give the leading instability, also contains\textemdash besides
the mass term $M(\hat{\boldsymbol{q}},\alpha_{0})$\textemdash the
anharmonic contribution $\cos^{2}(3\alpha_{0})$:

\begin{equation}
R(\hat{\boldsymbol{q}},\alpha_{0})=M(\hat{\boldsymbol{q}},\alpha_{0})+\frac{g^{2}\cos^{2}(3\alpha_{0})}{2u}.
\end{equation}
The last term favors the nematic director to align with the high-symmetry
directions $\alpha_{s}\in\{0,1,2,3,4,5\}\frac{\pi}{3}$. It is the
competition between these two terms that gives rise to the rich phase
diagram obtained numerically in Fig. \ref{fig:thresholds}.
Indeed, as shown in Fig. \ref{fig:MoverMstat2}, the mass term $M(\hat{\boldsymbol{q}},\alpha_{0})$
computed at the maximum of $R(\hat{\boldsymbol{q}},\alpha_{0})$ can
not overcome $M_{\mathrm{stat}}$. It reaches equality in the hexagonal
limit where $|c_{E3}|=0$, and above the upper threshold value when
the system approaches the pure structural phase transition $|c_{E3}|\rightarrow\sqrt{c_{E1}c_{E2}}$.
In the remainder of this section, we analyze these two regimes asymptotically,
as well as the intermediate regime perturbatively.

\subsubsection{Limit $|c_{E3}|_{\mathrm{max}}=\sqrt{c_{E1}c_{E2}}$\label{subsec:Limit_struc_instab}}

In the limit $|c_{E3}|_{\mathrm{max}}=\sqrt{c_{E1}c_{E2}}$, corresponding
to $\mathsf{d}_{E}=c_{E1}c_{E2}-c_{E3}^{2}=0$, the elastic action
(\ref{eq:S_el}) becomes unstable, see Eq. (\ref{eq:stab_cond1}).
In other words, one of the sound velocities $\omega_{j,\hat{\boldsymbol{q}}}$
of the dynamic matrix vanishes, and the system undergoes a pure structural
phase transition. In this limit, corresponding to the vicinity of
the top dashed light-blue line in figure \ref{fig:thresholds}, the
renormalized mass term (\ref{eq:M_func2}) diverges along certain
directions. These soft momentum directions are,
of course, the $12$ directions $\hat{\boldsymbol{q}}_{1}$ and $\hat{\boldsymbol{q}}_{2}$
defined in Sec. \ref{subsec:Static-limit} along which the renormalized
mass attains its maximum value $M_{\mathrm{stat}}$. The phonon mode
that becomes soft corresponds to the eigenvalue $\omega_{2,\hat{\boldsymbol{q}}_{1}}$
in Eq. (\ref{eq:omega2_q1}) and $\omega_{2,\hat{\boldsymbol{q}}_{2}}$
in Eq. (\ref{eq:omega2_q2}). Indeed, in the limit $\mathsf{d}_{E}\rightarrow0$,
they simplify to:
\begin{align}
\tilde{\omega}_{2,\boldsymbol{q}_{1}}^{2} & \approx\frac{1}{c_{E1}\!+\!c_{E2}}\mathsf{d}_{E},\label{eq:omega_q1_2}\\
\tilde{\omega}_{2,\hat{\boldsymbol{q}}_{2}}^{2} & =\frac{c_{E2}\kappa_{1}^{2}+c_{E1}\kappa_{2}^{2}-2c_{E3}\kappa_{1}\kappa_{2}}{\left(c_{E2}\kappa_{1}-c_{E3}\kappa_{2}\right)^{2}}\mathsf{d}_{E}.\label{eq:omega_q2_2}
\end{align}

Because $M_{\mathrm{stat}}\sim1/\mathsf{d}_{E}$,
the mass-term contribution to $R(\hat{\boldsymbol{q}},\alpha_{0})$
is much larger than the anharmonic contribution, which has coefficient
$g^{2}/2u$. As a result, the maxima of $R(\hat{\boldsymbol{q}},\alpha_{0})$
coincide with the maxima of $M(\hat{\boldsymbol{q}},\alpha_{0})$
in the regime where $\mathsf{d}_{E}\rightarrow0$. Hence, according
to the condition (\ref{eq:alpha_condition}), the nematic director
angle aligns with the ``anti-symmetry'' directions $\alpha_{0}=\alpha_{as}\in\frac{\pi}{6}\{1,3,5,7,9,11\}$,
in agreement with the findings depicted in Fig. \ref{fig:sol_kap1}(c,f).
Interestingly, for the elastic constants values reported for $\mathrm{Bi_{2}Se_{3}}$,
the soft polar angle (\ref{eq:q2}) approaches 
\begin{align*}
\cot\theta_{2} & \rightarrow\pm\sqrt{\frac{c_{E1}}{c_{E2}}}=\pm\sqrt{\frac{753}{754}}\approx\pm1,
\end{align*}
at the pure structural instability. Therefore, the soft polar angle
is very close to $\theta_{2}\approx\{\pi/4,\,3\pi/4\}$, as can be
seen in Fig. \ref{fig:sol_kap1}(c,f) or Fig. \ref{fig:max_evol}. 

It is important to note that when $\alpha_{0}=\alpha_{as}$, the cubic
term of the action (\ref{eq:S_prime_2}) vanishes and the nematic
transition becomes second-order\textemdash at least within our mean-field
approximation. This is reflected in our formalism by the fact that
the jump of the nematic order parameter in Eq. (\ref{eq:AbsC_val})
vanishes when $\alpha_{0}=\alpha_{as}$. Consequently, the condition
(\ref{eq:sign_cond}) does not need to be satisfied in this case.

\subsubsection{Limit $c_{E3}=0$\label{subsec:Limit_zero}}

In this limit, the elastic properties become the same as that of a
hexagonal lattice. The eigenvalues and eigenvectors
are given by Eqs. (\ref{eq:omega1_q2})-(\ref{eq:betapm}) with $c_{E3}=0$
and $\varphi_{2}\rightarrow\varphi$. Inserting these expressions
into the renormalized mass (\ref{eq:M_func2}) gives: \begin{align} 
M\hspace{-0.1em}(\hspace{-0.08em}\hat{\boldsymbol{q}} \hspace{-0.05em},\hspace{-0.08em} \alpha_{0}\hspace{-0.08em})\hspace{-0.05em}
& =\hspace{-0.08em} \kappa_{1}^{2}A_{\theta}^{-} \hspace{-0.12em} \cos(\hspace{-0.05em}2\alpha_{0}\hspace{-0.18em}+\!4\varphi \hspace{-0.08em})
\!+\!\kappa_{2}^{2}B_{\theta}^{-} \hspace{-0.12em} \cos(\hspace{-0.08em}2\alpha_{0}\hspace{-0.15em}-\!2\varphi \hspace{-0.08em})
\nonumber \\  
& \!\!\!\!\!\!\!\!\!\!\!\!\!\!\!\!\!+\!\kappa_{1}^{2}A_{\theta}^{+}
\!+\!\kappa_{2}^{2}B_{\theta}^{+}
\!-\!2\kappa_{1}\kappa_{2}\!\left(\hspace{-0.08em}C_{\theta}^{+}\!\sin(\hspace{-0.08em}2\alpha_{0}\hspace{-0.15em}+\hspace{-0.15em}\varphi\hspace{-0.08em})
\!-\!C_{\theta}^{-}\!\sin(\hspace{-0.08em}3\varphi\hspace{-0.08em})\hspace{-0.08em}\right)\!.
\label{eq:M_zero_cE3} \end{align} The functions $A_{\theta}^{\pm}$, $B_{\theta}^{\pm}$ and $C_{\theta}^{\pm}$
depend only on the polar angle $\theta$, and are given explicitly
in Appendix \ref{sec:Supplemental-to-analytical}. Now, in the limit
$c_{E3}=0$, for consistency one must also impose $\kappa_{2}=0$,
since the symmetry that enforces a vanishing $c_{E3}$ also makes
the out-of-plane and in-plane shear strain doublets in Eq. (\ref{eq:strain_doublets})
belong to different irreducible representations. Then, the function
$R(\hat{\boldsymbol{q}},\alpha_{0})$ in Eq. (\ref{eq:R_func2}) becomes

\begin{align}
R(\hat{\boldsymbol{q}},\alpha_{0}) & =\kappa_{1}^{2}A_{\theta}^{+}+\kappa_{1}^{2}A_{\theta}^{-}\cos(2\alpha_{0}\!+\!4\varphi)+\frac{g^{2}\cos^{2}(3\alpha_{0})}{2u}.\label{eq:R_ce30}
\end{align}
As demonstrated in Appendix \ref{sec:Supplemental-to-analytical},
the function (\ref{eq:R_ce30}) is maximized with respect to $\theta$
at $\theta_{0}=\pi/2$, leading to
\begin{align}
R(\varphi,\theta_{0},\alpha_{0}) & =\kappa_{1}^{2}\frac{c_{E1}\!+\!c_{A1}\sin^{2}(\alpha_{0}\!+\!2\varphi)}{\left(c_{A1}\!+\!c_{E1}\right)c_{E1}}\!+\!\frac{g^{2}\cos^{2}(3\alpha_{0})}{2u}.\label{eq:R2_ce30}
\end{align}
Note that the first term in the expression (\ref{eq:R2_ce30})
is the mass term, whose maximum is given by $\kappa_{1}^{2}/c_{E1}$.
This agrees with the expression for $M_{\mathrm{stat}}$ {[}Eq. (\ref{eq:M_stat}){]}
in the limit $\kappa_{2}=0$, $c_{E3}=0$. Moreover, in contrast to
the case where $c_{E3}\neq0$, the values of $\varphi$ that maximize
the mass term are not restricted to the discrete values given by Eqs.
(\ref{eq:q1}) and (\ref{eq:q2}). On the contrary, maximization of
the mass term alone only restricts the combination $\alpha_{0}+2\varphi=\frac{\pi}{2}n$,
with $n=\{1,3,5,7\}$. This is also related to the fact that $\theta_{2}=\theta_{1}=\pi/2$
in Eq. (\ref{eq:q2}).

Because of this peculiarity of the $c_{E3}=0$
term, the two terms that contribute to $R(\varphi,\theta_{0},\alpha_{0})$
in Eq. (\ref{eq:R2_ce30}) can be simultaneously maximized with respect
to $\varphi$ and $\alpha_{0}$. We obtain:
\begin{align}
\alpha_{0} & =\alpha_{s}, & \varphi & =\frac{\pi}{4}n-\frac{1}{2}\alpha_{0}.\label{eq:varphi_ce30}
\end{align}
Thus, the nematic director aligns with the high-symmetry directions,
and it is associated with the four momentum directions that correspond
to $n=\{1,3,5,7\}$. This result is in agreement with the findings
of Ref. \citep{Fernandes2020}, where the case of the $\mathsf{D_{6}}$
point group was considered. 

\subsubsection{Expansion in $|c_{E3}|$ \label{subsec:expansion}}

The fact that the two limiting cases $c_{E3}=0$ and $|c_{E3}|=\sqrt{c_{E1}c_{E2}}$
favor $\alpha_{0}=\alpha_{s}$ and $\alpha_{0}=\alpha_{as}$, respectively,
suggests that the nematic director has to rotate as $|c_{E3}|$ is
increased. We thus expand the renormalized mass in powers of small
$|c_{E3}|$ to elucidate how the rotation actually occurs. Formally,
we have: \begin{align} 
M\hspace{-0.1em}(\hspace{-0.08em}\hat{\boldsymbol{q}},\hspace{-0.08em} \alpha_{0}\hspace{-0.08em})\hspace{-0.05em} 
& =\!\kappa_{1}^{2}\hspace{-0.05em}\Big(\!M_{\hat{\boldsymbol{q}},\alpha_{0}}^{(0)}\!+\!M_{\hat{\boldsymbol{q}},\alpha_{0}}^{(2)}c_{E3}^{2}\!+\!M_{\hat{\boldsymbol{q}},\alpha_{0}}^{(4)}c_{E3}^{4}\hspace{-0.1em}+\hspace{-0.08em}\dots\hspace{-0.1em}\Big)\hspace{-0.15em}.
\label{eq:expand_ren_mass} \end{align}To keep the analysis transparent, we set $\kappa_{2}=0$. Additionally,
we expand the momentum directions according to
\begin{align}
\varphi & =\varphi^{(0)}+\varphi^{(1)}c_{E3}+\varphi^{(2)}c_{E3}^{2}+\varphi^{(3)}c_{E3}^{3}+\dots,\label{eq:phi_expand}\\
\cot\theta & =\Theta^{(0)}+\Theta^{(1)}c_{E3}+\Theta^{(2)}c_{E3}^{2}+\Theta^{(3)}c_{E3}^{3}+\dots.\label{cot_thethaexpand}
\end{align}
We are now in position to maximize the renormalized mass order by
order in $c_{E3}$.

The \textit{zeroth} order contribution to $M_{\hat{\boldsymbol{q}},\alpha_{0}}$
is given by Eq. (\ref{eq:M_zero_cE3}) with $A_{\theta}^{\pm}$ defined
in Appendix \ref{sec:Supplemental-to-analytical}. Also shown in the
Appendix \ref{sec:Supplemental-to-analytical} is the analysis to
determine the corresponding maxima at 
\begin{align}
\Theta^{(0)} & =0, & \varphi^{(0)} & =\frac{\pi}{4}N_{0}-\frac{1}{2}\alpha_{0},\label{eq:phi_0}
\end{align}
with $N_{0}=\{1,3,5,7\}$. This recovers the result that, for $c_{E3}=0$,
the maxima reside on the equator, $\text{\ensuremath{\theta=\pi/2}}$
{[}see Fig. \ref{fig:max_evol}(b){]}. Then, the zeroth order contribution
to the mass becomes 
\begin{align}
M_{\hat{\boldsymbol{q}},\alpha_{0}}^{(0)} & =\frac{1}{c_{E1}},\label{eq:M_zero}
\end{align}
which is independent of $\alpha_{0}$. 

To \textit{second} order, the renormalized mass is given by
\begin{align}
M_{\hat{\boldsymbol{q}},\alpha_{0}}^{(2)} & =\frac{-4c_{A1}\left(\varphi^{(1)}\right)^{2}}{\left(c_{A1}\!+\!c_{E1}\right)c_{E1}}-\frac{c_{E2}}{c_{E1}^{2}}\!\left(\!\Theta^{(1)}\!-\!\Theta_{0}^{(1)}\!\right)^{2}\!\!+\!\frac{1}{c_{E2}c_{E1}^{2}},\label{eq:M_2}
\end{align}
which is maximized by 
\begin{align}
\varphi^{(1)} & \!=\!0, & \Theta^{(1)} & \!=\Theta_{0}^{(1)}\!=\!\frac{1}{c_{E2}}\sin\!\Big(\frac{3\pi}{4}N_{0}-\frac{3\alpha_{0}}{2}\!\Big).\label{eq:cot_1}
\end{align}
Because Eq. (\ref{eq:M_2}) remains independent of the nematic director
angle $\alpha_{0}$, the latter is still determined solely by the
contribution arising from the bare nematic action, namely, the $\cos^{2}(3\alpha_{0})$
term in Eq. (\ref{eq:real_R}), which favors $\alpha_{0}=\alpha_{s}$.
Therefore, to describe the unlocking of the nematic director from
the high-symmetry directions, it is necessary to go to higher-order
in $c_{E3}$.

The \textit{fourth}-order contribution to the renormalized mass is
given by: 
\begin{align}
M_{\hat{\boldsymbol{q}},\alpha_{0}}^{(4)} & =-\frac{c_{E2}}{c_{E1}^{2}}\big(\Theta^{(2)}\big)^{2}-\frac{4c_{A1}\left(\varphi^{(2)}\!-\!\varphi_{0}^{(2)}\right)^{2}}{\left(c_{A1}\!+\!c_{E1}\right)c_{E1}}\nonumber \\
 & \quad-R_{1}^{(4)}\cos^{2}(3\alpha_{0})+\frac{1}{c_{E2}^{2}c_{E1}^{3}}.\label{eq:M_4}
\end{align}
Maximization leads to the second-order corrections to the angles:
\begin{align}
\Theta^{(2)} & =0, & \varphi^{(2)} & \!=\varphi_{0}^{(2)}\!=\!(-1)^{\frac{1\!+\!N_{0}}{2}}\frac{c_{A3}\cos(3\alpha_{0})}{4c_{E2}^{2}c_{A1}}.\label{eq:phi_2}
\end{align}
In these expressions, we defined 

\begin{align*}
R_{1}^{(4)} & =\frac{\mathsf{d}_{A}}{4c_{A1}c_{E2}^{4}c_{E1}^{2}}>0.
\end{align*}
Importantly, the fourth-order contribution, Eq. (\ref{eq:M_4}),
contains the same $\cos^{2}(3\alpha_{0})$ dependence as that arising
from the bare nematic action in Eq. (\ref{eq:real_R})\textemdash however,
with an opposite sign as $R_{1}^{(4)}>0$ is positive by definition.
Thus, while the bare nematic action favors the nematic director $\alpha_{0}=\alpha_{s}$
to be aligned with the high-symmetry direction, the fourth-order contribution
in Eq. (\ref{eq:M_4}) favors a director $\alpha_{0}=\alpha_{as}\in\frac{\pi}{6}\{1,3,5,7,9,11\}$
that aligns with an ``anti-symmetry'' direction. This demonstrates
the antagonistic contributions to the nematic director angle coming
from the phonons and from the bare nematic action. Nonetheless, the
contribution (\ref{eq:M_4}) is not sufficient to account for the
smooth evolution of the nematic director angle, and higher-order corrections
are required. 

The \textit{sixth}-order contribution to $M_{\hat{\boldsymbol{q}},\alpha_{0}}$
is given by
\begin{align}
M_{\hat{\boldsymbol{q}},\alpha_{0}}^{(6)} & =\frac{1}{c_{E2}^{3}c_{E1}^{4}}-\frac{4c_{A1}\left(\varphi^{(3)}\right)^{2}}{(c_{A1}\!+\!c_{E1})c_{E1}}-\frac{c_{E2}}{c_{E1}^{2}}\big(\!\Theta^{(3)}\!-\!\Theta_{0}^{(3)}\big)^{2}\nonumber \\
 & -R_{1}^{(6)}\cos^{2}(3\alpha_{0}),\label{eq:M_6}
\end{align}
whose maximization enforces the third-order corrections 
\begin{align}
\varphi^{(3)} & \!=0, & \Theta^{(3)} & \!=\Theta_{0}^{(3)},\label{eq:cot_3}
\end{align}
with
\begin{align*}
\Theta_{0}^{(3)} & =\!\Big(\frac{c_{A3}}{c_{A1}}\!-\!\frac{2\mathsf{d}_{A}}{c_{A1}c_{E2}}\Big)\frac{\cos(3\alpha_{0})}{4c_{E2}^{3}}\sin\!\Big(\frac{\pi}{4}N_{0}\!-\!\frac{3\alpha_{0}}{2}\!\Big),\\
R_{1}^{(6)} & =\frac{\mathsf{d}_{A}}{4c_{A1}c_{E2}^{6}c_{E1}^{2}}\Big(\frac{c_{A3}}{c_{A1}}\!+\!\frac{2c_{E2}}{c_{E1}}\!-\!\frac{\mathsf{d}_{A}}{c_{A1}c_{E2}}\Big)>0.
\end{align*}
Thus, like the fourth-order contribution, Eq. (\ref{eq:M_4}), the
sixth-order term (\ref{eq:M_6}) only reduces the prefactor of the
$\cos^{2}(3\alpha_{0})$ term in Eq. (\ref{eq:real_R}), which is
again maximized by $\alpha_{0}=\alpha_{s}$.

Eventually, it is the \textit{eighth}-order contribution to the renormalized
mass that unlocks the director from the high-symmetry directions.
We find:\begin{figure}[t] 
\raggedright{}
\ffigbox[][]
{\includegraphics[scale=0.48]{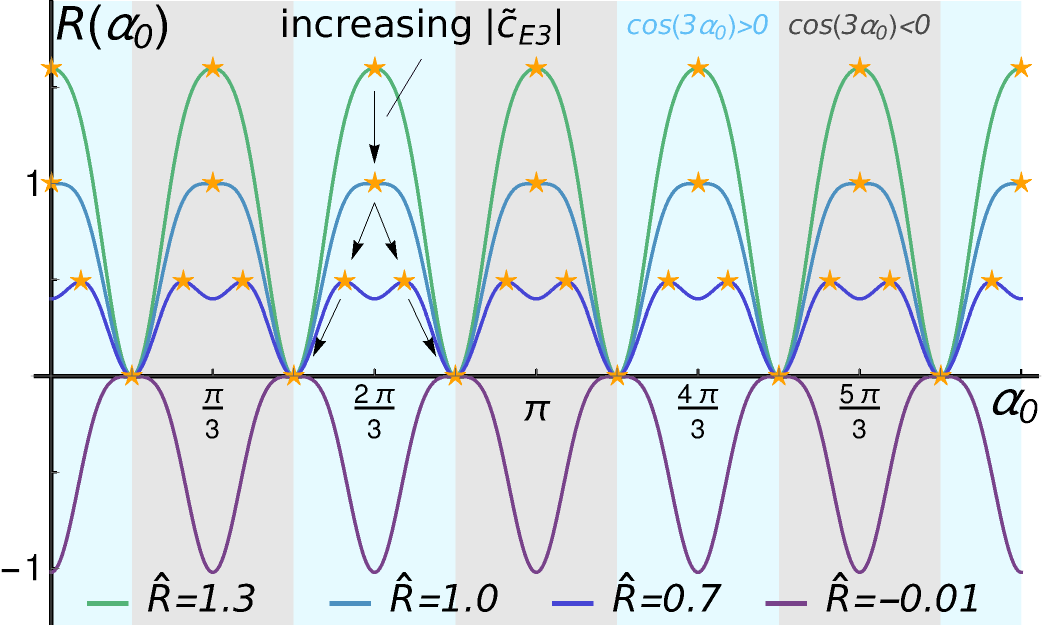}}   
{\caption{The analytical function $R(\alpha_0)$, given by Eq. (\ref{eq:R_func4}), for four distinct values of the parameter $\hat{R}=R_1/2R_2$. Due to the condition (\ref{eq:sign_cond}), relevant for $\hat{R} \geq 0$, the solution is restricted to either the blue or the gray regions. The evolution of the maximum is indicated by the black arrows as $\hat{R}$ is decreased---or, likewise, $|c_{E3}|$ is increased. The two threshold values are given by $\hat{R}=1$ and $\hat{R}=0$.} 
\label{fig:func_R_expand}}      
\end{figure}
\begin{align}
M_{\hat{\boldsymbol{q}},\alpha_{0}}^{(8)} & =\frac{1}{c_{E2}^{4}c_{E1}^{5}}-\frac{c_{E2}}{c_{E1}^{2}}\big(\Theta^{(4)}\big)^{2}-\frac{4c_{A1}\left(\varphi^{(4)}\!-\!\varphi_{0}^{(4)}\right)^{2}}{\left(c_{A1}\!+\!c_{E1}\right)c_{E1}}\nonumber \\
 & +R_{1}^{(8)}\cos^{2}(3\alpha_{0})-R_{2}^{(8)}\cos^{4}(3\alpha_{0}),\label{eq:M_8}
\end{align}
which enforces the fourth-order corrections 
\begin{align}
\Theta^{(4)} & \!=0, & \varphi^{(4)} & \!=\varphi_{0}^{(4)},\label{eq:phi_4}
\end{align}
with the definitions of $\varphi_{0}^{(4)}$, $R_{1}^{(8)}$ and $R_{2}^{(8)}$
shown in Appendix \ref{sec:Supplemental-to-analytical}. Crucially,
the eighth-order expression (\ref{eq:M_8}) has an additional $-R_{2}^{(8)}\cos^{4}(3\alpha_{0})$
dependence on $\alpha_{0}$ that is different from the bare term $\cos^{2}(3\alpha_{0})$.
The fact that $R_{2}^{(8)}>0$ is important as it allows for a smooth
$\alpha_{0}$ evolution. 

\begin{figure}[t] 
\raggedright{}
\ffigbox[][]
{\includegraphics[scale=0.48]{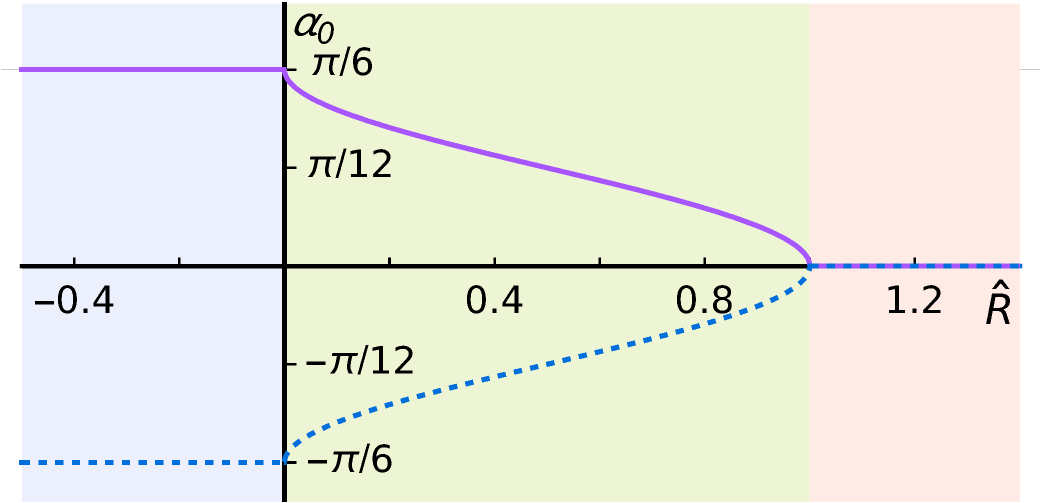}}   
{\caption{The nematic director $\alpha_0$ that maximizes Eq. (\ref{eq:R_func4}) as a function of $\hat{R}=R_1/2R_2$. The trends displayed here coincide with those obtained from the numerical solutions depicted in Fig. \ref{fig:max_evol}.} 
\label{fig:nematic_angle}}      
\end{figure}

To see this, we insert the expressions above into the function $R$,
Eq. (\ref{eq:real_R}), which then becomes 
\begin{align}
R(\alpha_{0}) & =R_{0}+R_{1}\cos^{2}(3\alpha_{0})-R_{2}\cos^{4}(3\alpha_{0})\label{eq:R_func3}\\
 & =R_{0}+\frac{R_{1}^{2}}{4R_{2}}-R_{2}\left[\cos^{2}(3\alpha_{0})-\frac{R_{1}}{2R_{2}}\right]^{2},\label{eq:R_func4}
\end{align}
with:
\begin{align}
R_{0} & =\frac{1}{c_{E1}}\sum_{n=0}^{4}\frac{c_{E3}^{2n}}{c_{E2}^{n}c_{E1}^{n}},\label{eq:R0_def}\\
R_{1} & =\frac{1}{2\tilde{\kappa}_{1}^{2}}-R_{1}^{(4)}c_{E3}^{4}-R_{1}^{(6)}c_{E3}^{6}+R_{1}^{(8)}c_{E3}^{8},\label{eq:R1_def}\\
R_{2} & =R_{2}^{(8)}c_{E3}^{8}.\label{eq:R2_def}
\end{align}
Note that a factor of $1/\kappa_{1}^{2}$ was absorbed into the function
$R(\alpha_{0})$ for convenience. Maximization of Eq. (\ref{eq:R_func4})
leads to three distinct regimes, depending on the value of the parameter
$\hat{R}=\frac{R_{1}}{2R_{2}}$. We find 
\begin{align}
1 & <\hat{R}, & \rightarrow &  & \cos(3\alpha_{0}) & =\pm1,\label{eq:reg1}\\
0\leq\hat{R} & \leq1, & \rightarrow &  & \cos(3\alpha_{0}) & =\pm\sqrt{\hat{R}},\label{eq:reg2}\\
\hat{R} & <0, & \rightarrow &  & \cos(3\alpha_{0}) & =0.\label{eq:reg3}
\end{align}
The function $R(\alpha_{0})$, Eq. (\ref{eq:R_func4}), is depicted
in Fig.~\ref{fig:func_R_expand} for four values of $\hat{R}$. Because
of the condition (\ref{eq:sign_cond}), valid solutions for the nematic
director in the case $\hat{R}\geq0$ lie either in the gray or in
the blue shaded regions, depending on the sign of the cubic parameter
$g$. As $\hat{R}$ is decreased\textemdash or $|c_{E3}|$ is increased\textemdash the
number of maxima doubles once $\hat{R}$ falls below the threshold
$\hat{R}=1$. We emphasize that the rotation can still happen within
the perturbative regime of $|c_{E3}|$, as long as $\tilde{\kappa}_{1}^{-1}\sim|c_{E3}|^{2}\ll1$.
The evolution of the nematic director angle $\alpha_{0}$ with $\hat{R}$
is plotted in Fig. \ref{fig:nematic_angle}, with the three phases
identified through the colored background. To make the comparison
with the numerical solution more transparent, we added in Fig. \ref{fig:thresholds}(a)
the curves $c_{E3}\left(\kappa_{1}\right)$ corresponding to the two
threshold values, $\hat{R}=0$ and $\hat{R}=1$, which separate the
three different regimes for the nematic director. Note that the analytic
results quantitatively capture the numerical ones when the threshold
value for $c_{E3}$ is small, which corresponds to larger $\tilde{\kappa}_{1}$.
Moreover, in agreement with the numerical solution, each nematic director
$\alpha_{0}$ is associated with four soft phonon directions given
by $N_{0}=\{1,3,5,7\}$. The actual momentum directions $\hat{\boldsymbol{q}}_{0}$
can be computed in a straightforward way via Eqs. (\ref{eq:phi_0}),
(\ref{eq:cot_1}), (\ref{eq:phi_2}), (\ref{eq:cot_3}) and (\ref{eq:phi_4}). 

\begin{figure*}[t]
\centering{}\ffigbox[][]{\includegraphics[scale=0.3]{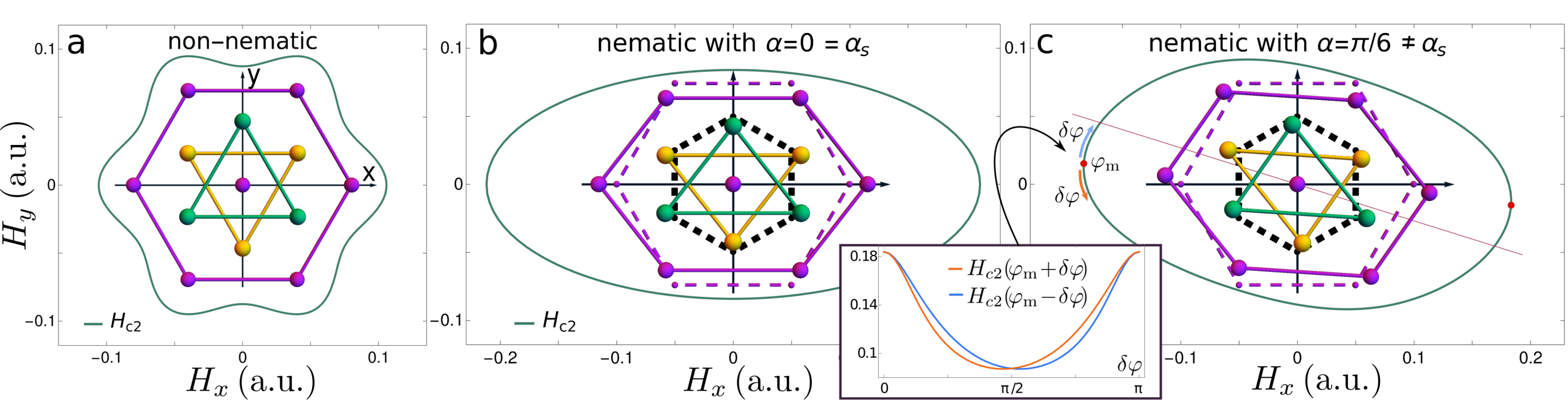}}  
{\caption{Shape of the in-plane upper critical field $H_{c2}$ superimposed to the schematic distortion of the unit cell. (a) Without nematic order, the upper critical field is sixfold symmetric and respects all of the trigonal point group symmetries. (b) With nematic order and $\alpha = \alpha_s$, the shape of $H_{c2}$ is a twofold symmetric ellipse, and the accompanying lattice distortion is monoclinic. (c) With a rotated nematic director ($\alpha \neq \alpha_s$), the residual in-plane twofold symmetry of $H_{c2}$ is lost, and the accompanying lattice distortion is triclinic. The main axis of $H_{c2}$ rotates by $\text{-}\alpha /2$ (red line), and the ellipse is deformed, as highlighted in the inset. The inset shows the behavior of the functions  $H_{c2}(\varphi_\mathrm{m}\pm \delta \varphi)$, which correspond to a clockwise and a counterclockwise sweeping of the $H_{c2}$ curve starting from the angle $\varphi_\mathrm{m}$ where $H_{c2}$ is maximum. If a residual twofold symmetry was present, the two curves would overlap.} \label{fig:Hc2}} 
\end{figure*}

\section{Experimental manifestations in doped $\mathrm{Bi}{}_{2}\mathrm{Se}{}_{3}$
\label{sec:Implications-for-doped}}

Having established that nemato-elastic interactions in trigonal lattices
tend to rotate the nematic director away from high-symmetry directions
($\alpha\neq\alpha_{s}$), we now discuss some of the experimentally
observable consequences in the context of the topological superconductor
$A\mathrm{_{x}Bi_{2}Se_{3}}$. As explained above, using the elastic
constant values extracted from first-principles for $\mathrm{Bi_{2}Se_{3}}$
\citep{Gao2016} (black dotted line in Fig. \ref{fig:thresholds}),
we expect the nematic director to be rotated in this compound. The
degree of rotation depends on the nemato-elastic coupling constants
$\tilde{\kappa}_{i}$, whose values are currently not known.

The first consequence of a rotated director is the breaking of the
residual $C_{2x}$ (twofold rotation with respect to an in-plane axis)
symmetry of the point group $\mathsf{D_{3d}}$. While any non-zero
$|\boldsymbol{\Phi}^{E_{g}}|$ breaks $C_{3z}$ (threefold rotation
with respect to an out-of-plane axis), the $C_{2x}$ symmetry is only
broken when $\alpha\neq\alpha_{s}$. To see this, we study the invariance
of a generic nematic order parameter $\boldsymbol{\Phi}^{E_{g}}$
upon the transformation of the group elements $g\in\mathsf{D_{3d}}$,
\begin{align}
\mathcal{R}_{E_{g}}(g)\,\boldsymbol{\Phi}^{E_{g}} & =\boldsymbol{\Phi}^{E_{g}},\label{eq:inv_cond}
\end{align}
with the $E_{g}$-transformation matrices $\mathcal{R}_{E_{g}}(g)$.
Depending on whether $\alpha$ aligns with a high-symmetry direction
or not, the residual symmetry group $\mathcal{G}$ is different. In
particular:

\begin{align}
\mathcal{G}_{\alpha=\alpha_{s}} & =\mathsf{C_{2h}}=\{E,I,C_{2x},IC_{2x}\},\label{eq:res_group1}\\
\mathcal{G}_{\alpha\neq\alpha_{s}} & =\mathsf{C_{I}}\;\;=\{E,I\}.\label{eq:res_group2}
\end{align}
Here, $E$ denotes the identity and $I$, inversion. Thus, for a rotated
director ($\alpha\neq\alpha_{s}$), as expected for doped $\mathrm{Bi_{2}Se_{3}}$,
there is no residual twofold symmetry axis. The first consequence
of this result is that the nematic transition triggers a triclinic
lattice distortion, rather than a monoclinic distortion as in the
case of $\alpha=\alpha_{s}$. While the lattice distortion may be
very small, it would be interesting to perform high-resolution x-ray
measurements to try to resolve between a monoclinic or a triclinic
lattice structure. The breaking of the $C_{2x}$ symmetry should also
be manifested in any physical quantity that depends on in-plane directions,
such as the in-plane $H_{c2}$, the penetration depth, and the thermal
conductivity. Perhaps the most accessible of these quantities is the
in-plane upper critical field $H_{c2}$.

\subsection{Upper critical field}

For a nematic director aligned along the high-symmetry directions,
$\alpha=\alpha_{s}$, the azimuthal function $H_{c2}(\varphi_{B})$,
where $\varphi_{B}$ is the in-plane angle with respect to the $x$-axis,
has an elliptical shape with the major axis oriented along (or perpendicular
to) $\text{-}\alpha/2$, see \citep{Venderbos2016,Hecker2020}. However,
once $\alpha\neq\alpha_{s}$, $H_{c2}(\varphi_{B})$ no longer has
a twofold symmetry axis. To illustrate this behavior, we follow Refs.
\citep{Venderbos2016,Hecker2020} and compute $H_{c2}$, with the
detailed derivation shown in Appendix \ref{sec:UpperCritField}. The
results are shown in the three panels of Fig. \ref{fig:Hc2}. In the
non-nematic phase {[}panel (a){]}, $H_{c2}(\varphi_{B})$ has a sixfold
symmetric shape. In the nematic phase with $\alpha=\alpha_{s}$, shown
in panel (b), $H_{c2}$ displays an approximately elliptical shape,
being invariant upon an in-plane twofold rotation about a high-symmetry
axis ($C_{2x}$). The orientation of the ellipse can be obtained from
the approximated analytical expression
\begin{align}
H_{c2} & \approx\frac{H_{c2}^{0}}{1+\frac{1}{2}\hat{\mathsf{d}}_{1}\cos(\alpha+2\varphi_{B})},\label{eq:Hc2_main}
\end{align}
with the details provided in the Appendix \ref{sec:UpperCritField}.
In Eq. (\ref{eq:Hc2_main}) the contributions associated with the
threefold rotational symmetry are neglected. Lastly, when the nematic
director unlocks from the high-symmetry directions ($\alpha\neq\alpha_{s}$),
as depicted in panel (c) and its inset, the elliptical shape of $H_{c2}$
is distorted and no longer symmetric under any in-plane twofold rotation.
The shape of $H_{c2}$ in panel (c) can be understood as a superposition
of a rotated ellipse {[}see Eq. (\ref{eq:Hc2_main}){]} and the underlying
sixfold symmetric pattern illustrated in panel (a). While the lack
of $C_{2x}$ symmetry is a robust prediction of the model, the degree
in which the ellipse of panel (b) is distorted when $\alpha\neq\alpha_{s}$
can be rather small. For instance, in panel (c), the nematic director
was chosen to be aligned with an ``anti-symmetry'' direction, where
the effect is the strongest. The absence of any twofold symmetry in
the $H_{c2}$ curve is emphasized in the inset of panel (c), where
the two curves $H_{c2}(\varphi_{\mathrm{m}}\pm\delta\varphi)$ are
plotted as functions of $\delta\varphi$, with $\varphi_{\mathrm{m}}$
denoting the angle where $H_{c2}$ is maximal. Since the ``clockwise''
and ``counterclockwise'' curves do not overlap, $H_{c2}$ lacks
twofold symmetry with respect to any in-plane axes. In all three panels,
we also show schematically the symmetry of the unit cell in each case,
corresponding to trigonal {[}non-nematic, panel (a){]}, monoclinic
{[}nematic with $\alpha=\alpha_{s}$, panel (b){]}, and triclinic
{[}nematic with $\alpha\neq\alpha_{s}$, panel (c){]}.

\begin{figure}[t] 
\raggedright{}
\ffigbox[][]
{\includegraphics[scale=1.1]{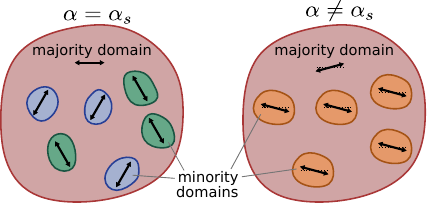}}   
{\caption{In the case $\alpha = \alpha_s$, the system is expected to form one majority nematic domain, e.g. $\alpha =0$ that is accompanied by minority domains randomly composed of the remaing two states, $\alpha \in \{\frac{2\pi}{3},\frac{4\pi}{3}\}$. In the case $\alpha \neq \alpha_s$, there are six non-equivalent nematic directors. The system establishes one majority domain, e.g. $\alpha = \delta$ with $0<\delta <\pi/3$. Then, the surface energy $\sigma(\delta,-\delta)$ between adjacent domains, $\alpha = \delta$ and $\alpha = -\delta$, is expected to be smaller than the surface energy between other domains, e.g. $\sigma(\delta,\frac{2\pi}{3}+\delta)$. As a result, the minority domains are expected to be dominated by the $\alpha = -\delta$ domains.} 
\label{fig:domains}}      
\end{figure}

\subsection{Domain formation}

The unlocking of the nematic director also has potential implications
for domain formation, as illustrated in Fig. \ref{fig:domains}. Consider
the surface energy cost $\sigma(\alpha_{1},\alpha_{2})$ to form two
neighboring domains with directors $\alpha_{1}$ and $\alpha_{2}$.
In the case $\alpha=\alpha_{s}$, the three possible directors have
the same angular separation, and we expect the surface energies between
any two domains to be equal. As a result, in the equilibrium state,
one director is chosen as the majority domain, with the other two
orientations randomly forming minority domains, see Fig. \ref{fig:domains}
(left panel). In the case $\alpha\neq\alpha_{s}$, however, there
are six degenerate directors that can be parametrized as $\alpha\in\{\pm\delta,\frac{2\pi}{3}\pm\delta,\frac{4\pi}{3}\pm\delta\}$,
with $0<\delta<\frac{\pi}{3}$. Given an arbitrary director (say $+\delta$)
there is an adjacent director that has a lower angular separation
than any other director (in this case, $-\delta$). As a result, we
expect the surface energy between domains with adjacent directors
to be smaller than the surface energy between domains with distant
directors, e.g. $\sigma(\delta,-\delta)<\sigma(\delta,\frac{2\pi}{3}+\delta)$.
As a result, for a given majority domain, e.g. $\alpha=\delta$, the
minority domains in equilibrium should be dominated by the $\alpha=-\delta$
domain, see Fig. \ref{fig:domains} (right panel). An interesting
question, which is however outside of the scope of this paper, is
if these results also affect the typical size of the nematic domains
in a macroscopic sample.

\subsection{Gap structure \label{subsec:Nodal-vs.-fully-gapped}}

The rotation of the nematic director angle also
has a direct impact on the gap structure, and particularly on the
presence or absence of nodal quasi-particles. As explained in the
Introduction, the nematic superconducting gap transforms according
to the two-dimensional irreducible representation $E_{u}$ and can
be parametrized according to:
\begin{align}
\boldsymbol{\Delta} & =\left(\begin{array}{c}
\Delta_{1}\\
\Delta_{2}
\end{array}\right)=|\boldsymbol{\Delta}|e^{\mathsf{i}\vartheta}\left(\begin{array}{c}
\cos\frac{\alpha}{2}\\
-\sin\frac{\alpha}{2}
\end{array}\right),\label{eq:Delta_def}
\end{align}
where $\vartheta\in[0,2\pi)$ is the global phase and $\alpha\in[0,2\pi)$,
the nematic director angle. Without the effects of the phonon renormalization,
the director angle $\alpha$ aligns with one of the high-symmetry
directions given by either $\alpha_{s}^{(1)}\in\{0,2,4\}\frac{\pi}{3}$
or $\alpha_{s}^{(2)}\in\{1,3,5\}\frac{\pi}{3}$. As discussed in Ref.
\citep{Fu2014}, $\boldsymbol{\Delta}(\alpha_{s}^{(1)})$ describes
a fully-gapped superconducting state whereas $\boldsymbol{\Delta}(\alpha_{s}^{(2)})$
is a nodal pairing state with point nodes protected by the $C_{2x}$
symmetry of the system. Following the arguments of Ref. \citep{Fu2014},
the stability of the nodes can be seen by noting that the $\boldsymbol{d}$-vector
that describes this triplet superconducting state is given in terms
of $\Delta_{1}$ and $\Delta_{2}$ according to:
\begin{align}
\boldsymbol{d}(\boldsymbol{k}) & =\Delta_{1}\boldsymbol{d}_{1}(\boldsymbol{k})+\Delta_{2}\boldsymbol{d}_{2}(\boldsymbol{k}),\label{eq:d_vector}
\end{align}
with the individual $\boldsymbol{d}_{j}$-vectors satisfying $\boldsymbol{d}_{j}(-\boldsymbol{k})=-\boldsymbol{d}_{j}(\boldsymbol{k})$
where $j=\{1,2\}$. Nodes emerge when all three components of $\boldsymbol{d}(\boldsymbol{k})$
vanish on either points or lines along the Fermi surface, which usually
only happens due to an underlying crystallographic symmetry \citep{Blount1985,Fu2014}.
Indeed, under the $C_{2x}$ symmetry operation, the $\boldsymbol{d}_{j}$-vectors
transform as 
\begin{align}
\boldsymbol{d}_{1}[\mathcal{R}_{v}^{\dagger}(C_{2x})\boldsymbol{k}] & =\mathcal{R}_{v}^{\dagger}(C_{2x})\boldsymbol{d}_{1}(\boldsymbol{k}),\label{eq:d1_C2x}\\
\boldsymbol{d}_{2}[\mathcal{R}_{v}^{\dagger}(C_{2x})\boldsymbol{k}] & =-\mathcal{R}_{v}^{\dagger}(C_{2x})\boldsymbol{d}_{2}(\boldsymbol{k}),\label{eq:d2_C2x}
\end{align}
with $\mathcal{R}_{v}^{\dagger}(C_{2x})$ denoting the transformation
matrix of the vector representation. Using the result $\mathcal{R}_{v}^{\dagger}(C_{2x})\boldsymbol{k}_{yz}=-\boldsymbol{k}_{yz}$,
which holds for any momentum $\boldsymbol{k}_{yz}=(0,k_{y},k_{z})^{T}$
in the $\left(k_{y},k_{z}\right)$ plane, i.e. the $IC_{2x}$ mirror
plane, Eqs. (\ref{eq:d1_C2x})-(\ref{eq:d2_C2x}) become
\begin{align*}
0 & =[\mathbbm{1}\!+\!\mathcal{R}_{v}^{\dagger}(C_{2x})]\,\boldsymbol{d}_{1}(\boldsymbol{k}_{yz}), & 0 & =[\mathbbm{1}\!-\!\mathcal{R}_{v}^{\dagger}(C_{2x})]\,\boldsymbol{d}_{2}(\boldsymbol{k}_{yz}).
\end{align*}
This results in $d_{1x}(\boldsymbol{k}_{yz})=0$ and $d_{2y}(\boldsymbol{k}_{yz})=d_{2z}(\boldsymbol{k}_{yz})=0$,
i.e. the vector $\boldsymbol{d}_{1}(\boldsymbol{k}_{yz})$ is parallel
to the $\left(k_{y},k_{z}\right)$ plane, whereas $\boldsymbol{d}_{2}(\boldsymbol{k}_{yz})$
is normal to it. This has important consequences for $\boldsymbol{d}_{2}(\boldsymbol{k}_{yz})$,
since the odd-parity constraint, $d_{2x}(-\boldsymbol{k}_{yz})=-d_{2x}(\boldsymbol{k}_{yz})$,
implies that $d_{2x}(\boldsymbol{k}_{yz})$ vanishes at least along
one line in the $\left(k_{y},k_{z}\right)$ plane. The intersection
of this line with the Fermi surface then leads to a point node \textendash{}
assuming, of course, that the Fermi surface also crosses this plane.
Thus, the twofold symmetry $C_{2x}$, via (\ref{eq:d2_C2x}), forces
the superconducting state described by the order parameter $\boldsymbol{\Delta}(\alpha_{s}^{(2)}=\pi)=-|\boldsymbol{\Delta}|e^{\mathsf{i}\vartheta}(0,1)^{T}$,
as well as its $\alpha_{s}^{(2)}$ partners, to have point nodes. 

Therefore, when the nematic director angle unlocks
from the high-symmetry directions, $\alpha\neq\alpha_{s}^{(1,2)}$,
the superconducting order parameter (\ref{eq:Delta_def}) rotates
accordingly and the system loses the symmetry element $C_{2x}$ that
protects the point nodes, see Eq. (\ref{eq:res_group2}). As a result,
the superconducting state becomes fully gapped \citep{Fu2014}. To
show this explicitly, we consider the well-established $\boldsymbol{k}\cdot\boldsymbol{p}$
Hamiltonian for Bi$_{2}$Se$_{3}$ \citep{Zhang2009,Liu2010,Venderbos2016a}
and write down an expression for the magnitude of the $\boldsymbol{d}$-vector
following the approach in Ref. \citep{Venderbos2016a} (see Appendix
\ref{sec:-model-Hamiltonian} for details). We obtain 
\begin{align}
\frac{|\boldsymbol{d}(\boldsymbol{k})|^{2}}{|\boldsymbol{\Delta}|^{2}} & =\big(\hat{f}_{\boldsymbol{k}}^{z}\big)^{2}+\hat{M}_{\boldsymbol{k}}^{2}\big(\hat{f}_{\boldsymbol{k}}^{C_{3}}\big)^{2}+\Big(\hat{f}_{\boldsymbol{k}}^{x}\sin\frac{\alpha}{2}+\hat{f}_{\boldsymbol{k}}^{y}\cos\frac{\alpha}{2}\Big)^{2}\nonumber \\
 & \!\!\!\!\!\!\!\!\!\!\!\!\!\!+\big(\hat{f}_{\boldsymbol{k}}^{C_{3}}\big)^{2}\frac{\hat{\boldsymbol{f}}_{\boldsymbol{k}}^{2}\!+\!2|\hat{M}_{\boldsymbol{k}}|(1\!+\!|\hat{M}_{\boldsymbol{k}}|)}{(1\!+\!|\hat{M}_{\boldsymbol{k}}|)^{2}}\Big(\hat{f}_{\boldsymbol{k}}^{x}\cos\!\frac{\alpha}{2}-\hat{f}_{\boldsymbol{k}}^{y}\sin\!\frac{\alpha}{2}\Big)^{2},\label{eq:gap_expr}
\end{align}
which is a sum of individually positive contributions. Hence, the
gap only vanishes when the three terms
\begin{align}
0 & =\hat{f}_{\boldsymbol{k}}^{z}, & 0 & =\hat{f}_{\boldsymbol{k}}^{C_{3}}, & 0 & =\hat{f}_{\boldsymbol{k}}^{x}\sin\frac{\alpha}{2}+\hat{f}_{\boldsymbol{k}}^{y}\cos\frac{\alpha}{2},\label{eq:conds}
\end{align}
are simultaneously equal to zero. Note that $\hat{M}_{\boldsymbol{k}}$
is generally different from zero. While the full expressions for these
functions are given in Appendix \ref{sec:-model-Hamiltonian}, the
important point is that $\hat{f}_{\boldsymbol{k}}^{z}$ transforms
as the $A_{2u}$ irreducible representation; $(\hat{f}_{\boldsymbol{k}}^{x},\,\hat{f}_{\boldsymbol{k}}^{y})$,
as $E_{u}$; and $\hat{f}_{\boldsymbol{k}}^{C_{3}}$, as $A_{1u}$.
Consequently, $\hat{f}_{\boldsymbol{k}}^{C_{3}}$ only vanishes along
the three momentum-space directions $k_{x}=0$ and $k_{x}=\pm\sqrt{3}k_{y}$,
which define the three mirror planes. Focusing on the $k_{x}=0$ plane,
the requirement $\hat{f}_{\boldsymbol{k}}^{z}=0$ implies $k_{z}\sim k_{y}^{3}$,
which defines a line within the $k_{x}=0$ plane. Along this line,
because $\hat{f}_{\boldsymbol{k}}^{x}=0$ and $\hat{f}_{\boldsymbol{k}}^{y}\neq0$,
the last condition in (\ref{eq:conds}) is satisfied only for $\alpha=\pi$.
Repeating the same steps for the other two planes, we find that the
last condition is satisfied for specific values of the nematic director
\begin{align*}
k_{x} & =0 & \rightarrow\alpha & =\pi,\\
k_{x} & =\sqrt{3}k_{y} & \rightarrow\alpha & =5\pi/3,\\
k_{x} & =-\sqrt{3}k_{y} & \rightarrow\alpha & =\pi/3,
\end{align*}
which coincide with the three high-symmetry directions $\alpha=\alpha_{s}^{(2)}$.
Any other director angle $\alpha\neq\alpha_{s}^{(2)}$ thus necessarily
leads to a full superconducting gap, as first shown in Ref. \citep{Fu2014}.
For the parameters of $\mathrm{Bi_{2}Se_{3}}$, because the director
is unlocked from the high-symmetry directions due to the phonon renormalization,
we expect always a fully gapped superconducting state.

\section{Concluding remarks \label{sec:Concluding-remarks}}

In lattices with threefold or sixfold rotational symmetry, the nematic
order parameter is defined not only by an amplitude, but also by the
orientation of the nematic director $\alpha$. Usually, one expects
this director to align with a high-symmetry direction of the crystal,
$\alpha=\alpha_{s}$. In this work, we have shown that the orientation
of the nematic director can be fundamentally changed by the nemato-elastic
coupling, due to the long-range nematic interactions mediated by the
acoustic phonons. This is the case for any $Z_{3}$-Potts nematic
order in trigonal lattices with point groups $\mathsf{D_{3d}}$, $\mathsf{D_{3}}$,
$\mathsf{C_{3v}}$, $\mathsf{S_{6}}$ and $\mathsf{C_{3}}$, but not
for hexagonal lattices with point groups $\mathsf{D_{6h}}$, $\mathsf{D_{6}}$,
$\mathsf{C_{6h}}$, $\mathsf{C_{6v}}$, $\mathsf{D_{3h}}$, $\mathsf{C_{3h}}$
and $\mathsf{C_{6}}$. This is a consequence of the fact that only
in the former groups the in-plane shear strain $\boldsymbol{\epsilon}^{E_{g},1}=(\epsilon_{11}-\epsilon_{22},-2\epsilon_{12})^{T}$
and out-of-plane shear strain $\boldsymbol{\epsilon}^{E_{g},2}=(2\epsilon_{23},-2\epsilon_{31})^{T}$
transform as the same two-dimensional irreducible representation,
which results in the emergence of a symmetry-allowed elastic constant
$c_{14}$. By minimizing the acoustic-phonon renormalized nematic
action, we found that when either $c_{14}$ or the nemato-elastic
coupling overcomes a threshold value, the nematic director unlocks
from the high-symmetry directions ($\alpha\neq\alpha_{s}$), resulting
in the breaking of a residual twofold rotational symmetry with respect
to an in-plane axis, $C_{2x}$. 

In doped $\mathrm{Bi_{2}Se_{3}}$, with point group $\mathsf{D_{3d}}$,
the value of $c_{14}$ extracted from first-principles calculations
place the system in the regime where the nematic director is rotated
with respect to the high-symmetry directions. In this regime, the
number of non-equivalent nematic directors doubles from three to six,
and each director is associated with four momentum-space directions
for which the nematic susceptibility is large. Moreover, we showed
that the breaking of $C_{2x}$ is manifested not only by a triclinic
distortion of the lattice, but also by an in-plane critical field
curve that retains only inversion symmetry and
by the complete removal of any point nodes that could otherwise exist
inside the superconducting state. Experimental verification of these
features would provide strong evidence for the fundamental impact
of the lattice on the nematic superconducting state of doped $\mathrm{Bi_{2}Se_{3}}$.
In this regard, we note that Refs. \citep{Du2016,Kuntsevich2018,Kuntsevich2019,Tao2018}
reported a mismatch between the long-axis of the in-plane $H_{c2}$
``ellipse'' and the lattice axes, which would be consistent with
our model.
\begin{acknowledgments}
We thank J. Schmalian, J. Venderbos and Z. Wang for fruitful discussions.
This work was supported by the U. S. Department of Energy, Office
of Science, Basic Energy Sciences, Materials Sciences and Engineering
Division, under Award No. DE-SC0020045.
\end{acknowledgments}

\bibliographystyle{apsrev4-1-titles}
\bibliography{phon_ren_nem_axis}

\appendix

\section{Irreducible representations of the shear strain doublets\label{sec:Strain-doublet-degeneracy}}

In this appendix, we discuss the conditions under which the two strain
doublets $\boldsymbol{\epsilon}^{(1)}=(\epsilon_{11}-\epsilon_{22},-2\epsilon_{12})^{T}$
(in-plane shear) and $\boldsymbol{\epsilon}^{(2)}=(2\epsilon_{23},-2\epsilon_{31})^{T}$
(out-of-plane shear) transform as the same irreducible representations
of a point group for which $C_{3z}$ is a symmetry element. To this
end, we consider the largest hexagonal point group $\mathsf{D_{6h}}$,
which has $24$ symmetry elements that can be conveniently written
as 
\begin{align}
\mathsf{D_{6h}} & =\{E,C_{3z}^{\pm1}\}\otimes\{E,C_{2x}\}\otimes\{E,C_{2z}\}\otimes\{E,I\}.\label{eq:D6h}
\end{align}
Among the elements in the brackets, the only element under which the
strain doublets $\boldsymbol{\epsilon}^{(1)}$ and $\boldsymbol{\epsilon}^{(2)}$
transform differently is $C_{2z}$. Hence, the presence of any element
that can be expressed in terms of $C_{2z}$ (e.g. $C_{2z}$, $IC_{2z}$,
$C_{2x}C_{2z}$, etc.) prohibits the two doublets from belonging to
the same irreducible representation. Given the representation (\ref{eq:D6h}),
it is straightforward to construct the corresponding subgroups. Note
that the block $\{E,C_{3z}^{\pm1}\}$ is responsible for the degeneracy
that enforces the existence of the doublets in the first place. Thus,
we will only consider subgroups that contain this main block $\{E,C_{3z}^{\pm1}\}$.
There are five subgroups that contain $12$ symmetry elements:
\begin{align*}
\mathsf{D_{6}} & =\{E,C_{3z}^{\pm1}\}\otimes\{E,C_{2x}\}\otimes\{E,C_{2z}\},\\
\boldsymbol{\mathsf{D_{3d}}} & =\{E,C_{3z}^{\pm1}\}\otimes\{E,C_{2x}\}\otimes\{E,I\},\\
\mathsf{C_{6h}} & =\{E,C_{3z}^{\pm1}\}\otimes\{E,C_{2z}\}\otimes\{E,I\},\\
\mathsf{D_{3h}} & =\{E,C_{3z}^{\pm1}\}\otimes\{E,C_{2x}\}\otimes\{E,IC_{2z}\},\\
\mathsf{C_{6v}} & =\{E,C_{3z}^{\pm1}\}\otimes\{E,IC_{2x}\}\otimes\{E,C_{2z}\}.
\end{align*}
The only group where $\boldsymbol{\epsilon}^{(1)}$ and $\boldsymbol{\epsilon}^{(2)}$
transform under the same IR (represented in boldface), i.e. where
the element $C_{2z}$ is absent, is the point group $\mathsf{D_{3d}}$.
The subgroups that contain $6$ elements are:
\begin{align*}
\boldsymbol{\mathsf{D_{3}}} & =\{E,C_{3z}^{\pm1}\}\otimes\{E,C_{2x}\},\\
\boldsymbol{\mathsf{S_{6}}} & =\{E,C_{3z}^{\pm1}\}\otimes\{E,I\},\\
\mathsf{C_{6}} & =\{E,C_{3z}^{\pm1}\}\otimes\{E,C_{2z}\},\\
\mathsf{C_{3h}} & =\{E,C_{3z}^{\pm1}\}\otimes\{E,IC_{2z}\},\\
\boldsymbol{\mathsf{C_{3v}}} & =\{E,C_{3z}^{\pm1}\}\otimes\{E,IC_{2x}\}.
\end{align*}
Lastly, there is only one subgroup with $3$ elements: 
\begin{align*}
\boldsymbol{\mathsf{C_{3}}} & =\{E,C_{3z}^{\pm1}\}.
\end{align*}
Note that the set of the five boldface point groups form the full
set of trigonal point groups.

\section{Dynamic matrix for a trigonal lattice \label{sec:Dynamical-matrix}}

The dynamic matrix $D_{ij}(\boldsymbol{q})=\sum_{i^{\prime},j^{\prime}}\mathcal{C}_{ii^{\prime}jj^{\prime}}q_{i^{\prime}}q_{j^{\prime}}$
introduced in Eq. (\ref{eq:S_el_2}), with $\{i,j,i^{\prime},j^{\prime}\}\in\{1,2,3\}$,
satisfies $D(-\boldsymbol{q})=D(\boldsymbol{q})$ and $D^{T}(\boldsymbol{q})=D(\boldsymbol{q})$.
In the $\mathsf{D_{3d}}$ point group, the matrix elements are given
by 
\begin{align*}
D_{11}(\boldsymbol{q}) & =F_{a,\boldsymbol{q}}^{A_{1g}}\!+\!F_{a,\boldsymbol{q}}^{E_{g},1}\!+\!\frac{1}{\sqrt{3}}F_{b,\boldsymbol{q}}^{A_{1g}}, & D_{12}(\boldsymbol{q}) & =-F_{a,\boldsymbol{q}}^{E_{g},2},\\
D_{22}(\boldsymbol{q}) & =F_{a,\boldsymbol{q}}^{A_{1g}}\!-\!F_{a,\boldsymbol{q}}^{E_{g},1}\!+\!\frac{1}{\sqrt{3}}F_{b,\boldsymbol{q}}^{A_{1g}}, & D_{13}(\boldsymbol{q}) & =-F_{b,\boldsymbol{q}}^{E_{g},2},\\
D_{33}(\boldsymbol{q}) & =F_{a,\boldsymbol{q}}^{A_{1g}}\!-\!\frac{2}{\sqrt{3}}F_{b,\boldsymbol{q}}^{A_{1g}}, & D_{23}(\boldsymbol{q}) & =F_{b,\boldsymbol{q}}^{E_{g},1}.
\end{align*}
Here, we defined 
\begin{align*}
F_{a,\boldsymbol{q}}^{A_{1g}} & =\frac{c_{A1}+2c_{E1}+c_{E2}}{3}f_{\boldsymbol{q}}^{A1}+\frac{c_{A2}+2c_{E2}}{3}f_{\boldsymbol{q}}^{A2},\\
F_{b,\boldsymbol{q}}^{A_{1g}} & =\frac{c_{A1}+2c_{E1}-2c_{E2}}{2\sqrt{3}}f_{\boldsymbol{q}}^{A1}+\frac{c_{E2}-c_{A2}}{\sqrt{3}}f_{\boldsymbol{q}}^{A2},\\
\boldsymbol{F}_{a,\boldsymbol{q}}^{E_{g}} & =\frac{c_{A1}}{2}\boldsymbol{f}_{\boldsymbol{q}}^{E1}+c_{E3}\boldsymbol{f}_{\boldsymbol{q}}^{E2},\\
\boldsymbol{F}_{b,\boldsymbol{q}}^{E_{g}} & =c_{E3}\boldsymbol{f}_{\boldsymbol{q}}^{E1}+\frac{c_{A3}+c_{E2}}{2}\boldsymbol{f}_{\boldsymbol{q}}^{E2},
\end{align*}
as well as 
\begin{align*}
f_{\boldsymbol{q}}^{A1} & =q_{x}^{2}+q_{y}^{2}, & f_{\boldsymbol{q}}^{A2} & =q_{z}^{2},\\
\boldsymbol{f}_{\boldsymbol{q}}^{E1} & =\left(\begin{array}{c}
q_{x}^{2}-q_{y}^{2}\\
-2q_{x}q_{y}
\end{array}\right), & \boldsymbol{f}_{\boldsymbol{q}}^{E2} & =\left(\begin{array}{c}
2q_{y}q_{z}\\
-2q_{x}q_{z}
\end{array}\right).
\end{align*}
We obtain the eigenvalues:
\begin{align*}
\omega_{j,\boldsymbol{q}}^{2} & =F_{a,\boldsymbol{q}}^{A_{1}}-\frac{1}{3}x_{j,\boldsymbol{q}},
\end{align*}
with 
\begin{align*}
\left(\begin{array}{c}
x_{1,\boldsymbol{q}}\\
x_{2,\boldsymbol{q}}\\
x_{3,\boldsymbol{q}}
\end{array}\right) & =-2\sqrt{3}F_{\boldsymbol{q}}\left(\begin{array}{c}
\cos\left(\frac{\eta_{\boldsymbol{q}}}{3}\right)\\
\cos\left(\frac{\eta_{\boldsymbol{q}}}{3}+\frac{2\pi}{3}\right)\\
\cos\left(\frac{\eta_{\boldsymbol{q}}}{3}-\frac{2\pi}{3}\right)
\end{array}\right),
\end{align*}
and 
\begin{align*}
F_{\boldsymbol{q}} & =\sqrt{(F_{b,\boldsymbol{q}}^{A_{1g}})^{2}+(\boldsymbol{F}_{a,\boldsymbol{q}}^{E_{g}})^{2}+(\boldsymbol{F}_{b,\boldsymbol{q}}^{E_{g}})^{2}},\\
\mathsf{d}_{\boldsymbol{q}} & =\det D(\boldsymbol{q})\Big|_{F_{a,\boldsymbol{q}}^{A_{1g}}=0}\\
 & =\frac{1}{\sqrt{3}}F_{b,\boldsymbol{q}}^{A_{1g}}\Big(2(\boldsymbol{F}_{a,\boldsymbol{q}}^{E_{g}})^{2}\!-\!(\boldsymbol{F}_{b,\boldsymbol{q}}^{E_{g}})^{2}\!-\!\frac{2}{3}(F_{b,\boldsymbol{q}}^{A_{1g}})^{2}\Big)\\
 & \quad-\boldsymbol{F}_{a,\boldsymbol{q}}^{E_{g}}\cdot\left(\begin{array}{c}
(F_{b,\boldsymbol{q}}^{E_{g},1})^{2}-(F_{b,\boldsymbol{q}}^{E_{g},2})^{2}\\
-2F_{b,\boldsymbol{q}}^{E_{g},1}F_{b,\boldsymbol{q}}^{E_{g},2}
\end{array}\right),\\
\eta_{\boldsymbol{q}} & =\arccos\left(\frac{3\sqrt{3}}{2}\frac{\mathsf{d}_{\boldsymbol{q}}}{F_{\boldsymbol{q}}^{3}}\right).
\end{align*}
The corresponding eigenvectors can be generically written as
\begin{align}
\hat{\boldsymbol{e}}_{\boldsymbol{q}}^{(j)} & =\frac{\mathrm{sign}(q_{j})}{|\boldsymbol{u}_{j,\boldsymbol{q}}|}\left(\begin{array}{c}
u_{j,\boldsymbol{q}}^{(1)}\\
u_{j,\boldsymbol{q}}^{(2)}\\
u_{j,\boldsymbol{q}}^{(3)}
\end{array}\right),\label{eq:eigenvector}
\end{align}
where we defined
\begin{align*}
u_{j,\boldsymbol{q}}^{(1)} & =3(F_{b,\boldsymbol{q}}^{E_{g},1})^{2}-2\sqrt{3}F_{b,\boldsymbol{q}}^{A_{1g}}F_{a,\boldsymbol{q}}^{E_{g},1}+2(F_{b,\boldsymbol{q}}^{A_{1}})^{2}\\
 & \quad+\Big(\frac{1}{\sqrt{3}}F_{b,\boldsymbol{q}}^{A_{1g}}+F_{a,\boldsymbol{q}}^{E_{g},1}\Big)x_{j,\boldsymbol{q}}-\frac{1}{3}x_{j,\boldsymbol{q}}^{2},\\
u_{j,\boldsymbol{q}}^{(2)} & =3F_{b,\boldsymbol{q}}^{E_{g},1}F_{b,\boldsymbol{q}}^{E_{g},2}\!+\!F_{a,\boldsymbol{q}}^{E_{g},2}\big(2\sqrt{3}F_{b,\boldsymbol{q}}^{A_{1g}}-x_{j,\boldsymbol{q}}\big),\\
u_{j,\boldsymbol{q}}^{(3)} & =3\big(F_{a,\boldsymbol{q}}^{E_{g},2}F_{b,\boldsymbol{q}}^{E_{g},1}\!+\!F_{b,\boldsymbol{q}}^{E_{g},2}F_{a,\boldsymbol{q}}^{E_{g},1}\big)\!\\
 & \quad-\!F_{b,\boldsymbol{q}}^{E_{g},2}\big(\!\sqrt{3}F_{b,\boldsymbol{q}}^{A_{1g}}\!+\!x_{j,\boldsymbol{q}}\big).
\end{align*}
The additional function $\mathrm{sign}(q_{j})$ in (\ref{eq:eigenvector})
is necessary to guarantee $\hat{\boldsymbol{e}}_{\boldsymbol{q}}^{(j)}=-\hat{\boldsymbol{e}}_{-\boldsymbol{q}}^{(j)}$.

\section{Symmetry-imposed degeneracies of the nematic director\label{sec:Symm_based_Deg}}

We determine here the set of symmetry-enforced degenerate maxima to
the function $R(\hat{\boldsymbol{q}},\alpha_{0})$ in Eq. (\ref{eq:R_func2}).
As discussed in Appendix \ref{sec:Strain-doublet-degeneracy}, the
$\mathsf{D_{3d}}$ point group consists of the $12$ symmetry elements:
\begin{align*}
\mathsf{D_{3d}} & =\{E,C_{3z}^{\pm1}\}\otimes\{E,C_{2x}\}\otimes\{E,I\}.
\end{align*}
For convenience, we list the transformation matrices of the two-dimensional
IRs $E_{g/u}$. For the following elements, the two IRs transform
identically,
\begin{align*}
\mathcal{R}_{E_{g/u}}(C_{3z}^{\pm1}) & =\frac{1}{2}\left(\begin{array}{cc}
-1 & \mp\sqrt{3}\\
\pm\sqrt{3} & -1
\end{array}\right),\\
\mathcal{R}_{E_{g/u}}(C_{2x}) & =\left(\begin{array}{cc}
1 & 0\\
0 & -1
\end{array}\right),\\
\mathcal{R}_{E_{g/u}}(C_{2\{A,B\}}) & =\frac{1}{2}\left(\begin{array}{cc}
-1 & \mp\sqrt{3}\\
\mp\sqrt{3} & 1
\end{array}\right),
\end{align*}
where $C_{2A}=C_{3z}^{-1}C_{2x}$ and $C_{2B}=C_{3z}C_{2x}$. With
$\mathcal{R}_{E_{g/u}}(I)=\pm\mathbbm{1}_{2}$, the remaining matrices
can be directly read from these expressions. The real-space and momentum-space
coordinates transform according to the vector representation $v$
with $\mathcal{R}_{v}(g)=\mathcal{R}_{E_{u}}(g)\oplus\mathcal{R}_{A_{2u}}(g)$
and $g\in\mathsf{D_{3d}}$. Let us rewrite the function $R(\hat{\boldsymbol{q}},\alpha_{0})$
that needs to be maximized, 
\begin{align}
R(\hat{\boldsymbol{q}},\alpha_{0}) & =M(\hat{\boldsymbol{q}},\alpha_{0})+\frac{g^{2}\cos^{2}(3\alpha_{0})}{2u},\label{eq:R_func_app}
\end{align}
and the mass function

\begin{align}
M(\hat{\boldsymbol{q}},\alpha_{0}) & =\Pi_{(\boldsymbol{q},0)}^{A_{1g}}+\boldsymbol{\Pi}_{(\boldsymbol{q},0)}^{E_{g}}\cdot\boldsymbol{\mathsf{b}}_{\alpha_{0}}^{E_{g}}\!,\label{eq:M_func_app}
\end{align}
where we introduced the abbreviated notation
\begin{align}
\boldsymbol{\mathsf{b}}_{\alpha_{0}}^{E_{g}} & =\left(\begin{array}{c}
\cos(2\alpha_{0})\\
-\sin(2\alpha_{0})
\end{array}\right),\label{eq:bilinearalpha}
\end{align}
that transforms according to $E_{g}$. We have
\begin{align}
\Pi_{(\mathcal{R}_{v}(g)\boldsymbol{q},0)}^{A_{1g}} & =\Pi_{(\boldsymbol{q},0)}^{A_{1g}},\label{eq:Pi_A1g_trafo}\\
\boldsymbol{\Pi}_{(\mathcal{R}_{v}(g)\boldsymbol{q},0)}^{E_{g}} & =\mathcal{R}_{E_{g}}(g)\,\boldsymbol{\Pi}_{(\boldsymbol{q},0)}^{E_{g}},\label{eq:Pi_Eg_trafo}
\end{align}
for any $g\in\mathsf{D_{3d}}$. Now, we establish the three constraints
that relate the degenerate maxima $R(\hat{\boldsymbol{q}}_{0},\alpha_{0})$
with each other. Let us assume $\{\alpha_{0},\hat{\boldsymbol{q}}_{0}\}$
to describe a given maximum.
\begin{itemize}
\item \textit{Inversion.} Since the functions $\Pi_{(\boldsymbol{q},0)}^{A_{1g}}$
and $\boldsymbol{\Pi}_{(\boldsymbol{q},0)}^{E_{g}}$ are invariant
upon the inversion operation, the whole function (\ref{eq:R_func_app})
is, such that $R(-\hat{\boldsymbol{q}}_{0},\alpha_{0})$ is a degenerate
maximum. 
\item \textit{Three-fold rotation.} A variable shift $\alpha_{0}\rightarrow\alpha_{0}\mp\frac{2\pi}{3}$
leaves the second term in (\ref{eq:R_func_app}) invariant, and it
shifts 
\begin{align}
\boldsymbol{\mathsf{b}}_{\alpha_{0}\mp\frac{2\pi}{3}}^{E_{g}} & =\mathcal{R}_{E_{g}}(C_{3z}^{\pm1})\,\boldsymbol{\mathsf{b}}_{\alpha_{0}}^{E_{g}}.\label{eq:alpha_shift}
\end{align}
As for Eqs. (\ref{eq:M_func_app}) and (\ref{eq:Pi_A1g_trafo})-(\ref{eq:Pi_Eg_trafo}),
the shift (\ref{eq:alpha_shift}) can be compensated by a momentum
rotation with $g=C_{3z}^{\pm1}$. Then, the function $R$ stays invariant,
and it holds 
\begin{align}
R(\mathcal{R}_{v}(C_{3z})\hat{\boldsymbol{q}}_{0},\alpha_{0}-\frac{2\pi}{3}) & =R(\hat{\boldsymbol{q}}_{0},\alpha_{0}),\label{eq:3fold_r}\\
R(\mathcal{R}_{v}^{-1}(C_{3z})\hat{\boldsymbol{q}}_{0},\alpha_{0}+\frac{2\pi}{3}) & =R(\hat{\boldsymbol{q}}_{0},\alpha_{0}).\label{eq:3fold_r2}
\end{align}
\item \textit{Reflection.} Let us define the nematic director angle $\alpha_{0}=\alpha_{s}+\delta$
with respect to a high-symmetry direction {[}recall $\alpha_{s}\in\frac{\pi}{3}\{0,1,2,3,4,5\}${]}.
Then, the second term in (\ref{eq:R_func_app}) is invariant upon
a sign change of the deviation $\delta\rightarrow-\delta$ as $\cos(3\alpha_{0})=\cos(3\alpha_{s})\cos(3\delta)$.
For the quantity (\ref{eq:bilinearalpha}), we obtain
\begin{align}
\boldsymbol{\mathsf{b}}_{\alpha_{s}-\delta}^{E_{g}} & =\begin{cases}
\mathcal{R}_{E_{g}}(IC_{2x})\boldsymbol{\mathsf{b}}_{\alpha_{s}+\delta}^{E_{g}} & ,\alpha_{s}\in\{0,3\}\frac{\pi}{3}\\
\mathcal{R}_{E_{g}}(IC_{2B})\boldsymbol{\mathsf{b}}_{\alpha_{s}+\delta}^{E_{g}} & ,\alpha_{s}\in\{1,4\}\frac{\pi}{3}\\
\mathcal{R}_{E_{g}}(IC_{2A})\boldsymbol{\mathsf{b}}_{\alpha_{s}+\delta}^{E_{g}} & ,\alpha_{s}\in\{2,5\}\frac{\pi}{3}
\end{cases}.\label{eq:bilinear_trafo}
\end{align}
Just like before, the appropriate momentum rotation with $g=IC_{2n_{s}}$
and $n_{s}\in\{x,B,A\}$ chosen according to (\ref{eq:bilinear_trafo})
compensates the transformation (\ref{eq:bilinear_trafo}). As a result,
we obtain the degenerate maxima $R(\mathcal{R}_{v}(IC_{2n_{s}})\hat{\boldsymbol{q}}_{0},\alpha_{s}-\delta)=R(\hat{\boldsymbol{q}}_{0},\alpha_{s}+\delta)$.
Clearly, this relation is also true for $\delta=0$.
\end{itemize}
Combined together, the above symmetry constraints lead to twelve degenerate
maxima of the function $R$. Importantly, a nematic director $\alpha_{0}=\alpha_{s}+\delta$
with a finite deviation $\delta$ necessarily induces a degenerate
maximum with $\alpha_{0}=\alpha_{s}-\delta$. Hence, a finite $\delta\neq0$
doubles the number of degenerate nematic directors. 

\section{Details of the analytical approach\label{sec:Supplemental-to-analytical}}

In Sec. \ref{subsec:Static-limit}, we introduced
the quantities $M_{1}$, $M_{\theta}^{s}$, and $M_{\theta}^{c}$
in Eqs. (\ref{eq:M_q1}) and (\ref{eq:M_q2}). The first expression
$M_{1}=\kappa_{1}^{2}/(c_{A1}\!+\!c_{E1})$ is obtained by inserting
the eigenvalues and eigenvectors associated with the $\hat{\boldsymbol{q}}_{1}$
direction, Eqs. (\ref{eq:omega2_q1})-(\ref{eq:eigs_q1}), into the
renormalized mass expression (\ref{eq:M_func2}). To show that $M_{\mathrm{stat}}>M_{1}$,
consider $M_{\mathrm{stat}}(c_{E3})$ as a function of $c_{E3}$:

\begin{equation}
M_{\mathrm{stat}}(c_{E3})=\frac{c_{E2}\kappa_{1}^{2}+c_{E1}\kappa_{2}^{2}-2c_{E3}\kappa_{1}\kappa_{2}}{c_{E1}c_{E2}-c_{E3}^{2}}.
\end{equation}
First, one finds $M_{1}<M_{\mathrm{stat}}(0)=\frac{\kappa_{1}^{2}}{c_{E1}}+\frac{\kappa_{2}^{2}}{c_{E2}}$
and $M_{1}<M_{\mathrm{stat}}(\pm|c_{E3}|_{\mathrm{max}})\rightarrow+\infty$.
Additionally, at the two zeros of the derivative $M_{\mathrm{stat}}^{\prime}(c_{E3})=\left(\kappa_{2}c_{E1}-\kappa_{1}c_{E3}\right)\left(\kappa_{2}c_{E3}-\kappa_{1}c_{E2}\right)/\mathsf{d}_{E}^{2}$,
defined through $c_{E3}^{(1)}=\kappa_{2}c_{E1}/\kappa_{1}$ and $c_{E3}^{(2)}=\kappa_{1}c_{E2}/\kappa_{2}$,
one similarly obtains $M_{\mathrm{stat}}(c_{E3}^{(1)})>M_{1}$ and
$M_{\mathrm{stat}}(c_{E3}^{(2)})>M_{1}$, proving that it always holds
$M_{\mathrm{stat}}>M_{1}$.

The quantities $M_{\theta}^{s}$ and $M_{\theta}^{c}$
can be derived in an analogous way using the eigenvalues and eigenvectors
associated with the $\hat{\boldsymbol{q}}_{\bar{2}}$ direction:
\begin{align*}
M_{\theta}^{c} & =\frac{\kappa_{1}\kappa_{2}\lambda_{2,\theta}}{\tilde{\omega}_{1,\hat{\boldsymbol{q}}_{\bar{2}}}^{2}\tilde{\omega}_{3,\hat{\boldsymbol{q}}_{\bar{2}}}^{2}}\Big[(\text{-}1)^{\frac{n_{2}+1}{2}}\!+\!\frac{\kappa_{2}}{\kappa_{1}}(\frac{\lambda_{1,\theta}}{\lambda_{2,\theta}}\!-\!\cot\!\theta)\\
 & \!\!\!\!\!\!\!\!+\!\Big(\!\frac{\kappa_{1}}{\kappa_{2}}\!+\!\frac{\kappa_{2}}{\kappa_{1}}\!+\!\frac{\kappa_{2}}{\kappa_{1}}\cot^{2}\!\theta\!-\!2(\text{-}1)^{\!\frac{n_{2}+1}{2}}\!\cot\!\theta\!\Big)\!\frac{c_{E2}\!+\!c_{A2}\!\cot^{2}\!\theta}{\lambda_{2,\theta}}\Big]\!,\\
M_{\theta}^{s} & =\frac{\kappa_{1}^{2}+2\kappa_{1}\kappa_{2}(\text{-}1)^{\frac{n_{2}+1}{2}}\cot\!\theta+\kappa_{2}^{2}\cot^{2}\!\theta}{c_{E1}\!+\!2c_{E3}(\text{-}1)^{\frac{n_{2}+1}{2}}\cot\!\theta\!+\!c_{E2}\cot^{2}\!\theta}.
\end{align*}
We note that $M_{\theta}^{s}$ originates from the second eigenvalue
$\tilde{\omega}_{2,\hat{\boldsymbol{q}}_{\bar{2}}}^{2}$, which is
also the one that vanishes at the pure structural phase transition.
Thus, it is $M_{\theta}^{s}$ that can attain the maximum value $M_{\mathrm{stat}}$.
Indeed, the solution of the equation $M_{\theta_{2}}^{s}=M_{\mathrm{stat}}$
is the polar angle $\theta_{2}$ given by Eq. (\ref{eq:q2}). 

In Sec. \ref{subsec:Limit_zero}, we introduced
the functions $A_{\theta}^{\pm}$, $B_{\theta}^{\pm}$ and $C_{\theta}^{\pm}$
that occur in the renormalized mass expression Eq. (\ref{eq:M_zero_cE3}),
and which are defined here. Let us recall that the eigenvalues and
eigenfrequencies of the dynamical matrix in the $c_{E3}=0$ case are
given through Eqs. (\ref{eq:omega1_q2})-(\ref{eq:betapm}) upon replacement
of $\hat{\boldsymbol{q}}_{2}\rightarrow\hat{\boldsymbol{q}}$ and
setting $c_{E3}=0$. A different way to express the eigenvalues $\tilde{\omega}_{i,\hat{\boldsymbol{q}}}$
is 
\begin{align}
\tilde{\omega}_{1,\hat{\boldsymbol{q}}}^{2} & =\frac{1}{2}\Big(\Omega_{1\theta}+\sqrt{\Omega_{1\theta}^{2}-4\Omega_{2\theta}}\,\Big),\label{eq:eigval1_app}\\
\tilde{\omega}_{2,\hat{\boldsymbol{q}}}^{2} & =c_{E1}+c_{E2}\cot^{2}\theta,\label{eq:eigval2_app}\\
\tilde{\omega}_{3,\hat{\boldsymbol{q}}}^{2} & =\frac{1}{2}\Big(\Omega_{1\theta}-\sqrt{\Omega_{1\theta}^{2}-4\Omega_{2\theta}}\,\Big),\label{eq:eigval3_app}
\end{align}
with 
\begin{align*}
\Omega_{1\theta} & =c_{A1}\!+\!c_{E1}\!+c_{E2}\!+\!(c_{E2}\!+\!c_{A2})\cot^{2}\theta,\\
\Omega_{2\theta} & =c_{E2}\left(c_{A1}+c_{E1}\right)+c_{E2}c_{A2}\cot^{4}\theta\\
 & +\left\{ c_{A2}\left(c_{A1}+c_{E1}\right)-c_{A3}(c_{A3}+2c_{E2})\right\} \cot^{2}\theta,
\end{align*}
which is used below. The functions $A_{\theta}^{\pm}$, $B_{\theta}^{\pm}$
and $C_{\theta}^{\pm}$ are then given by:
\begin{align*}
A_{\theta}^{\pm} & =\frac{1}{2}\left(A_{\theta}^{(1)}\pm A_{\theta}^{(2)}\right), & B_{\theta}^{\pm} & =\frac{1}{2}\left(B_{\theta}^{(1)}\pm B_{\theta}^{(2)}\right),\\
C_{\theta}^{\pm} & =\frac{1}{2}\left(C_{\theta}^{(1)}\pm C_{\theta}^{(2)}\right),
\end{align*}
with 
\begin{align*}
A_{\theta}^{(1)} & \!\!=\!\frac{c_{E2}+c_{A2}\cot^{2}\!\theta}{\tilde{\omega}_{1,\hat{\boldsymbol{q}}}^{2}\tilde{\omega}_{3,\hat{\boldsymbol{q}}}^{2}}, & A_{\theta}^{(2)} & \!\!=\!\frac{1}{\tilde{\omega}_{2,\hat{\boldsymbol{q}}}^{2}},\\
B_{\theta}^{(2)} & \!\!=\!\frac{c_{A1}\!+\!c_{E1}\!-\!2c_{A3}\!\cot^{2}\!\theta\!+\!c_{A2}\!\cot^{4}\!\theta}{\tilde{\omega}_{1,\hat{\boldsymbol{q}}}^{2}\tilde{\omega}_{3,\hat{\boldsymbol{q}}}^{2}}, & B_{\theta}^{(1)} & \!\!=\!\frac{\cot^{2}\!\theta}{\tilde{\omega}_{2,\hat{\boldsymbol{q}}}^{2}},\\
C_{\theta}^{(1)} & \!\!=\!\frac{c_{A2}\cot^{2}\!\theta\!-\!c_{A3}}{\tilde{\omega}_{1,\hat{\boldsymbol{q}}}^{2}\tilde{\omega}_{3,\hat{\boldsymbol{q}}}^{2}}\cot\!\theta, & C_{\theta}^{(2)} & \!\!=\!\frac{\cot\!\theta}{\tilde{\omega}_{2,\hat{\boldsymbol{q}}}^{2}}.
\end{align*}
In the following, we prove that the maxima of the function $R$ in
Eq. (\ref{eq:R_ce30}) and of the renormalized mass
\begin{align}
\frac{1}{\kappa_{1}^{2}}M(\hat{\boldsymbol{q}},\alpha_{0}) & =A_{\theta}^{(1)}\cos^{2}\bar{\varphi}+A_{\theta}^{(2)}\sin^{2}\bar{\varphi},\label{eq:M_appendix}
\end{align}
lie at $\theta_{0}=\frac{\pi}{2}$ and $\bar{\varphi}\equiv\alpha_{0}+2\varphi=\frac{\pi}{2}\{1,3,5,7\}$.
To find the maximum of Eq. (\ref{eq:M_appendix}) with respect to
$\theta$, we individually compute the maximum of $A_{\theta}^{(1)}$
and $A_{\theta}^{(2)}$. Beginning with $A_{\theta}^{(2)}$, we find

\begin{align*}
\frac{\partial A_{\theta}^{(2)}}{\partial\cot\theta} & =\frac{-2c_{E2}\cot\theta}{\left(c_{E1}+c_{E2}\cot^{2}\theta\right)^{2}},
\end{align*}
i.e. $A_{\theta}^{(2)}$ has only one maximum at $\theta=\frac{\pi}{2}$
giving:
\begin{align}
A_{\theta=\frac{\pi}{2}}^{(2)} & =\frac{1}{c_{E1}}.\label{eq:A2_equator}
\end{align}
For $A_{\theta}^{(1)}$ we compute 
\begin{align*}
\frac{\partial A_{\theta}^{(1)}}{\partial\cot\theta} & =2c_{E2}\frac{\cot\theta}{\Omega_{2\theta}^{2}}\,\Big\{\left(c_{A3}+2c_{E2}\right)c_{A3}\\
 & \qquad\qquad-2c_{E2}c_{A2}\cot^{2}\theta-c_{A2}^{2}\cot^{4}\theta\Big\}.
\end{align*}
For $\left(c_{A3}+2c_{E2}\right)c_{A3}<0$, i.e.
$-2c_{E2}<c_{A3}<0$, the function $A_{\theta}^{(1)}$ has a maximum
at $\theta=\frac{\pi}{2}$:
\begin{align}
A_{\theta=\frac{\pi}{2}}^{(1)} & =\frac{1}{c_{E1}\!+\!c_{A1}}.\label{eq:A1_th2}
\end{align}
For $\left(c_{A3}+2c_{E2}\right)c_{A3}>0$, i.e.
$c_{A3}>0$ or $c_{A3}<-2c_{E2}$, the maximum is at
\begin{align*}
\cot^{2}\theta_{>} & =\frac{c_{A3}}{c_{A2}}, & \mathrm{and} &  & \cot^{2}\theta_{<} & =-\frac{c_{A3}+2c_{E2}}{c_{A2}},
\end{align*}
respectively. The corresponding maxima are 
\begin{align}
A_{\theta_{>}}^{(1)} & =\frac{1}{c_{E1}+\frac{\mathsf{d}_{A}}{c_{A2}}},\label{eq:A1_thplus}\\
A_{\theta_{<}}^{(1)} & =\frac{1}{c_{E1}+\frac{\mathsf{d}_{A}-4c_{E2}\left(c_{A3}+c_{E2}\right)}{c_{A2}}}.\label{eq:A1_thminus}
\end{align}
In any case, Eqs. (\ref{eq:A1_th2}), (\ref{eq:A1_thplus}) or (\ref{eq:A1_thminus})
give $A_{\theta=\frac{\pi}{2}}^{(2)}>A_{\theta}^{(1)}$, and consequently,
the function (\ref{eq:M_appendix}) {[}or Eq. (\ref{eq:R_ce30}){]}
is maximized for $\theta_{0}=\frac{\pi}{2}$ and $\bar{\varphi}\equiv\alpha_{0}+2\varphi=\frac{\pi}{2}\{1,3,5,7\}$.

In Sec. \ref{subsec:expansion}, the functions $\varphi_{0}^{(4)}$,
$R_{1}^{(8)}$ and $R_{2}^{(8)}$ that occur in Eq. (\ref{eq:M_8})
are given by
\begin{align*}
\varphi_{0}^{(4)} & =\frac{(-1)^{\frac{1+N_{0}}{2}}\mathsf{d}_{A}}{8c_{A1}^{2}c_{E2}^{4}}\cos(3\alpha_{0})\quad\times\\
 & \Big(\frac{2c_{A3}^{2}-c_{A1}c_{A2}}{\mathsf{d}_{A}}\!-\!\frac{2c_{A3}}{c_{E2}}\!-\!\frac{3c_{A1}}{c_{E1}}\sin(3\alpha_{0})(-1)^{\frac{1\!+\!N_{0}}{2}}\Big),
\end{align*}
and

\begin{align*}
R_{1}^{(8)} & =\!\frac{\mathsf{d}_{A}^{2}\cos^{2}(3\alpha_{0})}{4c_{A1}^{2}c_{E2}^{9}c_{E1}^{2}}\Bigg(\!\frac{c_{E2}\!+\!2c_{A3}}{c_{A1}}+\frac{17}{4}\frac{c_{E2}}{c_{E1}}\\
 & \;\;-\frac{c_{E2}}{\mathsf{d}_{A}}\!\Big(\!\frac{3}{4}c_{A2}\!+\!2\frac{c_{E2}c_{A3}}{c_{E1}}\!+\!3\frac{c_{E2}^{2}c_{A1}}{c_{E1}^{2}}\!\Big)-\frac{\mathsf{d}_{A}}{c_{A1}c_{E2}}\!\Bigg),\\
R_{2}^{(8)} & =\!\frac{\mathsf{d}_{A}^{2}}{4c_{A1}^{2}c_{E2}^{9}c_{E1}^{2}}\Big(\!\frac{2c_{E2}}{c_{E1}}\!+\!\frac{c_{A3}^{2}c_{E2}}{2c_{A1}\mathsf{d}_{A}}\!+\!\frac{\mathsf{d}_{A}}{2c_{A1}c_{E2}}\!-\!\frac{c_{A3}}{c_{A1}}\!\Big).
\end{align*}

\section{Derivation of the upper critical field\label{sec:UpperCritField}}

We follow the same approach put forward in Ref. \citep{Venderbos2016}
to derive an expression for the upper critical field $H_{c2}$ in
the presence of an $E_{g}$ symmetry-breaking field (see also Ref.
\citep{Hecker2020}). In the theoretical model discussed in Refs.
\citep{Hecker2018,Hecker2020,Fernandes_review}, this symmetry-breaking
field is a vestigial nematic order described by the composite order
parameter $\boldsymbol{\Phi}^{E_{g}}$. In the presence of an in-plane
magnetic field $\boldsymbol{B}=\boldsymbol{B}^{E_{g}}\oplus B^{A_{2g}}$
with $\boldsymbol{B}^{E_{g}}=B_{0}(\cos\varphi_{B},\sin\varphi_{B})$
and $B^{A_{2g}}=0$, the Ginzburg-Landau expansion of the superconducting
action gives:
\begin{align}
\mathcal{S}_{\Delta} & =\int_{\boldsymbol{r}}\boldsymbol{\Delta}^{\dagger}\,\chi_{\Delta}^{-1}(\boldsymbol{r})\,\boldsymbol{\Delta},\label{eq:S_delta}\\
\chi_{\Delta}^{-1}(\boldsymbol{r}) & =\big(R_{0}+f_{D_{\boldsymbol{r}}}^{A_{1g}}\big)\tau^{0}+\Big(\boldsymbol{f}_{D_{\boldsymbol{r}}}^{E_{g}}+\boldsymbol{\Phi}_{0}^{E_{g}}\Big)\cdot\boldsymbol{\tau}^{E_{g}}\nonumber \\
 & \quad+\mathsf{i}\kappa_{A2}\big(\boldsymbol{B}^{E_{g}}\tau^{y}\boldsymbol{\Phi}_{0}^{E_{g}}\big)\tau^{y},\label{eq:chi_Delta}
\end{align}
with the covariant derivative $D_{j}=-\mathsf{i}\partial_{j}-qA_{j}(\boldsymbol{r})$,
the vector potential $\boldsymbol{A}(\boldsymbol{r})=-\boldsymbol{r}\times\boldsymbol{B}/2$
and the charge of a Cooper pair $q=2|e|$. The covariant derivatives
satisfy the commutation relations $[D_{i},D_{j}]=\mathsf{i}q\sum_{k}\epsilon_{ijk}B_{k}$.
The gradient functions are given by
\begin{align}
f_{D_{\boldsymbol{r}}}^{A_{1g}} & =\mathsf{d}_{\parallel}(D_{x}^{2}+D_{y}^{2})+\mathsf{d}_{z}D_{z}^{2},\label{eq:fA1g_r}\\
\boldsymbol{f}_{D_{\boldsymbol{r}}}^{E_{g}} & =\mathsf{d}_{1}\left(\begin{array}{c}
D_{x}^{2}-D_{y}^{2}\\
-[D_{x},D_{y}]_{+}
\end{array}\right)+\mathsf{d}_{2}\left(\begin{array}{c}
[D_{y},D_{z}]_{+}\\
-[D_{x},D_{z}]_{+}
\end{array}\right),\label{eq:fEg_r}
\end{align}
with four stiffness coefficients $\mathsf{d}_{\parallel}$, $\mathsf{d}_{z}$,
$\mathsf{d}_{1}$ and $\mathsf{d}_{2}$. Here, $[D_{x},D_{y}]_{+}$
denotes the anticommutator of the corresponding operators. The last
term in Eq. (\ref{eq:chi_Delta}), which can be rewritten as $\kappa_{A2}|\boldsymbol{\Phi}_{0}^{E_{g}}|B_{0}\sin(\alpha_{0}-\varphi_{B})$,
is a symmetry allowed coupling with coefficient $\kappa_{A2}$. As
we show below, its main effect is to enhance the value of $H_{c2}$
when the field is applied perpendicular to the nematic director. 

The superconducting transition occurs when the susceptibility, Eq.
(\ref{eq:chi_Delta}), diverges. In the absence of a magnetic field
this happens when the renormalized superconducting mass inside the
nematic phase $r_{\Delta}=R_{0}-|\boldsymbol{\Phi}_{0}^{E_{g}}|$
vanishes. In the presence of a magnetic field, instead of treating
the whole problem self-consistently, we employ
a mean-field like assumption where we treat the renormalized fields
$R_{0}$ and $|\boldsymbol{\Phi}_{0}^{E_{g}}|$ as externally given
values, and in particular, for temperatures $T\lesssim T_{c}$ it
holds $r_{\Delta}\lesssim0$.

To derive the upper critical field, we first rotate the coordinate
system such that the $x^{\prime}$-axis aligns with the magnetic field
$\boldsymbol{B}$. Formally, we define $\boldsymbol{r}^{\prime}=R_{3}(-\varphi_{B})\boldsymbol{r}$
with the rotation matrix $R_{3}(\varphi_{B})=R_{2}(\varphi_{B})\oplus1$,
where
\begin{align}
R_{2}(\varphi_{B}) & =\left(\begin{array}{cc}
\cos\varphi_{B} & -\sin\varphi_{B}\\
\sin\varphi_{B} & \cos\varphi_{B}
\end{array}\right).
\end{align}
The covariant derivative $D_{\boldsymbol{r}}=(D_{x},D_{y},D_{z})$
transforms as $D_{\boldsymbol{r}^{\prime}}=R_{3}(-\varphi_{B})D_{\boldsymbol{r}}$,
which leads to the commutation relations $[D_{x^{\prime}},D_{y^{\prime}}]=[D_{x^{\prime}},D_{z^{\prime}}]=0$
and $[D_{y^{\prime}},D_{z^{\prime}}]=\mathsf{i}qB_{0}$. As a result,
the gradient functions are written, in the rotated coordinates system,
as
\begin{align}
\!\!f_{D_{\boldsymbol{r}^{\prime}}}^{A_{1g}}\! & =\mathsf{d}_{\parallel}(D_{x^{\prime}}^{2}+D_{y^{\prime}}^{2})+\mathsf{d}_{z}D_{z^{\prime}}^{2},\label{eq:fA1g_rp}\\
\!\!\boldsymbol{f}_{D_{\boldsymbol{r}^{\prime}}}^{E_{g}}\! & =\mathsf{d}_{1}\!\!\left(\!\!\begin{array}{c}
\cos(2\varphi_{B})\,(D_{x^{\prime}}^{2}\!-\!D_{y^{\prime}}^{2})-\sin(2\varphi_{B})\,[D_{x^{\prime}},D_{y^{\prime}}]_{+}\\
-\sin(2\varphi_{B})\,(D_{x^{\prime}}^{2}\!-\!D_{y^{\prime}}^{2})
\end{array}\!\!\!\right)\nonumber \\
 & \quad-\mathsf{d}_{2}[D_{x^{\prime}},D_{z^{\prime}}]_{+}\hat{\boldsymbol{e}}_{\varphi_{B}}^{\phi}+\mathsf{d}_{2}[D_{y^{\prime}},D_{z^{\prime}}]_{+}\hat{\boldsymbol{e}}_{\varphi_{B}}^{r},\label{eq:fEg_rp}
\end{align}
where we have introduced the vectors 
\begin{align*}
\hat{\boldsymbol{e}}_{\beta}^{r} & =\left(\begin{array}{c}
\cos\beta\\
\sin\beta
\end{array}\right), & \hat{\boldsymbol{e}}_{\beta}^{\phi} & =\left(\begin{array}{c}
-\sin\beta\,\\
\cos\beta
\end{array}\right).
\end{align*}
Additionally, we rotate the superconducting field $\boldsymbol{\Delta}_{B}=R_{2}(-\varphi_{B})\boldsymbol{\Delta}$,
which transforms the susceptibility (\ref{eq:chi_Delta}) into 
\begin{align}
\chi_{\Delta_{B}}^{-1}(\boldsymbol{r}^{\prime}) & =R_{2}(-\varphi_{B})\chi_{\Delta}^{-1}(\boldsymbol{r}^{\prime})R_{2}(\varphi_{B})\nonumber \\
 & =\big(R_{0}+f_{D_{\boldsymbol{r}^{\prime}}}^{A_{1g}}\big)\tau^{0}+\Big(\boldsymbol{f}_{D_{\boldsymbol{r}^{\prime}}}^{E_{g},B}+\boldsymbol{\Phi}_{0}^{E_{g},B}\Big)\cdot\boldsymbol{\tau}^{E_{g}}\nonumber \\
 & \qquad+\mathsf{i}\kappa_{A2}\big(\boldsymbol{B}^{E_{g}}\tau^{y}\boldsymbol{\Phi}_{0}^{E_{g}}\big)\tau^{y},\label{eq:chi_Delta_rp}
\end{align}
with $\boldsymbol{\Phi}_{0}^{E_{g},B}=\boldsymbol{\Phi}_{0}^{E_{g}}R_{2}(-2\varphi_{B})$
and $\boldsymbol{f}_{D_{\boldsymbol{r}^{\prime}}}^{E_{g},B}=\boldsymbol{f}_{D_{\boldsymbol{r}^{\prime}}}^{E_{g}}R_{2}(-2\varphi_{B})$.
The saddle-point equation with respect to $\boldsymbol{\Delta}_{B}$
gives the Schr{\"o}dinger-type equation 
\begin{align}
0 & =\left(R_{0}\mathbbm{1}+\mathcal{H}_{0}+\mathcal{H}_{\Phi}\right)\boldsymbol{\Delta}_{B0},\label{eq:Schr_eqn}\\
\mathcal{H}_{0} & =f_{D_{\boldsymbol{r}^{\prime}}}^{A_{1g}}\tau^{0}+\boldsymbol{f}_{D_{\boldsymbol{r}^{\prime}}}^{E_{g},B}\cdot\boldsymbol{\tau}^{E_{g}},\label{eq:H0}\\
\mathcal{H}_{\Phi} & =\!|\boldsymbol{\Phi}_{0}^{E_{g}}|\!\left(\hat{\boldsymbol{e}}_{\alpha_{0}+2\varphi_{B}}^{r}\!\!\cdot\!\boldsymbol{\tau}^{E_{g}}+\kappa_{A2}B_{0}\sin(\alpha_{0}\!-\!\varphi_{B})\tau^{y}\right).\label{eq:H_phi}
\end{align}
where $\boldsymbol{\Delta}_{B0}$ is the saddle-point value. The superconducting
state is stabilized when the eigenvalue equation (\ref{eq:Schr_eqn})
has a non-trivial solution for the first time. Thus, we need to find
the smallest eigenvalue $\lambda_{\text{min}}$ of the matrix $\mathcal{H}_{0}+\mathcal{H}_{\Phi}$.
Then, the upper critical field $H_{c2}(\varphi_{B})$ is given by
the condition
\begin{align}
-R_{0} & =\lambda_{\text{min}}(H=B_{0},\varphi_{B}).\label{eq:-R0}
\end{align}

To determine the smallest eigenvalue of the spatial Hamiltonian (\ref{eq:H0}),
we note that modulations in the saddle-point value $\boldsymbol{\Delta}_{B0}$
along the magnetic field axis increase the energy. As a result, we
can set $D_{x^{\prime}}\boldsymbol{\Delta}_{B0}=-\mathsf{i}\partial_{x^{\prime}}\boldsymbol{\Delta}_{B0}=0$,
and the Hamiltonian simplifies to: 
\begin{align}
\mathcal{H}_{0} & =D_{y^{\prime}}^{2}\left(\mathsf{d}_{\parallel}\tau^{0}-\mathsf{d}_{1}\tau^{z}\right)+\mathsf{d}_{z}D_{z^{\prime}}^{2}\tau^{0}\nonumber \\
 & \qquad+\mathsf{d}_{2}\hat{\boldsymbol{e}}_{3\varphi_{B}}^{r}\cdot\boldsymbol{\tau}^{E_{g}}\,[D_{y^{\prime}},D_{z^{\prime}}]_{+}.\label{eq:H0_2}
\end{align}
Given the commutation relation $[D_{y^{\prime}},D_{z^{\prime}}]=\mathsf{i}qB_{0}$,
it is convenient to introduce ``creation'' and ``annihilation''
operators 
\begin{align*}
a^{\dagger} & =\frac{\sqrt{\mathsf{d}_{\parallel}}D_{y^{\prime}}\!-\!\mathsf{i}\sqrt{\mathsf{d}_{z}}D_{z^{\prime}}}{\sqrt{2qB_{0}\sqrt{\mathsf{d}_{\parallel}\mathsf{d}_{z}}}}\,, & a & =\frac{\sqrt{\mathsf{d}_{\parallel}}D_{y^{\prime}}\!+\!\mathsf{i}\sqrt{\mathsf{d}_{z}}D_{z^{\prime}}}{\sqrt{2qB_{0}\sqrt{\mathsf{d}_{\parallel}\mathsf{d}_{z}}}}\,,
\end{align*}
satisfying $[a,a^{\dagger}]=1$. Then, we can expand the superconducting
order parameter in the basis of the unperturbed harmonic oscillator
\begin{align*}
\boldsymbol{\Delta}_{B0} & =\sum_{n=0}^{\infty}\boldsymbol{v}_{n}|n\rangle\;,
\end{align*}
with the expansion coefficients $\boldsymbol{v}_{n}=(a_{n},b_{n})^{T}$
and the operator relations $a|n\rangle=\sqrt{n}\,|n-1\rangle$ and
$a^{\dagger}|n\rangle=\sqrt{n+1}\,|n+1\rangle$. Inserting this expansion,
we find that there is only coupling between the coefficients $\boldsymbol{v}_{n}$
and $\boldsymbol{v}_{n\pm2}$, i.e. the resulting matrix block-diagonalizes
with respect to even and odd numbers $n$. It is then convenient to
introduce the basis vectors 
\begin{align*}
\boldsymbol{V}_{e} & =\left(\begin{array}{c}
\boldsymbol{v}_{0}\\
\boldsymbol{v}_{2}\\
\vdots
\end{array}\right), & \boldsymbol{V}_{d} & =\left(\begin{array}{c}
\boldsymbol{v}_{1}\\
\boldsymbol{v}_{3}\\
\vdots
\end{array}\right).
\end{align*}

The lowest eigenvalue lies in the even sector, such that Eq. (\ref{eq:Schr_eqn})
can be re-expressed in terms of $\boldsymbol{V}_{e}$, 
\begin{align}
-\hat{R}_{0}\boldsymbol{V}_{e} & =\mathcal{M}\boldsymbol{V}_{e},\label{eq:Schr_Eqn_M}
\end{align}
with the matrix 
\begin{align}
\mathcal{M} & =\left(\begin{array}{cccc}
\mathcal{M}_{d,0} & \mathcal{M}_{o,0} & 0 & 0\\
\mathcal{M}_{o,0}^{\dagger} & \mathcal{M}_{d,2} & \mathcal{M}_{o,2} & 0\\
0 & \mathcal{M}_{o,2}^{\dagger} & \mathcal{M}_{d,4} & \ddots\\
0 & 0 & \ddots & \ddots
\end{array}\right),\label{eq:mat_M}
\end{align}
that contains the $2\times2$ matrices 
\begin{align*}
\mathcal{M}_{d,n} & =B_{0}\left(2n+1\right)\big[2\tau^{0}-\hat{\mathsf{d}}_{1}\tau^{z}\big]\\
 & +|\hat{\boldsymbol{\Phi}}_{0}^{E_{g}}|\!\left(\hat{\boldsymbol{e}}_{\alpha_{0}+2\varphi_{B}}^{r}\!\cdot\boldsymbol{\tau}^{E_{g}}\!+\kappa_{A2}B_{0}\sin(\alpha_{0}\!-\!\varphi_{B})\tau^{y}\right)\!,\\
\mathcal{M}_{o,n} & =-B_{0}\sqrt{(n\!+\!2)(n\!+\!1)}\left[\hat{\mathsf{d}}_{1}\tau^{z}\!+2\mathsf{i}\,\hat{\mathsf{d}}_{2}\,\hat{\boldsymbol{e}}_{3\varphi_{B}}^{r}\!\cdot\boldsymbol{\tau}^{E_{g}}\right]\!,
\end{align*}
where $\{\hat{R}_{0},|\hat{\boldsymbol{\Phi}}_{0}^{E_{g}}|\}=\{R_{0},|\boldsymbol{\Phi}_{0}^{E_{g}}|\}2/q\sqrt{\mathsf{d}_{z}\mathsf{d}_{\parallel}}$.
We numerically evaluated the minimal eigenvalue $\hat{\lambda}_{\mathrm{min}}$
of the matrix (\ref{eq:mat_M}) to obtain the upper critical field
curves $H_{c2}(\varphi_{B})$ in Fig. \ref{fig:Hc2}. The free parameters
were set to $\hat{\mathsf{d}}_{1}=-0.49$, $\hat{\mathsf{d}}_{2}=0.53$,
$\hat{R}_{0}=1$, and $|\hat{\boldsymbol{\Phi}}_{0}^{E_{g}}|=1.2\hat{R}_{0}$,
corresponding to a temperature below the superconducting transition.
Only in panel (a)\textemdash where nematicity is absent\textemdash we
set $\hat{R}_{0}=-0.1$. 

An approximate expression for $H_{c2}(\varphi_{B})$ can be derived
in the limit $\hat{\mathsf{d}}_{1}\ll1$ and $\hat{\mathsf{d}}_{2}=0$,
in which case the lowest eigenvalue is dominated by $\mathcal{M}_{d,0}$.
Then, diagonalization leads to
\begin{align*}
H_{c2}(\varphi_{B}) & =\frac{h_{1}(\varphi_{B})}{h_{2}(\varphi_{B})}\left[\sqrt{1\!+\!\big(|\hat{\boldsymbol{\Phi}}_{0}^{E_{g}}|^{2}\!-\!\hat{R}_{0}^{2}\big)\frac{h_{2}(\varphi_{B})}{[h_{1}(\varphi_{B})]^{2}}}-1\right],\\
h_{1}(\varphi_{B}) & =2\hat{R}_{0}+\hat{\mathsf{d}}_{1}|\hat{\boldsymbol{\Phi}}_{0}^{E_{g}}|\cos(\alpha_{0}+2\varphi_{B}),\\
h_{2}(\varphi_{B}) & =4-\hat{\mathsf{d}}_{1}^{2}-\kappa_{A2}^{2}|\hat{\boldsymbol{\Phi}}_{0}^{E_{g}}|^{2}\sin^{2}(\alpha_{0}-\varphi_{B}),
\end{align*}
which, at the superconducting transition $\hat{r}_{\Delta}=\hat{R}_{0}-|\hat{\boldsymbol{\Phi}}_{0}^{E_{g}}|\lesssim0$,
simplifies to
\begin{align}
H_{c2}(\varphi_{B}) & \approx\frac{-\hat{r}_{\Delta}}{2+\hat{\mathsf{d}}_{1}\cos(\alpha_{0}+2\varphi_{B})}.\label{eq:Hc2_app}
\end{align}
As expected, in this perturbative analysis in $\hat{\mathsf{d}}_{1}$,
the upper critical field has the shape of an ellipse with the long
axis along $-\alpha_{0}/2$. Contributions arising from $\hat{\mathsf{d}}_{2}$
will distort this ellipse and remove any symmetry with respect to
an $180^{\circ}$ rotation when $\alpha_{0}\neq\alpha_{s}$. Moreover,
the additional contribution from $\kappa_{A2}$ is only sub-leading
at the transition, and its effect is to enhance $H_{c2}$ for angles
$\varphi_{B}$ orthogonal to the nematic axis, i.e. to effectively
make the elliptical shape less pronounced. 

\section{Model Hamiltonian for $\mathrm{Bi_{2}Se_{3}}$
\label{sec:-model-Hamiltonian}}

In this Appendix, we write down the expression
for the superconducting $\boldsymbol{d}$-vector used in Section \ref{subsec:Nodal-vs.-fully-gapped}.
This derivation is based on the work of Ref. \citep{Venderbos2016a},
and uses the notation introduced in Ref. \citep{Hecker2020}. We start
from the mean-field decoupled superconducting Hamiltonian \citep{Zhang2009,Liu2010}
\begin{align}
\hat{\mathcal{H}} & =\sum_{\boldsymbol{k}}\!(\hat{\boldsymbol{c}}_{\boldsymbol{k}}^{\dagger})^{T}h_{\boldsymbol{k}}\,\hat{\boldsymbol{c}}_{\boldsymbol{k}}+\!\sum_{\boldsymbol{k}}\!\!\Big[(\hat{\boldsymbol{c}}_{\boldsymbol{k}}^{\dagger})^{T}\Delta(\boldsymbol{k})\,\hat{\boldsymbol{c}}_{-\boldsymbol{k}}^{\dagger}+H.c.\Big],\label{eq:H_BCS}
\end{align}
written in the electronic basis $\hat{\boldsymbol{c}}_{\boldsymbol{k}}=(\hat{c}_{\boldsymbol{k}1\uparrow},\hat{c}_{\boldsymbol{k}1\downarrow},\hat{c}_{\boldsymbol{k}2\uparrow},\hat{c}_{\boldsymbol{k}2\downarrow})^{T}$
in terms of the orbital $(1,2)$ and spin $(\uparrow,\downarrow)$
degrees of freedom. The non-interacting Hamiltonian is given by
\begin{align}
h_{\boldsymbol{k}} & =\sigma^{0}\mathfrak{s}^{0}\left(-\mu+C_{\boldsymbol{k}}\right)-\sigma^{y}\mathfrak{s}^{0}f_{\boldsymbol{k}}^{z}+\sigma^{z}\mathfrak{s}^{0}M_{\boldsymbol{k}}\nonumber \\
 & \quad+\sigma^{x}\left(\mathfrak{s}^{y}f_{\boldsymbol{k}}^{x}-\mathfrak{s}^{x}f_{\boldsymbol{k}}^{y}\right)+\sigma^{x}\mathfrak{s}^{z}f_{\boldsymbol{k}}^{C_{3}}\,,\label{eq:SP_Ham}
\end{align}
with the Pauli matrices $\sigma^{j}$, $\mathfrak{s}^{j}$ acting
in orbital and spin space, respectively. The first term in the second
line of Eq. (\ref{eq:SP_Ham}) represents a Rashba spin-orbit coupling,
whereas the last term, $f_{\boldsymbol{k}}^{C_{3}}$, accounts for
the threefold rotational symmetry of the crystal. Note that the band
dispersion has non-trivial topology as long as $M_{\boldsymbol{k}=0}<0$.
In the continuum description, the  functions in Eq. (\ref{eq:SP_Ham})
are given by:
\begin{align}
A_{1g} & : & M_{\boldsymbol{k}} & =M_{0}+M_{2}(\tilde{k}_{x}^{2}+\tilde{k}_{y}^{2})+M_{1}\tilde{k}_{z}^{2},\label{eq:Mk}\\
A_{1g} & : & C_{\boldsymbol{k}} & =C_{0}+C_{2}(\tilde{k}_{x}^{2}+\tilde{k}_{y}^{2})+C_{1}\tilde{k}_{z}^{2},\label{eq:Ck}\\
A_{2u} & : & f_{\boldsymbol{k}}^{z} & =v_{z}\tilde{k}_{z}+R_{2}\left(\tilde{k}_{y}^{3}-3\tilde{k}_{y}\tilde{k}_{x}^{2}\right),\label{eq:fzk}\\
E_{u} & : & \left(\begin{array}{c}
f_{\boldsymbol{k}}^{x}\\
f_{\boldsymbol{k}}^{y}
\end{array}\right) & =v_{0}\!\left(\!\!\!\begin{array}{c}
\tilde{k}_{x}\\
\tilde{k}_{y}
\end{array}\!\!\!\right)+\mathsf{d}_{2}^{E_{u}}\tilde{k}_{z}\!\left(\!\!\!\begin{array}{c}
2\tilde{k}_{x}\tilde{k}_{y}\\
\tilde{k}_{x}^{2}-\tilde{k}_{y}^{2}
\end{array}\!\!\!\right),\label{eq:fxyk}\\
A_{1u} & : & f_{\boldsymbol{k}}^{C_{3}} & =R_{1}\left(\tilde{k}_{x}^{3}-3\tilde{k}_{x}\tilde{k}_{y}^{2}\right),\label{eq:fC3k}
\end{align}
where we defined the dimensionless momentum $\tilde{\boldsymbol{k}}=\left(k_{x}a,k_{y}a,k_{z}c\right)$
and the lattice constants $a$ and $c$. For convenience, we include
above the irreducible representations according to which each function
transforms. While specific set of parameter values are available,
see Ref. \citep{Hecker2020}, they are not essential for our purposes. 

The superconducting gap function in Eq. (\ref{eq:H_BCS})
is assumed to be in the $E_{u}$ symmetry channel. As a result, it
is described in terms of the order parameter $\boldsymbol{\Delta}=(\Delta_{1},\Delta_{2})^{T}$
according to:
\begin{align}
\Delta(\boldsymbol{k}) & =\Delta_{1}\sigma^{y}\mathsf{i}\mathfrak{s}^{0}+\Delta_{2}\sigma^{y}\mathfrak{s}^{z}.\label{eq:Delta_mat}
\end{align}
In the presence of inversion and time-reversal symmetries, it is convenient
to change basis to the band space $\hat{\boldsymbol{\psi}}_{\boldsymbol{k}}=(\hat{\psi}_{\boldsymbol{k}c+},\hat{\psi}_{\boldsymbol{k}c-},\hat{\psi}_{\boldsymbol{k}v+},\hat{\psi}_{\boldsymbol{k}v-})$.
Here, the index $\pm$ behaves like a pseudospin $\frac{1}{2}$, whereas
the subscripts $c,v$ denote conduction and valence bands, respectively
\citep{Fu2015,Venderbos2016a}. The corresponding unitary matrix $U_{b}(\boldsymbol{k})$
that defines $\hat{\boldsymbol{\psi}}_{\boldsymbol{k}}=U_{b}^{\dagger}(\boldsymbol{k})\hat{\boldsymbol{c}}_{\boldsymbol{k}}$
is explicitly given in Ref. \citep{Hecker2020}. In this band basis,
the non-interacting Hamiltonian and the gap function become:
\begin{align}
h_{b}(\boldsymbol{k}) & =U_{b}^{\dagger}(\boldsymbol{k})h_{\boldsymbol{k}}U_{b}(\boldsymbol{k})=\text{diag}(E_{\boldsymbol{k}}^{+},E_{\boldsymbol{k}}^{+},E_{\boldsymbol{k}}^{-},E_{\boldsymbol{k}}^{-}),\label{eq:h_b}\\
\Delta_{b}(\boldsymbol{k}) & =U_{b}^{\dagger}(\boldsymbol{k})\Delta(\boldsymbol{k})U_{b}^{*}(-\boldsymbol{k}),\label{eq:Delta_b}
\end{align}
where $E_{\boldsymbol{k}}^{\pm}=-\mu+C_{\boldsymbol{k}}\pm\lambda_{\boldsymbol{k}}$,
$\lambda_{\boldsymbol{k}}=\sqrt{M_{\boldsymbol{k}}^{2}+\boldsymbol{f}_{\boldsymbol{k}}^{2}+(f_{\boldsymbol{k}}^{C_{3}})^{2}}$,
and $\boldsymbol{f}_{\boldsymbol{k}}=(f_{\boldsymbol{k}}^{x},f_{\boldsymbol{k}}^{y},f_{\boldsymbol{k}}^{z})^{T}$.
For $\mathrm{Bi_{2}Se_{3}}$ doped with $\mathrm{Cu}$, $\mathrm{Ni}$
or $\mathrm{Sr}$, the chemical potential moves into the conduction
band. As a result, the low-energy physics is well-described by the
conduction band states only. Thus, we employ the $2\times4$ projection
matrix $P_{c}=(\mathbbm{1}_{2},0_{2})$ to obtain the gap function
projected onto the conduction band:
\begin{align}
\Delta_{c}(\boldsymbol{k}) & =P_{c}\Delta_{b}(\boldsymbol{k})P_{c}^{T}=\boldsymbol{d}(\boldsymbol{k})\cdot\tilde{\mathfrak{\boldsymbol{s}}}(\mathsf{i}\tilde{\mathfrak{s}}^{y}).\label{eq:Delta_c}
\end{align}
In this expression, the $\boldsymbol{d}$-vector is given by
\begin{align}
\boldsymbol{d}(\boldsymbol{k}) & =\Delta_{1}\boldsymbol{d}_{1}(\boldsymbol{k})+\Delta_{2}\boldsymbol{d}_{2}(\boldsymbol{k}),\label{eq:d_vec_app}
\end{align}
with the two components 
\begin{align}
\boldsymbol{d}_{1}(\boldsymbol{k}) & =\left(\begin{array}{c}
\hat{M}_{\boldsymbol{k}}\hat{f}_{\boldsymbol{k}}^{C_{3}}\\
-\hat{f}_{\boldsymbol{k}}^{z}\\
\hat{f}_{\boldsymbol{k}}^{y}
\end{array}\right)+\mathrm{sign}(\hat{M}_{\boldsymbol{k}})\frac{\hat{f}_{\boldsymbol{k}}^{C_{3}}\hat{f}_{\boldsymbol{k}}^{x}}{1+|\hat{M}_{\boldsymbol{k}}|}\,\hat{\boldsymbol{f}}_{\boldsymbol{k}},\label{eq:d1_expr}\\
\boldsymbol{d}_{2}(\boldsymbol{k}) & =\left(\begin{array}{c}
\hat{f}_{\boldsymbol{k}}^{z}\\
\hat{M}_{\boldsymbol{k}}\hat{f}_{\boldsymbol{k}}^{C_{3}}\\
-\hat{f}_{\boldsymbol{k}}^{x}
\end{array}\right)+\mathrm{sign}(\hat{M}_{\boldsymbol{k}})\frac{\hat{f}_{\boldsymbol{k}}^{C_{3}}\hat{f}_{\boldsymbol{k}}^{y}}{1+|\hat{M}_{\boldsymbol{k}}|}\,\hat{\boldsymbol{f}}_{\boldsymbol{k}}.\label{eq:d2_expr}
\end{align}
In the equations above, we defined $\{\hat{M}_{\boldsymbol{k}},\hat{f}_{\boldsymbol{k}}^{j}\}=\{M_{\boldsymbol{k}},f_{\boldsymbol{k}}^{j}\}/\sqrt{M_{\boldsymbol{k}}^{2}+\boldsymbol{f}_{\boldsymbol{k}}^{2}}$
with $j=\{x,y,z\}$ and $\hat{f}_{\boldsymbol{k}}^{C_{3}}=f_{\boldsymbol{k}}^{C_{3}}/\lambda_{\boldsymbol{k}}$.
\end{document}